\documentclass[aps,
               pra,
               twocolumn,
               showpacs,
               amsmath,
               amssymb,
               superscriptaddress,
               reprint,
               10pt
              ]{revtex4-1}

\usepackage[colorlinks=true,breaklinks=true,allcolors=blue]{hyperref}
\usepackage{url}
\usepackage{txfonts}

\usepackage{amsmath, amssymb}

\usepackage{pstricks}
\usepackage{graphicx}
\usepackage{tikz}
\usepackage[utf8x]{inputenc}
\usepackage{color}
\usepackage{hyperref}

\usepackage{pstool}
\usepackage{pgf}
\usepackage{import}

\usetikzlibrary{snakes}
\usetikzlibrary{arrows}

\usepackage{tabularx}
\usepackage{multirow}
\usepackage{array}
\newcolumntype{L}[1]{>{\raggedright\let\newline\\\arraybackslash\hspace{0pt}}m{#1}}
\newcolumntype{C}[1]{>{\centering\let\newline\\\arraybackslash\hspace{0pt}}m{#1}}
\newcolumntype{R}[1]{>{\raggedleft\let\newline\\\arraybackslash\hspace{0pt}}m{#1}}

\DeclareMathOperator{\diag}{diag}
\DeclareMathOperator{\sgn}{sgn}

\renewcommand{\i}{\mathrm{i}}

\renewcommand{\vec}[1]{\mathbf{#1}}

\DeclareMathOperator{\Trace}{Tr}
\newcommand{\Tr}[1]{\Trace\left[{#1}\right]}

\newcommand{\eq}[1]{(\ref{eq:#1})}
\newcommand{\Eq}[1]{Eq.\,\eqref{eq:#1}}

\newcommand{\Fig}[1]{Fig.~\ref{fig:#1}}

\newcommand{\Sect}[1]{Sect.~\ref{sec:#1}}

\newcommand{\App}[1]{App.~\ref{app:#1}}

\makeatletter
\let\cat@comma@active\@empty
\makeatother

\begin{document}

\title{Low-energy effective theory of non-thermal fixed points in a multicomponent Bose gas}

\author{Aleksandr N. Mikheev}
\email{aleksandr.mikheev@kip.uni-heidelberg.de}
\affiliation{Kirchhoff-Institut f\"ur Physik, 
             Ruprecht-Karls-Universit\"at Heidelberg,
             Im Neuenheimer Feld 227, 
             69120 Heidelberg, Germany}

\author{Christian-Marcel Schmied}
\email{christian-marcel.schmied@kip.uni-heidelberg.de}
\affiliation{Kirchhoff-Institut f\"ur Physik, 
             Ruprecht-Karls-Universit\"at Heidelberg,
             Im Neuenheimer Feld 227, 
             69120 Heidelberg, Germany}
\affiliation{Dodd-Walls Centre for Photonic and Quantum Technologies,
             Department of Physics, 
             University of Otago, 
             Dunedin 9016, 
             New Zealand}

\author{Thomas~Gasenzer}
\email{t.gasenzer@uni-heidelberg.de}
\affiliation{Kirchhoff-Institut f\"ur Physik, 
             Ruprecht-Karls-Universit\"at Heidelberg,
             Im Neuenheimer Feld 227, 
             69120 Heidelberg, Germany}

\date{\today}

\begin{abstract}
Non-thermal fixed points in the evolution of a quantum many-body system quenched far out of equilibrium manifest themselves in a scaling evolution of correlations in space and time. We develop a low-energy effective theory of non-thermal fixed points in a bosonic quantum many-body system by integrating out long-wavelength density fluctuations. The system consists of $N$ distinguishable spatially uniform Bose gases with $\mathrm{U}(N)$-symmetric interactions. The effective theory describes interacting Goldstone modes of the total and relative-phase excitations. It is similar in character to the non-linear Luttinger-liquid description of low-energy phonons in a single dilute Bose gas, with the markable difference of a universal non-local coupling function depending, in the large-$N$ limit, only on momentum, single-particle mass, and density of the gas. Our theory provides a perturbative description of the non-thermal fixed point, technically easy to apply to experimentally relevant cases with a small number of fields $N$. Numerical results for $N=3$ allow us to characterize the analytical form of the scaling function and confirm the analytically predicted scaling exponents. The predicted and observed exponentially suppressed coherence at short distances takes the form of that of a quasicondensate in low-dimensional equilibrium systems. The fixed point which is dominated by the relative phases is found to be Gaussian, while a non-Gaussian fixed point is anticipated to require scaling evolution with a distinctly lower power of time.
\end{abstract}


\maketitle

\section{Introduction}
\label{sec:Intro}
Relaxation of quantum many-body systems after a quench far out of equilibrium has been  studied intensely during recent years. Little is known about the general structure of possible evolutions. Various scenarios have been proposed and observed, including prethermalization \cite{Gring2011a,Berges:2004ce}, generalized Gibbs ensembles \cite{Langen2015b.Science348.207,Jaynes1957a,Jaynes1957b}, many-body localization \cite{Schreiber2015a.Science349.842}, critical and prethermal dynamics \cite{Braun2014a.arXiv1403.7199B,Nicklas:2015gwa,Navon2015a.Science.347.167N,Eigen2018a.arXiv180509802E}, decoherence and revivals \cite{Rauer2017a.arXiv170508231R.Science360.307},  turbulence \cite{Navon2016a.Nature.539.72}, and the approach to a non-thermal fixed point \cite{Prufer:2018hto,Erne:2018gmz}.
The rich spectrum of different possible phenomena highlights the capabilities of quantum dynamics as compared to what is possible in classical statistical ensembles.

An important difference concerns the phase angle of the quantum mechanical wave function and the associated superposition principle. In the case of a quantum many-body system, the phase angle gives rise to interference effects and decoherence and encodes the collective dynamics of the fundamental field degrees of freedom. For example, long-range coherence and thus stiffness of the phase forms the basis of sound excitations on the top of a Bose-Einstein condensate of (weakly) interacting particles. This is related to local (quasi) particle number conservation reflecting a $\mathrm{U}(1)$ symmetry of the underlying model description.

Here we focus on universal scaling dynamics in the relaxation of a dilute Bose gas quenched far out of equilibrium. Universal dynamics depends on a few basic symmetry properties only and thus can be classified independently of the details of microscopic properties and initial conditions. Scaling dynamics has been discussed for classical systems almost as long as spatial scaling alone. From dynamical critical phenomena \cite{Hohenberg1977a,Janssen1979a} the discussion extended to coarsening and phase-ordering kinetics \cite{Bray1994a.AdvPhys.43.357}, glassy dynamics and ageing \cite{Calabrese2005a.JPA38.05.R133}, (wave-)turbulence \cite{Frisch1995a,Zakharov1992a,Nazarenko2011a}, and its variants in the quantum realm of superfluids \cite{Svistunov1991a,Kagan1992a,Kagan1994a,Kagan1995a,Semikoz1995a.PhysRevLett.74.3093,Semikoz1997a,Berloff2002a,Kozik2004a.PhysRevLett.92.035301,Kozik2005a.PhysRevLett.94.025301,Kozik2005a.PhysRevB.72.172505,Kozik2009a,Tsubota2008a, Vinen2006a}. Different types of pre\-thermal and universal dynamics after quenches of quantum many-body systems far out of equilibrium have been studied recently 
\cite{Berges:2004ce,Aarts2000a.PhysRevD.63.025012,Lamacraft2007.PhysRevLett.98.160404,Rossini2009a.PhysRevLett.102.127204,DallaTorre2013.PhysRevLett.110.090404,Gambassi2011a.EPL95.6,Sciolla2013a.PhysRevB.88.201110,Smacchia2015a.PhysRevB.91.205136,Smacchia2015a.PhysRevB.91.205136,Maraga2015a.PhysRevE.92.042151,Maraga2016b.PhysRevB.94.245122,Chiocchetta2015a.PhysRevB.91.220302,Chiocchetta2016a.PhysRevB.94.134311,Chiocchetta:2016waa.PhysRevB.94.174301,Chiocchetta2016b.161202419C.PhysRevLett.118.135701,Marino2016a.PhysRevLett.116.070407,Marino2016PhRvB..94h5150M,Damle1996a.PhysRevA.54.5037,Mukerjee2007a.PhysRevB.76.104519,Williamson2016a.PhysRevLett.116.025301,Hofmann2014PhRvL.113i5702H,Williamson2016a.PhysRevA.94.023608,Bourges2016a.arXiv161108922B.PhysRevA.95.023616}, many of them in the context of ultracold Bose gases.
Non-thermal fixed points have been proposed, without \cite{Berges:2008wm,Berges:2008sr,Scheppach:2009wu,Berges:2010ez,Orioli:2015dxa,Berges:2015kfa,Walz:2017ffj.PhysRevD.97.116011,Chantesana:2018qsb.PhysRevA.99.043620} and with \cite{Nowak:2010tm,Nowak:2011sk,Schole:2012kt,Nowak:2012gd,Karl:2013mn,Karl:2013kua,Karl2017b.NJP19.093014,Karl2017b.NJP19.093014,SchmiedPhysRevA.99.033611,Schmied:2019abm} reference to order-parameters, topological defects, and ordering kinetics, paving the way to a unifying description of universal dynamics.

\begin{figure*}
\includegraphics[width=0.96\textwidth]{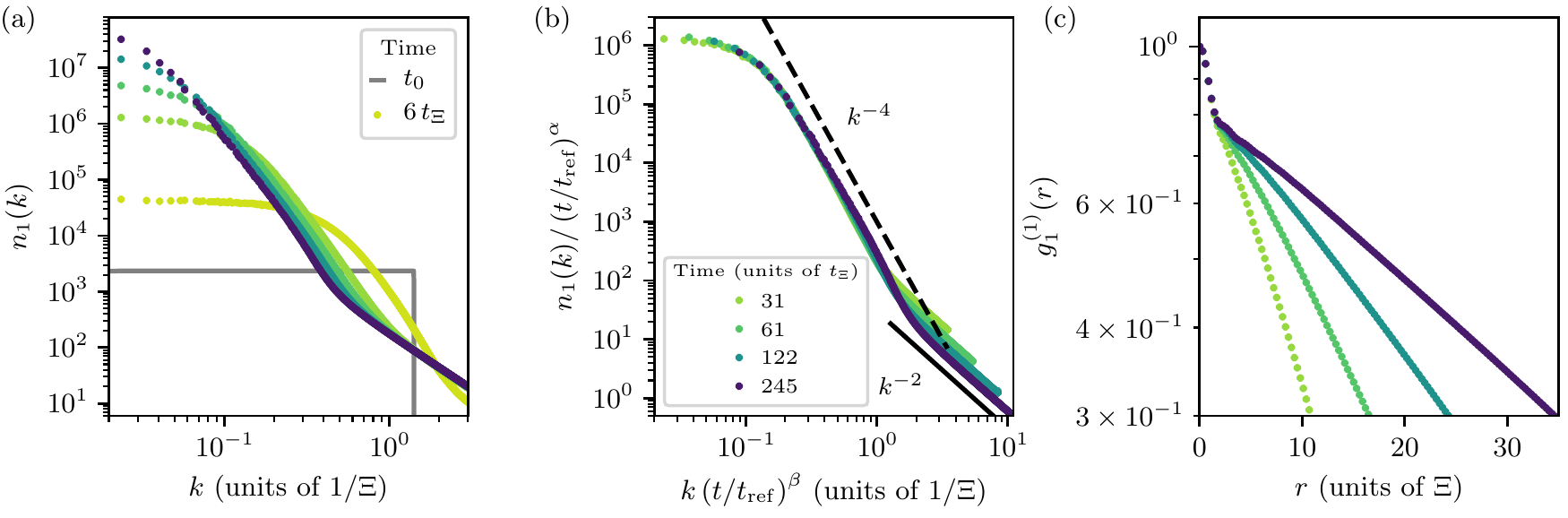}
\caption{\label{fig:NTFPEvolution} 
(a) Universal scaling dynamics of the single-component occupation number $n_{1}(k)\equiv n_{1}(\mathbf{k},t)$ according to \eq{NTFPscaling}, for a $(d=3)$-dimensional gas with $N=3$ components.
For details of the Truncated-Wigner simulations see \Sect{TWA}.
The time evolution is starting from the initial distribution $n_{1}(\mathbf{k},t_{0})=n_{0}\,\Theta(k_\mathrm{q}-|\mathbf{k}|)$, identical in all three components $a$ (grey solid line), with $n_{0}=(4\pi k_{q}^{3})^{-1}\rho^{(0)}$, $k_\mathrm{q}=1.4\,k_{\Xi}$
($k_{\Xi} = \Xi^{-1}=[2 m \, \rho^{(0)} g]^{1/2}$).
Colored dots show $n_1$ at five different times.
(b) The collapse of the data to the universal scaling function $f_{\mathrm{S},1}(\mathbf{k})=n_{1}(\mathbf{k},t_\mathrm{ref})$, with reference time $t_\mathrm{ref}/t_{\Xi}=31$ (units of $t_{\Xi}^{-1}=g\rho^{(0)}/2\pi$), shows the scaling \eq{NTFPscaling} in space and time.
For the time window $t_{\mathrm{ref}}=200\,t_{\Xi} < t <  350\,t_{\Xi}$, we extract exponents $\alpha=1.62\pm0.37$, $\beta=0.53\pm0.09$.
(c) Universal scaling dynamics of the single-component first-order coherence function $g^{(1)}_{1}(r)=g^{(1)}_{1}(\mathbf{r},t)$ (colored dots), for the same system and the same color encoding of $t$ as in (a) and (b).
At short distances $r \gtrsim \Xi$ the first-order coherence function takes an exponential form, reminiscent of a quasicondensate, which is characterized by a correlation length rescaling in time. Note the semi-log scale.
}
\end{figure*}

A major part of the analytical work on non-thermal fixed points is based on scalar model systems with quartic interactions between $N$-component Bose fields, $\mathrm{O}(N)$-symmetric under orthogonal transformations in the space of field components. A non-perturbative large-$N$ approximation \cite{Berges:2001fi,Berges:2004yj} allows for a description of scaling at non-thermal fixed-points \cite{Berges:2008wm,Berges:2008sr,Scheppach:2009wu,Berges:2010ez,Orioli:2015dxa,Berges:2015kfa,Chantesana:2018qsb.PhysRevA.99.043620}.

Here we develop a low-energy effective field theory (EFT) of such systems. We describe the linear phase-angle excitations around ground states with broken $\mathrm{U}(N)$ symmetry. Their non-linear bare interactions are, in general, non-local and, at the fixed point, characterized by the momentum-dependent coupling which scales analogously to the resummed couplings in the non-perturbative theory. We use this to outline a complementary approach to non-thermal fixed points which is based on a leading-order coupling expansion, practically applicable for small $N\ge1$, and discuss consequences for $N=1$. 

We furthermore study numerically universal scaling dynamics in an $N=3$ component $(d=3)$-dimensional dilute Bose gas to test our analytical predictions. Our simulations corroborate the predicted scaling exponents and at the same time point to modifications of the pure scaling behavior at relatively early evolution times. While the non-thermal fixed point is conjectured to be approached asymptotically in time, prescaling of the short-distance correlations which are more easily accessible in experiment, can be seen at much earlier evolution times \cite{Schmied:2018upn.PhysRevLett.122.170404}.

Our paper is organized as follows:
In the remainder of this section, we introduce the model system under consideration. In \Sect{LEEFT}, we develop the non-linear low-energy effective theory and discuss the limit of infinitely many number of components $N \to \infty$. In this limit, in \Sect{Scaling}, we make predictions for the scaling at a non-thermal fixed point using the kinetic equation derived in the same section. Finally, in \Sect{TWA}, we show the numerical results for $N = 3$ component Bose gas in $d = 3$ dimensions, partially reproducing data from \cite{Schmied:2018upn.PhysRevLett.122.170404}, and discuss the central properties of the fixed point. We summarize and draw our conclusions in \Sect{Conclusions}. The appendices contain further details.
  
\subsection{Model}
We consider a system of $N$ spatially uniform Bose gases of identical particles. The different gases are distinguished by, e.g., the hyperfine magnetic level the gas atoms are in.
They are described by a $\mathrm{U}(N)$-symmetric Gross-Pitaevskii (GP) model with quartic contact interaction in the total density,
\begin{equation}
 H_{\mathrm{U}(N)} =   \int  \mathrm{d}^dx \, \left[ -\Phi_{a}^\dag\frac{\nabla^2}{2m}\Phi_{a}  +  \frac{g}{2} \, (\Phi_{a}^\dag\Phi_{a})^{2} \right] .
\label{eq:ONGPH}
\end{equation}
Here and in the following, we use units implying $\hbar=1$, space-time field arguments are suppressed, and $m$ is the particle mass.
Summation over the $a=1,...,N$ Bose fields which are obeying standard commutators $[\Phi_{a}(\mathbf{x},t),\Phi_{b}^\dag(\mathbf{y},t)]=\delta_{ab}\delta(\mathbf{x}-\mathbf{y})$ is implied.
The identical interspecies and intraspecies contact interactions are parametrized by the coupling $g$. As a result, the model exhibits a full $\mathrm{U}(N)$ symmetry under unitary transformations $\mathcal{U}\in \mathrm{U}(N)$ of the fields, $\Phi_{a}\to \mathcal{U}_{ab}\Phi_{b}$. The generalization to inhomogeneous systems in a trap is possible but disregarded here.

\subsection{Universal scaling dynamics at a non-thermal fixed point}
The nearly condensed gas is assumed to exhibit long-range order in the total phase, while domain walls and topological defects of any kind are assumed to be subdominant during the time evolution. 
Relative phases, though, can be strongly excited, representing particlelike and holelike Goldstone excitations with single-particle dispersion as further discussed below. 

This can be achieved in a system with sufficiently large $N$ by, e.g., a strong cooling quench or an instability: 
An extreme version of a cooling quench would be to first tune adiabatically to a chemical potential $0 < -\mu \ll k_B T_c$, with the condensation temperature $T_c$, and then remove all particles with energies higher than a certain energy scale $\omega(k_q) \sim |\mu |$ \cite{Chantesana:2018qsb.PhysRevA.99.043620}. 
Such an extreme initial condition, in experiment, can alternatively be prepared by means of an instability \cite{Berges:2008wm,Prufer:2018hto}.
In both cases, the crucial condition is to build up strong over\-occupation such that the majority of particles and energy is around the small but non-zero momentum scale $k_{q}$, ideally close to the healing-length scale set by the chemical potential $\mu$ \cite{Chantesana:2018qsb.PhysRevA.99.043620}. 
This induces a far-from-equilibrium evolution starting from modes with the comparatively high excitation energy $\omega(k_{q})$. 

During this evolution, the majority of particles is transported to lower momenta while the major part of the energy is deposited by a few particles being scattered to even higher-momentum modes, where they eventually form a thermal tail.
The highly occupied modes which take up most of the particles have excitations energies $\omega(k)$ below the scale set by $\mu$. This allows us to use a low-energy effective theory description.

From this type of initial conditions, the system can approach a non-thermal fixed point and show universal scaling evolution \cite{Orioli:2015dxa,Berges:2015kfa,Chantesana:2018qsb.PhysRevA.99.043620}. This universal scaling in space and time at the fixed point is expected, e.g., in the occupation numbers $n_{a}(\mathbf{k},t)=\langle\Phi_{a}^{\dag}(\mathbf{k},t)\Phi_{a}(\mathbf{k},t)\rangle$ of the Bose field excitations in each component $a$ in momentum space, in the form of
\begin{equation}
n_{a}(\mathbf{k},t) = (t/t_\mathrm{ref})^{\alpha}n_{\mathrm{S},a}([t/t_\mathrm{ref}]^{\beta}\mathbf{k}).
 \label{eq:NTFPscaling}
\end{equation}
The scaling function $n_{\mathrm{S},a}(\mathbf{k})=n_{a}(\mathbf{k},t_\mathrm{ref})$ depends only on a single argument, and defines the scaling form together with the exponents $\alpha$, $\beta$. 
The reference time $t_\mathrm{ref}$ is chosen within the scaling regime.

\begin{figure}
\centering
\includegraphics[width=0.7\columnwidth]{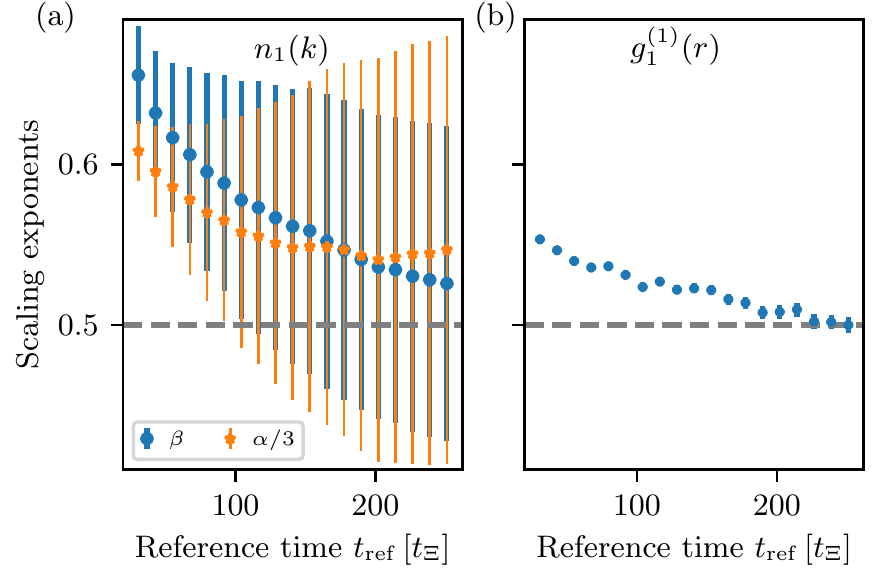}
\caption{Scaling exponents $\alpha/3$ (orange stars) and $\beta$ (blue dots) obtained from least-square rescaling fits of the occupancy spectra $n_{1}(k)\equiv n_1(\mathbf{k},t)$ shown in \Fig{NTFPEvolution}b. 
The exponents correspond to the mean required to collapse the spectra within the time window $[t_\mathrm{ref},t_\mathrm{ref}+\Delta t]$ with $\Delta t = 146\,t_{\Xi}$.
Error bars denote the least-square fit error.
\label{fig:WindowFit1}
}
\end{figure}

As an example, we study, in \Fig{NTFPEvolution}, the evolution of a Bose gas with $N=3$ components in $d=3$ dimensions, starting from a far-from-equilibrium initial condition (grey `box' in (a)) with momentum distributions $n_{a}(\mathbf{k},t_{0})=n_{0}\Theta(k_\mathrm{q}-|\mathbf{k}|)$, $a=1,2,3$, while all phases $\theta_{a}(\mathbf{k},t_{0})$ are random.
The data shown here and in the following are partially reproduced from \cite{Schmied:2018upn.PhysRevLett.122.170404}. The figure shows that, for times $t\gtrsim t_\mathrm{ref}=31\,t_{\Xi}$, and within a limited range of low momenta, the evolution of the momentum distribution exhibits approximate scaling according to \eq{NTFPscaling}.  Rescaling the momentum occupation spectra at different times they fall onto a single universal scaling function as shown in \Fig{NTFPEvolution}b, in accordance with the findings of \cite{Orioli:2015dxa} for the case $N=1$. At late evolution times, $t_\mathrm{ref}=200\,t_{\Xi}<t<350\,t_{\Xi}$, we extract $\alpha=1.62\pm0.37$, $\beta=0.53\pm0.09$, see \Fig{WindowFit1}.

The scaling function is characterized by a plateau up to an inverse coherence-length scale $k_{\Lambda}(t)$, which rescales in time according to $k_{\Lambda}(t) \sim t^{-\beta}$. At momenta larger than this inverse coherence-length scale, $|\mathbf{k}|=k\gg k_{\Lambda}(t)$, the scaling function takes the power-law form $n_{a}(k)\sim k^{-\zeta}$, with $\zeta\simeq d+1=4$, confirming earlier predictions \cite{Orioli:2015dxa,Walz:2017ffj.PhysRevD.97.116011,Chantesana:2018qsb.PhysRevA.99.043620}.

Taking the Fourier transform of the momentum distribution $n_a(k)$ one obtains the first-order spatial coherence function $g^{(1)}_{a}(\mathbf{r},t)=\langle\Phi_{a}^{\dag}(\mathbf{x}+\mathbf{r},t)\Phi_{a}(\mathbf{x},t)\rangle$,
which, at short distances $\Xi\lesssim r\ll V^{-1/3}$ (with volume $V$), takes an exponential form $g^{(1)}(\mathbf{r},t) \sim e^{-k_{\Lambda}(t)\,|\mathbf{r}|}$ (see \Fig{NTFPEvolution}c). Here, $\Xi=  [2 m \, \rho^{(0)} g]^{-1/2}$ denotes the healing length corresponding to the total density. The exponential form at comparatively short distances is reminiscent of the build-up of a quasicondensate with a rescaling coherence length scale.

The scaling evolution in the vicinity of a non-thermal fixed point in a dilute Bose gas has been described in terms of kinetic equations with many-body $T$-matrices derived from a non-perturbative approximation of the underlying field-theoretic equations of motion \cite{Orioli:2015dxa,Chantesana:2018qsb.PhysRevA.99.043620}. 
As shown in \cite{Chantesana:2018qsb.PhysRevA.99.043620}, the non-perturbative collisional properties of the bosons become relevant in the low-energy region of momenta below the healing-length scale $k_{\Xi} = \Xi^{-1}$, i.e., for momenta $k\ll k_{\Xi}$, where the occupation number is strongly overoccupied (cf.~Figs.~\ref{fig:NTFPEvolution}a and \ref{fig:NTFPEvolution}b) and exhibits momentum scaling $n_{a}(k)\sim k^{-\zeta}$ for momenta above $k_{\Lambda}$ which defines the transition from the plateau to the power-law fall-off.

Here, we present an alternative approach in which we first reformulate the theory in terms of phase excitations only which are the relevant degrees of freedom at momentum scales $k\ll k_{\Xi}$. 
This leads to a low-energy effective theory which  takes the form of a non-linear Luttinger liquid, with density fluctuations integrated out at quadratic order, inducing cubic and quartic interactions of the phase excitations. 
The theory is used to obtain a Boltzmann-type kinetic equation with a $T$-matrix evaluated in leading perturbative order.
Close to the non-thermal fixed point, where the far-from-equilibrium dynamics is dominated by the over-occupied low-energy excitations at $k\ll k_{\Xi}$, this perturbative approximation remains valid because the resulting $T$-matrix is power-law suppressed at the low momenta transferred between these modes.
Similar to equilibrium systems in $d<3$ dimensions we find quasicondensate-type coherence close to the non-thermal fixed point, characterized by $g^{(1)}(\mathbf{r},t) \sim e^{-k_{\Lambda}(t)\,|\mathbf{r}|}$ of the time evolving system in three spatial dimensions.

\section{Low-energy effective theory}
\label{sec:LEEFT}
In this section, we provide details of the derivations of the low-energy effective theory for the $\mathrm{U}(N)$-component dilute Bose gas \eqref{eq:ONGPH} with $\mathrm{U}(N)$-symmetric contact interactions, which forms the central result of this paper.
%
\subsection{Collective excitations}
\label{sec:Classical}
We start our analytical discussion of the universal scaling dynamics with a brief summary of the collective low-energy linear excitations of the model \eq{ONGPH}. From \Eq{ONGPH}, one obtains the classical action in terms of fluctuating complex fields $\varphi_{a}(x)$, with $x\equiv(\mathbf{x},x^0)\equiv(\mathbf{x},t)$, 
\begin{align}
\label{eq:ONGPA}
  S[\varphi] =& \int_{x} \Bigg\lbrace\frac{\i}{2}\Big[\varphi_a^{*} (x) \partial_t \varphi_a (x) - \varphi_a (x) \partial_t \varphi_a^{*} (x) \Big] 
  \nonumber\\
  &  
  - \frac{1}{2 m}\Big[\nabla \varphi_a^{*} (x) \Big]\cdot\Big[\nabla \varphi_a (x) \Big] 
  - \frac{g}{2}\Big[\varphi_a^{*} (x) \varphi_a (x) \Big]^2\Bigg\rbrace \,,
\end{align}
where $\int_{x}=\int\mathrm{d}^{d+1}x$ and sums over double indices are implied.
Using the Madelung field representation in terms of polar coordinates \cite{Madelung1926a},
\begin{equation}
\label{eq:Madelung}
{\varphi}_a (\mathbf{x},t) = \sqrt{\rho_a(\mathbf{x},t)}\, \exp\big\{\i \theta_a (\mathbf{x},t)\big\},
\end{equation}
with densities $\rho_{a}$ and phases $\theta_{a}$, the Lagrangian reads as
\begin{align}
\label{eq:Madelung_Lagr}
  \mathcal{L} 
  = -\sum_a \Bigg\lbrace  \rho_a \partial_t \theta_a 
   + \frac{1}{2m} \left[ \rho_a (\nabla \theta_a)^2  + (\nabla \!  \sqrt{\rho_a})^2 \right] \Bigg\rbrace 
   - \frac{g}{2} \rho^{2},
\end{align}
where $\rho=\sum_{a}\rho_{a}$.
From this, a continuity equation relating the density to the particle current $\mathbf{j}_{a}=\rho_{a}\partial_{\mathbf{x}}\theta_{a}/m$, and an equation for the phase $\theta_{a}$ follow, which, in the limit of small fluctuations $\theta_{a}$ and $\delta\rho_{a}=\rho_{a}-\rho_{a}^{(0)}$ about the uniform ground-state densities $\rho_{a}^{(0)}= \langle \Phi_{a}^{\dagger}(x)\Phi_{a}(x) \rangle$~\footnote{Expanding around the time-independent homogeneous solution has to be distinguished from expanding around the condensate mean field which, during the evolution in the vicinity of an NTFP, typically grows algebraically. The density corresponding to the field expectation value squared needs to be distinguished from the zero mode of the total density which is proportional to the total particle number and given by the sum of the condensate density and the integral over all non-zero momentum occupancies, \protect{$\rho^{(0)}_a = |\langle\Phi_a\rangle|^2 + \sum_{\vec k}\langle\Phi^{\dagger}_a (\vec k) \Phi_a(\vec k)\rangle$}. Hence, conservation of the total particle number in a closed system ensures the constancy of $\rho^{(0)}_a$.
Furthermore, one should note that, in the presence of a condensate in the ground state of, e.g., a 3D system, one would have \protect{$\rho_{a}^{(0)}= |\langle \Phi_{a}(x) \rangle|^{2}$}, while the approach here also applies to lower-dimensional systems where, in equilibrium, the ground state bears a quasi-condensate.},
reduce to the linearized equations of motion
\begin{align}
\label{app:dtdeltarhoa}
  \partial_t \theta_a &= \frac{1}{4 m \rho_a^{(0)}} \nabla^2 \delta \rho_a - g \sum_b \delta \rho_b\,, \quad
  \partial_t \delta \rho_a = -\frac{\rho_a^{(0)}}{m} \nabla^2 \theta_a.
\end{align}  
In Fourier space, taking a further time derivative, they can be combined to the Bogoliubov-type wave equation for the $\theta_{a}$,
\begin{equation}
\partial_t^2 \theta_a(\mathbf{k},t) + \frac{\mathbf{k}^2}{2m} \left(\frac{\mathbf{k}^2}{2m} \delta^{ab} + 2 g \rho^{(0)}_b \right) \theta_b (\mathbf{k},t) = 0\,,
\end{equation}
where Einstein's sum convention is implied. 
While, for $N=1$, we recover the Bogoliubov dispersion, for general $N$, diagonalization of the coefficient matrix yields the eigenfrequencies of $N-1$ Goldstone and one Bogoliubov mode~\footnote{Obviously, all of these modes are gapless Goldstone excitations. Nevertheless, in order to distinguish them, hereinafter we will only refer to the quadratic modes as Goldstone ones, whereas the linear one will be addressed as Bogoliubov.}, 
\begin{align}
  \omega_c(\mathbf{k}) 
  &\equiv\omega_{\mathrm{G}}(\mathbf{k}) 
  = \frac{\mathbf{k}^2}{2m}, \quad c=1,\ldots,N-1\,,
  \label{eq:GoldstoneFreq}
  \\
  \omega_N (\mathbf{k}) 
  &\equiv\omega_{\mathrm{B}}(\mathbf{k})
  = \sqrt{\frac{\mathbf{k}^2}{2m} \left(\frac{\mathbf{k}^2}{2m} + 2 g \rho^{(0)} \right)}\,,
  \label{eq:BogHiggsFreq}
\end{align}
where $\rho^{(0)}=\sum_{a}\rho_{a}^{(0)}$ is the total condensate density (see Appendix~\ref{app:Mab} for details). Note that the Goldstone theorem \cite{Goldstone:1961eq} predicts, due to the spontaneous breaking of the $\mathrm{U}(N) \to \mathrm{U}(N-1)$, $2N - 1$ gapless Goldstone modes. However, only $N$ of these modes, with frequencies \eq{GoldstoneFreq} and \eq{BogHiggsFreq}, are independent due to the absence of Lorentz invariance and thus particle-hole symmetry \cite{Watanabe2012a.PhysRevLett.108.251602,Watanabe2014a.PhysRevX.4.031057,Pekker2015aARCMP.6.269P}.

The general solution can be written as a linear combination of the eigenmodes, 
\begin{equation}
  \theta_{a}(\mathbf{k},t)
  = \sum_{b=1}^{N} C^{(a)}_{b}\,{e}_{ba}\,\cos \left(\omega_b (\mathbf{k})\, t + \vartheta^{(a)}_b \right)\,,
\end{equation}
where ${e}_{ba}=(\mathbf{e}_{b})_{a}$ are the eigenvector components \eqref{eq:app:GoldstoneEV} and \eqref{eq:app:GoldstoneBog}, and $C^{(a)}_{b}$ and $\vartheta^{(a)}_{b}$ are real constants defined by the initial conditions. We find that the free-particle excitations are the relative phases between the different components, while the Bogoliubov excitation is in the total phase. 

We can always choose the eigenbasis such that all modes (equations) decouple due to the $\mathrm{U}(N)$ symmetry of the action. On the other hand, in experimentally realistic scenarios, this symmetry may be explicitly broken by the initial density matrix $\hat{\rho}_0$ which fixes the mean densities $\rho^{(0)}_a$. Therefore, it is suggestive to stick with the initial ``physical'' basis (see also the discussion in Appendix~\ref{app:SSB}). The solution for the density fluctuations has the same form, i.e., 
\begin{equation}
  \delta \rho_{a}(\mathbf{k},t)
  = \sum_{b=1}^{N} D^{(a)}_{b}\,{e}_{ba}\,\cos \left(\omega_b (\mathbf{k})\, t + \phi^{(a)}_b \right).
  \label{eq:deltarhoasolution}
\end{equation}
Taking the time derivative of \eq{deltarhoasolution}, we notice that, at low energies, i.e., for $k \ll k_\Xi$, where 
\begin{align}
\label{eq:app:kXi}
  k_{\Xi} = [2 m \, \rho^{(0)} g]^{1/2}
\end{align}
is a momentum scale set by the inverse healing length corresponding to the total density, the Bogoliubov mode contribution to the time derivative of fluctuations dominates, i.e.,
\begin{equation}
\partial_t \delta \rho_a (\mathbf{k},t) \sim \omega_N (\mathbf{k},t) \delta \rho_a (\mathbf{k},t).
\end{equation}  
Then, according to \eqref{app:dtdeltarhoa}, below $k_\Xi$, 
\begin{equation}
  \frac{\delta \rho_a(\mathbf{k})}{\rho^{(0)}_a} 
  \sim \frac{|\mathbf{k}|}{k_\Xi} \theta_a(\mathbf{k}) 
  \ll \theta_a(\mathbf{k})
  \quad (k \ll k_\Xi).
  \label{eq:CondLEEFT}
\end{equation}
Here we assumed that, at low energies, 
\begin{equation}
\left|D^{(a)}_N \omega_N(\mathbf{k}) \sin \phi^{(a)}_N \right| \gtrsim \left|D^{(a)}_b \omega_b(\mathbf{k}) \sin \phi^{(a)}_b \right| \,,
\end{equation}
i.e., that either both terms in $\partial_t \delta \rho_a$ are of the same order or that the Bogoliubov term dominates, which  depends, in general, on the initial conditions. 

\subsection{Luttinger-liquid-type effective action}
\label{sec:luttingerleeft}
%
\subsubsection{Derivation of the effective action}
In order to demonstrate the implications of the collective excitations introduced above for the possible non-equilibrium dynamics of the system we use the density-phase representation to obtain a low-energy effective field theory of the model \eq{ONGPH} with all $\rho^{(0)}_{a}>0$.
As for a single-component GP system, the quartic interactions proportional to the square of the local total density $\rho=\sum_{a}\varphi_{a}^*\varphi_{a}$ give rise to a reduction of density fluctuations at low energies. While this does not necessarily imply a suppression of the fluctuations of the single-species densities below some energy scale, such a suppression may be generated dynamically for certain initial conditions. In this section, we will derive the effective theory with a single-species density fluctuations suppression, while the more general case is considered in App.~\ref{app:NLSM}.

To begin with, we again adopt the Madelung representation and expand the Lagrangian \eqref{eq:Madelung_Lagr} up to the second order in density fluctuations. The resulting action reads:
\begin{align}
\label{eq:app:Sthetadelrho}
  S[\theta, \delta \rho] 
  = S_\mathrm{G}[\theta, \delta \rho]+ {S}_\mathrm{nG}[\theta, \delta \rho]\,,
\end{align}
where
\begin{align}
  S_\mathrm{G}
  = -\int_{x} \Bigg\lbrace \mathcal{J}_a \delta \rho_a& 
  + \frac{1}{2} \delta\rho_a g^{ab} \delta\rho_b 
  + \rho_a^{(0)} \left[\partial_t \theta_a 
  + \frac{\left(\nabla  \theta_a \right)^2}{2 m}\right]  \Bigg\rbrace\,,
  \\
  \mathcal{J}_a 
   &= g \rho^{(0)} + \partial_t \theta_a + \frac{1}{2 m} \left(\nabla \theta_a \right)^2\,, 
  \label{eq:app:J_a}
  \\
  g^{ab} 
    &= g - \frac{\delta^{ab}}{4 m \rho_a^{(0)}} \nabla^2\,, 
  \label{eq:app:g_ab}
  \\
  {S}_\mathrm{nG} 
   &= \int_x \frac{(\nabla \delta\rho_a)^2}{8 m\rho_a^{(0)}}  \left(\frac{\delta\rho_a}{\rho_a^{(0)}}\right)
   +\text{h.o.t.}\,
   \label{eq:app:SNG}
\end{align}
and we introduced $\rho^{(0)} \equiv \sum_a \rho_a^{(0)}$.
Note that the neglected higher-order terms (h.o.t.) involve powers of the density fluctuations only, while all phase-dependent terms have been taken into account. Hence, there is no approximation made concerning the size of the phase fluctuations. This is relevant, e.g., for Bose gases in $d < 3$ dimensions where even below the BEC crossover the quasicondensate exhibits large phase fluctuations.

As it was mentioned above, the derivation of the low-energy effective action consists of integrating out density fluctuations. This can be formally done adopting a Feynman-Vernon influence functional approach \cite{Feynman:1963fq} explained in App.~\ref{app:EFT_general}. In this paper, for the sake of simplicity, we assume that the initial state can be well-described by a product state of a (Gaussian) density matrix of phase fluctuations and a ground-state density fluctuations matrix, i.e., 
\begin{equation}
\hat{\rho}(t_0) \approx \hat{\rho}_{\delta \rho} (t_0) \times \hat{\rho}_{\theta} (t_0)\,, \quad \rho_{\delta \rho} (t_0) \approx |\Omega \rangle \langle \Omega |.
\end{equation}
While this approximation might look extreme, it should be
noted that the kinetic theory which we are going to use in the following disdains details of the initial conditions regardless.

With the aforementioned approximations, the derivation of the low-energy effective action is reduced to the standard procedure within zero-temperature quantum field theory:
\begin{equation}
\label{eq:Seff_def}
Z = \int D \theta \, D \delta \rho e^{\i S[\theta,\delta \rho]} \equiv \int D \theta \, e^{\i S_{\mathrm{eff}}[\theta]}.
\end{equation}
In general, the integral \eqref{eq:Seff_def} is non-Gaussian and thus cannot be computed exactly. We can, nevertheless, compute it perturbatively,
\begin{equation}
\int D \delta \rho \, e^{\i S_{\mathrm{G}} + \i S_{\mathrm{nG}}} = Z_0 \sum_{n=0}^{\infty} \frac{\i^n \langle (S_{\mathrm{nG}})^n \rangle}{n!},
\end{equation}
where $Z_0 = \int D \delta \rho \, e^{\i S_{\mathrm{G}}}$, and restrict ourselves to some finite number of terms in the above expression. For instance, at the lowest order:
\begin{equation}
\label{eq:Seff_gaus}
\exp \left(\i S_{\mathrm{eff}}[\theta]\right) \approx \int D \delta \rho \, \exp \left(\i S_{\mathrm{G}}[\theta,\delta \rho]\right).
\end{equation}
Equivalently, one can truncate the expansion \eqref{eq:app:Sthetadelrho} at the Gaussian order from the very beginning, also resulting in \eqref{eq:Seff_gaus}.

Since the kernel \eq{app:g_ab} contains a spatial derivative, it is more convenient to proceed in Fourier space (see \App{Not} for details of the notation):
\begin{align}
\label{eq:app:SGp}
  &S_\mathrm{G}[\theta,\delta\rho]
  = -\frac{1}{2}\int_{\substack{\mathbf{k},t \\ \mathbf{k}',t'}}  
   \delta\rho_a(\mathbf{k},t) g^{ab}(\mathbf{k},t;\mathbf{k}',t') \delta\rho_b(\mathbf{k}',t')   
  \nonumber\\
  &
  - \int_{\mathbf{k},t} \left\lbrace \mathcal{J}_a (\mathbf{k},t) \delta\rho_a(\mathbf{k},t) 
  + \frac{\rho_a^{(0)}}{2 m} \mathbf{k}^2 \theta_a(\mathbf{k},t) \theta_a(-\mathbf{k},t) \right\rbrace\,,
\end{align}
where the total-derivative term $\rho^{(0)} \partial_t \theta_{a}$ is dropped, and where
\begin{align}
  \mathcal{J}_a(\mathbf{k},t) 
  &=\  g\rho^{(0)} \, (2 \pi)^d \delta (\mathbf{k}) + \partial_t \theta_a (-\mathbf{k},t)  
  \nonumber\\
  + \frac{1}{2 m}&\int_{\mathbf{k}'} \mathbf{k}' ( \mathbf{k} 
  + \mathbf{k}') \theta_a(\mathbf{k}',t) \theta_a(-\mathbf{k} - \mathbf{k}',t)\,,
  \label{eq:app:Jk}
  \\
  g^{ab}(\mathbf{k},t;\mathbf{k}',t') 
  =&\ g\left(1 +  \frac{\delta^{ab}\,\mathbf{k}^2}{2k_{\Xi,a}^2}\right)(2\pi)^d \delta (\mathbf{k} + \mathbf{k}') \,\delta (t-t').
  \label{eq:app:gkk}
\end{align}
Here $k_{\Xi,a} = [2 m \, \rho_a^{(0)} g]^{1/2}$ is a momentum scale taking the form of the inverse healing length of a single component.
We recall that no condensate is required such that the interpretation of $1/k_{\Xi,a}$ as a healing length is in general not applicable.

We can absorb the first term in the definition of $\mathcal{J}_a$, \Eq{app:Jk}, by going over to a (grand-)canonical formulation, effectively shifting the energy of the zero-mode by a constant, 
\begin{equation}
  g\rho^{(0)} \, (2 \pi)^d \delta (\mathbf{k}) + \partial_t \theta_a (-\mathbf{k},t)
  \to \partial_t \theta_a (-\mathbf{k},t).
\end{equation}
In our path-integral formulation in terms of the fundamental Bose fields this can be achieved by shifting all densities by a background term, see App.~\ref{app:SSB}. As a result of the above, the current field simplifies to
\begin{align}
  \mathcal{J}_a(\mathbf{k},t) 
  &=\  \partial_t \theta_a (-\mathbf{k},t)  
  \nonumber\\
  + \frac{1}{2 m}&\int_{\mathbf{k}'} \mathbf{k}' ( \mathbf{k} 
  + \mathbf{k}') \theta_a(\mathbf{k}',t) \theta_a(-\mathbf{k} - \mathbf{k}',t).
  \label{eq:app:Jkshifted}
\end{align}

Note that the operator \eq{app:gkk} is diagonal both in $\mathbf{k}$-space and in $t$-space so that the only non-trivial step is to invert the matrix part of the kernel.

Before we perform the Gaussian integral over the $\delta\rho_{a}$ to obtain the effective action, we rescale the density fluctuations,
\begin{equation}
  \delta \rho_a \rightarrow \sqrt{\frac{N \rho_a^{(0)}}{\rho^{(0)}}} \delta \rho_a'\,,
\end{equation}
which multiplies the path-integral measure by an irrelevant constant. 
With this, we obtain a modified kernel in \eq{app:SGp}, 
\begin{align}
\label{eq:app:G_mod}
  \tilde{g}^{ab}(\mathbf{k},t;\mathbf{k}',t') 
  =\ &N g \left[\frac{({\rho_a^{(0)} \rho_b^{(0)}})^{1/2}}{\rho^{(0)}} +  \frac{\delta^{ab}\,\mathbf{k}^2}{2 k^2_{\Xi}}\right]
  \nonumber\\
  &\times\ (2\pi)^d \delta (\mathbf{k} + \mathbf{k}') \,\delta (t-t')\,
\end{align}
and also need to take into account the rescaling in the linear term, $\mathcal{J}_a(\mathbf{k},t) \delta \rho_a(\mathbf{k},t) \rightarrow \mathcal{J}_a(\mathbf{k},t) (N\rho_a^{(0)}/\rho^{(0)})^{1/2}\delta \rho_a'(\mathbf{k},t)$.

\begin{widetext}
To absorb these factors which take into account the deviation from the mean density per component $\rho^{(0)}/N$, we also rescale the remaining fluctuating phase-angle fields as 
\begin{equation}
  \theta_a' = \sqrt{N \rho_a^{(0)}/\rho^{(0)}} \, \theta_a\,,
\end{equation}
with the condition
\begin{equation}
\int \mathcal{D} \theta \, \mathcal{O} (\theta) \, e^{\i S_{\mathrm{eff}} [\theta]} 
= \int \mathcal{D} \theta' \, \mathcal{O} (\theta) \, e^{\i S'_{\mathrm{eff}} [\theta']},
\label{eq:RescCond}
\end{equation}
where the argument of the operator $\mathcal{O}$ is understood to be expressed in terms of the rescaled field. Equation \eq{RescCond} yields $\exp\left(\i S_{\mathrm{eff}}[\theta]\right) = \mathrm{const.} \times \exp\left(\i S'_{\mathrm{eff}}[\theta']\right)$, with an insignificant multiplicative constant. This implies a corresponding rescaling of $\mathcal{J}_a \rightarrow \mathcal{J}_a'$.
As discussed in detail in \App{Inversegab}, the inverse of $\tilde{g}^{ab}$ is 
\begin{equation}
\label{eq:app:G_inv}
  (\tilde{g}^{-1})^{ab}(\mathbf{k},t;\mathbf{k}',t') 
  = \frac{1}{N} \frac{2 k_{\Xi}^2}{g \mathbf{k}^2} \left(\delta^{ab} - \frac{\sqrt{\rho_a^{(0)} \rho_b^{(0)}}/\rho^{(0)}}{1 + \mathbf{k}^2/2 k_{\Xi}^2}\right)
(2 \pi)^d \delta (\mathbf{k} + \mathbf{k}') \delta (t - t').
\end{equation}
After completing the squares and Gaussian integration over the $\delta \rho'$, and taking into account the rescaling of $\mathcal{J}_{a}$, this kernel enters the contribution
\begin{align}
  \frac{1}{2}\int_{\mathbf{k}\mathbf{k}'tt'}\mathcal{J}'_a(\mathbf{k},t)
  (\tilde{g}^{-1})^{ab}(\mathbf{k},t;\mathbf{k}',t') 
  \mathcal{J}'_b(\mathbf{k}',t') 
\end{align}
to the effective action which is quadratic in the $\mathcal{J}_a$. The effective action finally becomes:
\begin{equation}
  S'_{\mathrm{eff}}[\theta'] 
  = \int_{\mathbf{k},t} 
  \frac{1}{2 N} \Bigg\lbrace\frac{1}{g_{\mathrm{1/N}} 
  (\mathbf{k})} \left(\delta^{ab} - \frac{k_{\Xi,a} k_{\Xi,b}/k^2_{\Xi}}{1 + \mathbf{k}^2/2 k^2_{\Xi}} \right) 
  \times\ \mathcal{J}'_a (\mathbf{k},t) \mathcal{J}'_{b} (-\mathbf{k},t) 
 - \frac{\rho^{(0)} \mathbf{k}^2}{m} \theta'_a(\mathbf{k},t) \theta'_a (-\mathbf{k},t)\Bigg\rbrace,
 \label{eq:Seffthetas}
\end{equation}
where
\begin{equation}
\mathcal{J}'_a(\mathbf{k},t)
  =  
  \partial_t \theta'_a (-\mathbf{k},t)  
+ \left(\frac{\rho^{(0)}}{N \rho^{(0)}_a}\right)^{1/2}\int_{\mathbf{k}'} 
  \frac{\mathbf{k}' ( \mathbf{k} + \mathbf{k}')}{2 m} 
  \theta'_a(\mathbf{k}',t) \theta'_a(-\mathbf{k} - \mathbf{k}',t)\,,
\end{equation}
with momentum-dependent ``coupling function''
\begin{align}
  g_{\mathrm{1/N}}(\mathbf{k}) = \frac{g \mathbf{k}^2}{2 k^2_{\Xi}}.
  \label{eq:app:g1N}
\end{align}

The effective action can be split into a Gaussian and a non-Gaussian interaction part.
Omitting primes in denoting the fields and the action in what follows, we get
\begin{equation}
\label{eq:Seffsum}
  S_{\mathrm{eff}}=S_{\mathrm{eff,G}}+S_{\mathrm{eff,nG}}^{(3)}+S_{\mathrm{eff,nG}}^{(4)}\,,
\end{equation}
with
\begin{equation}
 S_{\mathrm{eff,G}}[\theta] 
  = \int_{\mathbf{k},t} 
  \frac{1}{2 N} \Bigg\lbrace\frac{1}{g_{\mathrm{1/N}} 
  (\mathbf{k})} \left(\delta^{ab} - \frac{k_{\Xi,a} k_{\Xi,b}/k^2_{\Xi}}{1 + \mathbf{k}^2/2 k^2_{\Xi}} \right) 
  \partial_{t}\theta_a (\mathbf{k},t) \partial_{t}\theta_b (-\mathbf{k},t) 
 - \frac{\rho^{(0)} \mathbf{k}^2}{m} \theta_a(\mathbf{k},t) \theta_a (-\mathbf{k},t)\Bigg\rbrace,
 \label{eq:SeffGthetaa}
\end{equation}
and the three- and four-wave interaction parts
\begin{align}
  S_{\mathrm{eff,nG}}^{(3)}[\theta] 
  &= \int_{\mathbf{k}\mathbf{k}',t} 
  \frac{1}{N^{3/2}} \frac{1}{g_{\mathrm{1/N}} 
  (\mathbf{k})} \left(\delta^{ab} - \frac{k_{\Xi,a} k_{\Xi,b}/k^2_{\Xi}}{1 + \mathbf{k}^2/2 k^2_{\Xi}} \right) 
  \frac{k_{\Xi}}{k_{\Xi,b}} 
  \frac{\mathbf{k}' ( \mathbf{k}' - \mathbf{k})}{2 m}
  \partial_{t}\theta_a (-\mathbf{k},t) \theta_b(\mathbf{k}',t) \theta_b(\mathbf{k} - \mathbf{k}',t),
 \label{eq:SeffnG3thetaa}
 \\
  S_{\mathrm{eff,nG}}^{(4)}[\theta] 
  &= \int_{\mathbf{k}\mathbf{k}'\mathbf{k}'',t} 
  \frac{1}{2N^{2}} \frac{1}{g_{\mathrm{1/N}} 
  (\mathbf{k})} \left(
 \frac{ \delta^{ab} k_{\Xi}^{2}}{k_{\Xi,a}^{2}}
  - \frac{1}{1 + \mathbf{k}^2/2 k^2_{\Xi}} \right) 
  \frac{\mathbf{k}' ( \mathbf{k}' + \mathbf{k})}{2 m}
  \frac{\mathbf{k}'' ( \mathbf{k}'' - \mathbf{k})}{2 m}
  \nonumber\\
  &
  \times\ 
  \theta_a(\mathbf{k}',t) \theta_a(-\mathbf{k} - \mathbf{k}',t)  
  \theta_b(\mathbf{k}'',t) \theta_b(\mathbf{k} - \mathbf{k}'',t).
 \label{eq:SeffnG4thetaa}
\end{align}
\end{widetext}
Note that the quadratic term in \Eq{SeffGthetaa}, for a single-component gas ($N=1$), reduces to the standard Luttinger-liquid  action \footnote{The Hamiltonian of the (1D) (Tomonaga-)Luttinger-liquid model \cite{Cazalilla2011a} reads $H_\mathrm{TLL}=(c_{s}/2\pi)\int dx\,\{K^{-1}[\pi\Pi(x,t)]^{2}+K[\partial_{x}\theta(x,t)]^{2}\}$, with speed of sound $c_{s}=\sqrt{v_{J}v_{N}}$, Luttinger parameter $K=\sqrt{v_{J}/v_{N}}$, phase stiffness $v_{J}=\pi\rho_{0}/m$, and density stiffness $v_{J}=g/\pi$. 
The canonically conjugate momentum is defined in terms of the density fluctuation as $\Pi(x,t)=\delta\rho(x,t)$.
This corresponds to the Lagrangian $L_\mathrm{TLL}=(2g)^{-1}\int dx\,\{[\partial_{t}\theta(x,t)]^{2}-[c_{s}\partial_{x}\theta(x,t)]^{2}\}$, and thus to the canonically conjugate field $\Pi(x,t)=g^{-1}\partial_{t}\theta(x,t)$.}
accounting for non-interacting sound modes on the background of a (quasi) condensate, useful for describing equilibrium low-temperature Bose gases in one and two spatial dimensions \cite{Cazalilla2011a}.
In the following, the non-linear terms \eq{SeffnG3thetaa} and \eq{SeffnG4thetaa}, however, will be crucial to describe the kinetic transport of phase excitations close to the non-thermal fixed point.

In equilibrium, the Luttinger liquid serves as a low-energy effective theory for near-zero-temperature Bose gases in less than three dimensions, where no spontaneously broken field is available as a Bogoliubov mean field to expand around  \cite{Cazalilla2011a}. 
Here, we will exploit that the same low-energy effective theory, with non-linear terms taken into account, serves to describe transport in momentum space within a kinetic formulation of the time evolution.
We emphasize that the formulation of the low-energy effective theory is independent of the particular state the system is in, provided the assumption of having a constant high mean density in each field component to expand about is well satisfied.
Hence, far-from-equilibrium phenomena can be described by our theory provided the mode occupancies of the state are dominated by the low-energy range of momenta below $k_\Xi$.

\subsubsection{Large-$N$ limit of the Luttinger-liquid-type effective action}
\label{sec:LargeN}
To allow comparisons with previous formulations of scaling at a non-thermal fixed point, including \cite{Berges:2008wm,Berges:2008sr,Scheppach:2009wu,Berges:2010ez,Orioli:2015dxa,Berges:2015kfa,Chantesana:2018qsb.PhysRevA.99.043620}, it is necessary to consider the limit $N \rightarrow \infty$. Using the fact that $k_{\Xi,a}/k_{\Xi} \sim 1/\sqrt{N}$ one easily obtains that all the ``tensor structures'' in \eq{SeffGthetaa}, \eq{SeffnG3thetaa}, and \eq{SeffnG4thetaa} become diagonal to the leading order in $1/N$~\footnote{One has to be careful, however. Even though each off-diagonal term is of the order $O(1/N)$, there are $N - 1$ of them. Therefore, the total contribution is of the order $O(1)$. Furthermore, the very presence of the Bogoliubov mode is due to these off-diagonal elements, and hence neglecting them will result in losing the linear dispersion mode, which we will see immediately. Thus, strictly speaking, the following effective action accounts only for a part of the full set
of possible dynamical excitations.}.

Summarizing the above derivations and going back to the non-rescaled phase fields, we obtain, in the large-$N$ limit, the following low-energy effective action (see  \App{vertices} for details)
\begin{align}
  &S_{\mathrm{eff}}[\theta] 
  = \int\limits_{\vec k, \vec k', \mathcal{C}} \frac{1}{2} \,\theta_a (\vec k, t) \i D_{{ab}}^{-1}(\vec k, t; \vec k', t') \theta_b(\vec k', t')
  \nonumber\\
  &- \int\limits_{\lbrace \vec k_i \rbrace, \mathcal{C}} 
  \frac{k_{\Xi,a}^2}{k_{\Xi}^2} \frac{\vec k_1 \cdot \vec k_2}{2m\,g_{\mathrm{1/N}} (\vec k_3)} \, 
  \theta_a (\vec k_1, t)\, \theta_a (\vec k_2, t) \partial_t \theta_a (\vec k_3, t) \,\delta \Big(\sum_{i=1}^{3} \vec k_i\Big) 
  \nonumber\\
  &+ \int\limits_{\lbrace \vec k_i \rbrace, \mathcal{C}} 
  \frac{k_{\Xi,a}^2}{k_{\Xi}^2} \frac{(\vec k_1 \cdot \vec k_2) \, (\vec k_3 \cdot \vec k_4)}{8m^2 \,g_{\mathrm{1/N}} (\vec k_1 - \vec k_2)} \, \theta_a(\vec k_1,t) \cdots \theta_a(\vec k_4,t)\, \delta \Big(\sum_{i=1}^{4} \vec k_i\Big)\,,
  \label{eq:Seff4}
\end{align} 
with free inverse propagator, defining $\delta_{\vec k, -\vec k'} \equiv (2 \pi)^d \, \delta (\vec k + \vec k')$,
\begin{align}
\label{eq:freeinverse}
\i D_{{ab}}^{-1}(\vec k, t; \vec k', t') 
= \frac{\delta_{\vec k, -\vec k'}}{Ng_{\mathrm{1/N}} (\vec k)} \, \delta_{{ab}} \delta_{\mathcal{C}} (t - t') \left(-\partial_t^2 - (\vec k^2/2m)^2 \right).
\end{align}

We note that the large-$N$ limit can be taken either for a fixed total density $\rho^{(0)}$, or keeping the single-component densities $\rho_{a}^{(0)}$ constant. 
In the former case, the single-component densities vanish for $N\to\infty$, such that one eventually obtains a classical gas of distinguishable bosons. As a result, the condition $\delta\rho_{a}/\rho_{a}^{(0)}\ll\theta_{a}$, cf.~\eq{CondLEEFT}, breaks down asymptotically as the particles become distinguishable and lose mutual coherence.

Hence, the low-energy effective theory, in the large-$N$ limit, rather applies to the situation of fixed, i.e., $N$-independent, densities $\rho_{a}^{(0)}$ equal in each of the $N$ effectively decoupled copies of field components.
To avoid, in this case, that the gas loses its diluteness and becomes a dense fluid as the total density $\rho^{(0)} = N\rho_{a}^{(0)}$~\footnote{Hereinafter, we assume, for simplicity, that all components are equally occupied, $\rho^{(0)}  = N \rho^{(0)}_a$. Since in the large-$N$ limit there is no coupling between different modes, this does not effect the following analysis.} increases, one needs to adjust the coupling $g$.
A sensible choice is to keep the relevant Gross-Pitaevskii coupling, i.e., the healing-length energy scale corresponding to the total density, $k_{\Xi}^{2}/(2m)=g\rho^{(0)}$, fixed and thus the rescaled coupling $\lambda=gN$.
For this choice, the low-energy effective action becomes invariant under changes of $N$. 
Moreover, its regime of validity, with excitations on scales longer than the healing length [cf.~\eq{CondLEEFT}], remains fixed at large $N$.

We remark that rescaling the density with $N$ and the coupling with $1/N$ implies that, in $d>1$ spatial dimensions, the diluteness parameter of the total gas scales as 
\begin{align}
  \zeta_\mathrm{d} = [\rho^{(0)}]^{1/d}a_{d}\sim N^{1/d-1}\,,
\end{align}
where $a_{d}\sim g$ is the length scale associated with the coupling $g$ in $d$ dimensions.
For example, in $d=3$ dimensions, the diluteness of the total gas increases with $N$ such that, in the large-$N$ limit, the theory becomes perturbative, and quantum fluctuations can be neglected.

Taking the single-component densities fixed as discussed above, we will write all expressions, in what follows, in terms of the Goldstone coupling, 
\begin{equation}
\label{eq:g_Gold}
g_{\mathrm{G}} (\mathbf{k})\equiv N\,g_{\mathrm{1/N}}(\mathbf{k}).
\end{equation}

\begin{figure}
\centering
\includegraphics[scale=1.0]{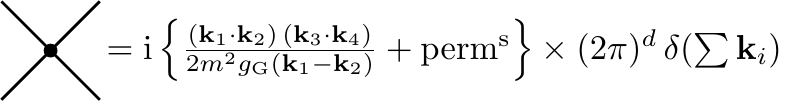}
\includegraphics[scale=1.0]{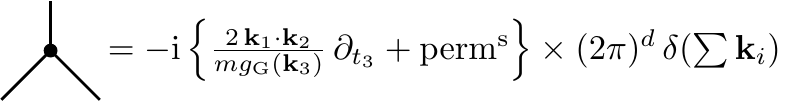}
\caption{Diagrammatic representation of the interaction terms.}
\label{fig:verts}
\end{figure}

Finally, we note that the interaction terms can be conveniently represented by the diagrams depicted in Fig.~\ref{fig:verts}

\section{Kinetic theory and scaling analysis}
\label{sec:Scaling}
In the previous section, we have derived the effective action \eqref{eq:Seffsum} to describe dynamics of the $\mathrm{U}(N)$-symmetric Gross-Pitaevskii model below the scale $k_{\Xi}$. To do so, we integrated out density fluctuations which, as was shown in Sect.~\ref{sec:Classical}, are suppressed by the factor $k/k_{\Xi}$ as compared to phase fluctuations. Furthermore, in the limit $N \to \infty$, the effective description reduces to $N$ independent and identical copies described by the action \eqref{eq:Seff4}. 

 Our aim now is to apply the low-energy effective theory developed in \Sect{luttingerleeft} to universal scaling dynamics at non-thermal fixed points. A first step in this direction is to derive a kinetic description of transport in momentum space, as has been done previously within the non-perturbative approach based on the fundamental Bose fields \cite{Berges:2008wm,Berges:2008sr,Scheppach:2009wu,Berges:2010ez,Orioli:2015dxa,Berges:2015kfa,Chantesana:2018qsb.PhysRevA.99.043620}.
We first set up kinetic equations and scattering integrals, in analogy to standard derivations (see, e.g., \cite{Berges:2004yj,Berges:2015kfa,Lindner:2005kv,Lindner:2007am} for details).

The procedure is then to state a scaling hypothesis of the form \eq{NTFPscaling} for the momentum distribution function whose time evolution is governed by the kinetic equation.
Hence, while the distribution evolves by rescaling in time and space, it is assumed to keep a stationary scaling form which is governed by a fixed-point equation corresponding to a rescaled form of the kinetic equation.
As a result, the analysis focuses, mathematically speaking, on the fixed-point configuration, providing the properties of the quasi-stationary scaling form and the scaling exponents governing the time evolution.

Most of the discussion is kept within the large-$N$ limit, except for the case of $N=1$ discussed separately in \Sect{SingleCompScaling}. 

\subsection{Kinetic description}
\label{sec:Kinetic}
The kinetic description focuses on the evolution of equal-time two-point correlators, specifically on occupation-number distributions of quasiparticles in momentum space (assuming identical excitations in all modes, e.g., in a large-$N$ approximation),
\begin{equation}
\label{eq:DefPhaseAngleCorr}
f_a(\mathbf{k},t) = \langle \theta_a(\mathbf{k},t) \theta_a(-\mathbf{k},t) \rangle.
\end{equation}

One starts with Kadanoff-Baym dynamic equations for the Schwinger-Keldysh time-ordered two-point Greens functions $G_{ab}(x,y)=\langle\mathcal{T}_{\mathcal{C}}\theta_{a}(x) \theta_{b}(y)\rangle$. After decomposing into spectral and Keldysh statistical components, one introduces a quasiparticle ansatz for the spectral functions.
Finally, performing a gradient expansion, in Wigner coordinates, along the relative-time direction, and sending the initial time $t_0 \rightarrow -\infty$, one obtains, in two-loop approximation of the two-particle irreducible effective action (Luttinger-Ward functional), a kinetic equation for the time evolution of a suitably defined momentum spectrum $f_{\mathbf{k}}\equiv f(\mathbf{k},t)\equiv f_{a}(\mathbf{k},t)$ of phase excitations,
\begin{equation}
  \partial_{t}f(\mathbf{k},t)
  = I[f](\mathbf{k},t)\,,
  \label{eq:QKinEq}
\end{equation}
where $I[f]$ is a scattering integral to be specified below.

From the action \eq{Seff4}, we infer (see Appendix~\ref{app:Boltzmann} for details) that the scattering integral has the form
\begin{align}
  I[f](\mathbf{k},t)=&\ I_{3}(\mathbf{k},t)+I_{4}(\mathbf{k},t)\,,
\end{align}
with
\begin{align}
  I_3 (\mathbf{k},t) 
  \sim& \int_{k_{0},p,q} |\gamma(\mathbf{k},\mathbf{p},\mathbf{q})|^2  
  \tilde{\rho}_k \tilde{\rho}_p \tilde{\rho}_q\,\delta (k + p - q) 
  \nonumber\\ 
  &\times \Big[ (f_k + 1) (f_p + 1) f_q  -  f_k f_p (f_r + 1) \Big]\,,
  \label{eq:I_3}
  \\
  \label{eq:I_4}
  I_4 (\mathbf{k},t) 
  \sim& \int_{k_{0},p,q,r} |\lambda(\mathbf{k},\mathbf{p},\mathbf{q},\mathbf{r})|^2
  \tilde{\rho}_k \tilde{\rho}_p \tilde{\rho}_q \tilde{\rho}_r\, \delta (k + p - q - r) 
  \nonumber\\
  &\times\ \Big[ (f_k + 1) (f_p + 1) f_q f_r -f_k f_p (f_q + 1) (f_r + 1) \Big] \,,
\end{align}
where the $(d+1)$-dimensional delta distributions imply energy and momentum conservation, with $k_{0}=\omega(k)=k^{2}/2m$ denoting the quasiparticle frequencies, and where the scattering matrices are defined as
\begin{align}
  \gamma(\mathbf{k},\mathbf{p},\mathbf{q}) 
  &= \frac{(\mathbf{k} \cdot \mathbf{p})\, \omega(\mathbf{q})}{m\,g_{\mathrm{G}} (\mathbf{q})} + \text{perm}^{\text{s}}\,,
  \label{eq:gamma3}
  \\
  \lambda(\mathbf{k},\mathbf{p},\mathbf{q},\mathbf{r}) 
  &= \frac{(\mathbf{k} \cdot \mathbf{p}) (\mathbf{q} \cdot \mathbf{r})}{2m^{2}\,g_{\mathrm{G}} (\mathbf{k} - \mathbf{p})} 
  + \text{perm}^{\text{s}}\,,
  \label{eq:lambda4}
\end{align}
with momentum-dependent coupling function which, in the large-$N$ limit, takes the form $g_{\mathrm{G}}(\mathbf{k}) = {Ng \mathbf{k}^2}/({2 k^2_{\Xi}})$ [cf.~Eqs.~\eq{app:g1N} and \eq{g_Gold}].
Assuming that the quasiparticle (on-shell) approximation is valid, i.e., that the spectral functions $\tilde\rho(k)=\tilde\rho(k_{0},\mathbf{k})$ describe free stable quasiparticles, we find that for our EFT the bare retarded Green's function has the form
\begin{align}
  G_R^{(0)} &(\mathbf{k}, \omega) 
  = \frac{g_{\mathrm{G}} (\mathbf{k})}{(\omega + \i 0^{+})^2 - \omega^2(\mathbf{k})} 
  \nonumber \\ 
  &= \frac{g_{\mathrm{G}} (\mathbf{k})}{2 \omega (\mathbf{k})} \left[ \frac{1}{\omega - \omega(\mathbf{k}) + \i 0^{+}} 
  - \frac{1}{\omega + \omega(\mathbf{k}) + \i 0^{+}} \right].
  \label{eq:GR}
\end{align}
Using the definition of the spectral function in terms of $G_{R}$, $\tilde{\rho} (\mathbf{k},\omega) = -2 \,\text{Im}\, G_R (\mathbf{k},\omega)$, as well as the relations
\begin{align}
  \text{Im} \frac{1}{x + \i \epsilon} 
  = - \frac{\epsilon}{x^2 + \epsilon^2}\,, \quad
  \lim_{\epsilon \to 0} \frac{\epsilon}{x^2 + \epsilon^2} 
  = \pi \delta (x)\,, 
\end{align}
we obtain the bare spectral function of our EFT \eq{Seff4},
\begin{equation}
\tilde{\rho}^{(0)} (\mathbf{k}, \omega) = \frac{\pi g_{\mathrm{G}} (\mathbf{k})}{\omega(\mathbf{k})} \left[ \delta (\omega - \omega(\mathbf{k})) - \delta (\omega + \omega(\mathbf{k})) \right]\,,
\end{equation}
so that the $T$-matrices adopt the following form:
\begin{equation}
\label{eq:T_3}
   |T_3(\mathbf{k},\mathbf{p},\mathbf{q})|^2 = |\gamma(\mathbf{k},\mathbf{p},\mathbf{q})|^2 
   \frac{g_{\mathrm{G}} (\mathbf{k})\, g_{\mathrm{G}} (\mathbf{p})\, g_{\mathrm{G}} (\mathbf{q})}
   {8 \,\omega (\mathbf{k}) \, \omega (\mathbf{p}) \, \omega (\mathbf{q})}\,,
\end{equation}
\begin{equation}
\label{eq:T_4}
  |T_4(\mathbf{k},\mathbf{p},\mathbf{q},\mathbf{r})|^2 = |\lambda(\mathbf{k},\mathbf{p},\mathbf{q},\mathbf{r})|^2   
  \frac{g_{\mathrm{G}} (\mathbf{k}) \cdots g_{\mathrm{G}} (\mathbf{r})}
  {2 \omega (\mathbf{k})  \cdots 2 \omega (\mathbf{r})}.
\end{equation}

\subsection{Scaling at a non-thermal fixed point}
\label{sec:ScalingLEEFT}
We have now the tools at hand to proceed to the main physical results of this paper, predictions for the scaling behavior of the considered model at a non-thermal fixed point.
The procedure is as follows.
The kinetic equation derived above is assumed to describe the transport process in momentum space of low-energy phase excitations towards infrared scales.
As illustrated by our numerical data obtained for the evolution after a strong quench (see \Fig{NTFPEvolution}), this transport assumes a universal scaling form after having passed a brief period of fast initial evolution.

The main task of the analytical work presented here is to formulate and evaluate fixed-point equations which allow one to predict the scaling functions and exponents characterizing this universal dynamics.
We focus on the analysis at the fixed point, i.e., we assume that the scaling evolution has already been fully established.
This allows us to insert a scaling hypothesis for the solution into the kinetic equation and apply power-counting techniques to determine the unknown scaling exponents \cite{Svistunov1991a,Zakharov1992a,Berges:2010ez,Orioli:2015dxa,Chantesana:2018qsb.PhysRevA.99.043620}.
In this way we will re-derive previously known scaling exponents and complement them with results for situations which have not been considered before, in particular for the experimentally relevant case of a single field component, $N=1$, including in particular the three-vertex term $I_{3}$.

\subsubsection{Spatio-temporal scaling}
\label{sec:SpatioTemporal}
Using the $T$-matrices given above, we can perform a scaling analysis of universal dynamics of the system at a non-thermal fixed point. 
To this end, we consider the scattering integrals corresponding to each vertex. 
According to the scaling hypothesis, the quasiparticle spectral distribution satisfies the scaling form
\begin{equation}
\label{eq:distr_scaling}
f (\mathbf{k},t) = s^{\alpha/\beta} f(s \mathbf{k}, s^{-1/\beta}t),
\end{equation}
and so do the $T$-matrices,
\begin{equation}
\label{eq:T_scaling}
  |T_l(\mathbf{k}_{1},\dots,\mathbf{k}_{l};t)| 
  = s^{-m_l} |T_l(s \mathbf{k}_{1},\dots,s \mathbf{k}_{l};s^{-1/\beta}t)|\,, 
\end{equation}
where the index $l\in\{3,4\}$ denotes the $l$-point vertex.
From the above, one obtains the spatio-temporal scaling of the scattering integral \cite{Chantesana:2018qsb.PhysRevA.99.043620},
\begin{equation}
I_{l}[f](\mathbf{k},t) = s^{-\mu_{l}} I_{l}[f](s \mathbf{k},s^{-1/\beta} t).
\label{eq:Iltimemomentumscaling}
\end{equation}
Using \eq{I_4} and \eq{I_3} together with \eq{distr_scaling} and \eq{T_scaling}, one derives the expression for a scaling exponent $\mu_l$ for each collision term:
\begin{equation}
\mu_l = (l - 2)  d - z + 2m_{l} - (l  - 1) \frac{\alpha}{\beta},
\end{equation}
where 
\begin{equation}
2m_{l} =  2\tilde m_l + l\, (\gamma-z),
\end{equation}
with 
\begin{align}
\tilde m_3 &= z + 2 - \gamma,\\
\tilde m_4 &= 4 - \gamma,
\end{align}
and $\gamma$ being defined as the scaling exponent of $g_{\mathrm{G}}$:
\begin{equation}
g_{\mathrm{G}} (\mathbf{k}) = s^{-\gamma} g_{\mathrm{G}} (s \mathbf{k}).
\end{equation}
Note that we here assume that the scaling of the coupling $g_\mathrm{G}$ appearing in the retarded Green's function \eq{GR}, and in the couplings, \eq{gamma3} and \eq{lambda4}, is the same, i.e., the momenta as well as their differences are in the same scaling regime.
This should be guaranteed if the external momentum $\mathbf{k}$ of the scattering integrals is in the infrared (IR) scaling limit and thus  the integrals are dominated by momenta in this regime.

If the distribution function $f$ obeying the scaling form \eq{distr_scaling} is to solve the kinetic equation \eq{QKinEq} for a given $\mu=\mu_{l}$, the scaling exponents need to satisfy the relation
\begin{equation}
\label{eq:alphabeta}
\alpha = 1 - \beta \mu.
\end{equation}
In addition, in the presence of global conservation laws for the integral $\int_{\mathbf k}f(\mathbf{k},t)$ (quasiparticle number) or the integral $\int_{\mathbf k}\omega(\mathbf{k})f(\mathbf{k},t)$ (quasiparticle energy), the following constraints apply for the scaling exponents:
\begin{align}
\label{eq:qp_cons}
\alpha  &= \beta d, &\text{number conservation}; \\
\label{eq:en_cons}
\alpha' &= \beta' (d + z), &\text{energy conservation}.
\end{align}
We note that the conservation of quasiparticle number is, in general, not expected.
In the low-energy regime of a Bose gas considered here, the quasiparticles are gapless phonon and relative-phase modes as described above whose number can change due to the cubic interaction terms appearing in the action \eq{Seff4}.
It appears nevertheless as one of the key properties of non-thermal fixed points that the transport underlying the universal scaling dynamics is related to conservation laws which can include that of quasiparticle number.

Here and in the following, primed exponents $\alpha'$, $\beta'$ refer to transport conserving energy.
Collecting the above results, we obtain
\begin{align}
\label{eq:mu3}
\mu_3 &= d + 4 - 2 z + \gamma - 2 {\alpha}/{\beta}\,,\\
\label{eq:mu4}
\mu_4 &= 2 d + 8 - 5 z + 2\gamma - 3 {\alpha}/{\beta}\,,
\end{align}
which, together with (\ref{eq:alphabeta}), gives:
\begin{align}
\label{eq:mu3rel}
&l = 3:& \qquad \beta \cdot (d + 4 - 2 z + \gamma) = 1 + \alpha\,,\\
\label{eq:mu4rel}
&l = 4:& \qquad \beta \cdot (2 d + 8 - 5 z + 2 \gamma) = 1 + 2 \alpha.
\end{align}
One should note that, in principle, \eq{mu3rel} and \eq{mu4rel} provide the closed system of equations allowing to determine $\alpha$ and $\beta$:
\begin{align}
\alpha &= ({d + 4 - 3 z + \gamma})/{z}\,,\\
\beta &= {1}/{z}.
\end{align}
However, since, depending on the dimensionality $d$ and the momentum region of interest, one type of interaction can dominate over the other one, it is more reasonable to analyze the two terms independently. 
In order to close the system in that case, an additional relation is required, which, in fact, is provided by either  quasiparticle number conservation \eq{qp_cons} or energy conservation \eq{en_cons} within the scaling regime. 
For particle number conservation, we obtain
\begin{align}
\label{eq:betaQ3}
&l = 3: &\beta =&\ \frac{1}{4 - 2 z + \gamma}\,,\\
\label{eq:betaQ4}
&l = 4: &\beta =&\ \frac{1}{8 - 5 z + 2 \gamma}\,,
\end{align}
while, for energy conservation, one gets
\begin{align}
&l = 3:  &\beta' =&\ \frac{1}{4 - 3 z + \gamma}\,,\\
&l = 4:  &\beta' =&\ \frac{1}{8 - 7 z + 2 \gamma}.
\end{align}
Substituting, in the large-$N$ limit, the scaling exponent $z = 2$ of the free Goldstone dispersion \eq{app:GoldstoneFreq} and $\gamma = 2$ of the coupling function \eq{app:g1N}, the $T$-matrices scale with
\begin{align}
m_{3} = 2\,,\qquad
m_{4} = 2\,,
\end{align}
and the resulting scaling exponents read as
\begin{align}
\beta &= {1}/{2}\,,\qquad(z=2)
\label{eq:betalargeN}
\\
\alpha &= {d}/{2}\,,
\label{eq:alphalargeN}
\end{align}
for both three- and four-point vertices, and
\begin{align}
\beta' &= -{1}/{2}\,,\\
\alpha' &= -({d + z})/{2}\,,
\end{align}
for the four-point vertex, while, at the same time, for the three-point vertex, no valid solution exists.
We point out that the above exponents are equivalent to the respective exponents derived in \cite{Orioli:2015dxa,Chantesana:2018qsb.PhysRevA.99.043620} using the large-$N$ resummed kinetic theory for the fundamental Bose fields, for the case of a dynamical exponent $z=2$, and a vanishing anomalous dimension $\eta=0$ (see also \Sect{ScalingFunctionBosePhase} below).

The question which arises here is, whether both, three- and four-wave interactions are equally relevant.
To answer this, we need to compare the spatio-temporal  scaling properties of the scattering integrals \eq{I_3} and \eq{I_4}, for given fixed-point solutions $f(\mathbf{k},t)$.
Focusing on the infrared transport of quasiparticles, for which $\alpha/\beta=d$, we obtain, from \eq{mu3} and \eq{mu4}, inserting also the relation $\gamma=2(z-1)$ valid for both, free ($z=2$) and Bogoliubov ($z=1$) quasiparticles, that
\begin{align}
\label{eq:mu3QP}
-\mu_3 &= d - 2 \,,\\
\label{eq:mu4QP}
-\mu_4 &= d - 4 +  z  .
\end{align}
Hence, we find, for $z=2$ in the large-$N$ limit, that $\mu_3=\mu_4$, which implies that the relative importance of the scattering integrals $I_{3}$ and $I_{4}$ is not expected to change in time.

\subsubsection{Scaling function}
\label{sec:ScalingFunction}
A further important question concerns the form of the scaling function.
This form is determined by the stationary fixed-point equation, i.e., self-consistently, by the infrared scaling properties of the scattering integrals \eq{I_3} and \eq{I_4}, for given fixed-point solutions $f(\mathbf{k},t_\mathrm{ref})$ at a fixed reference time $t_\mathrm{ref}$.
Inserting $s=(t/t_\mathrm{ref})^{\beta}$ into the scaling hypothesis \eq{distr_scaling} yields the scaling form
\begin{equation}
\label{eq:fScalingForm}
f (\mathbf{k},t) = (t/t_\mathrm{ref})^{\alpha} f_\mathrm{S}([t/t_\mathrm{ref}]^{\beta} \mathbf{k})\,,
\end{equation}
with universal scaling function
\begin{equation}
\label{eq:fScalingFunction}
f_\mathrm{S}(\mathbf{k})=f (\mathbf{k},t_\mathrm{ref}).
\end{equation}
For the spatial form of the scaling function we make the scaling ansatz
\begin{equation}
\label{eq:fScalingFunctionForm}
f_\mathrm{S}(\mathbf{k})=s^{\kappa}f_\mathrm{S}(s\mathbf{k})\, ,
\end{equation}
which we will motivate in more detail later on.
Inserting \eq{fScalingFunctionForm} into the kinetic equation \eq{QKinEq} yields the stationary fixed-point equation
\begin{align}
  (\alpha - \beta\kappa)f_\mathrm{S}(p) &= t_\mathrm{ref}I[f_\mathrm{S}](p).
  \label{eq:FPEqkappa}
\end{align}
If both sides of \Eq{FPEqkappa} are non-zero, they must scale in the same way in $p$.
Taking into account the fixed-time scaling
\begin{equation}
I_{l}[f](\mathbf{k},t_\mathrm{ref}) = s^{-\mu_{\kappa,l}} I_{l}[f](s \mathbf{k},t_\mathrm{ref}).
\end{equation}
of the scattering integral $I_{l}$ defined by the scaling exponent $\mu_{\kappa,l}$, one finds that the momentum exponent $\kappa$ of the scaling function $f_\mathrm{S}$  is 
\begin{equation}
\kappa=-\mu_{\kappa,l}\,,
\label{eq:kappafrommukappa}
\end{equation}
provided that the scaling of $I_{l}$ determines that of $I$.

Hence, we need to determine the exponents $\mu_{\kappa,l}$, which, in general, can take different values for the three-wave and the four-wave collision integrals.
As we will show in the following, $\kappa>0$ such that, in the scaling limit of small momenta, all distribution functions in the scattering integrals can be assumed to be in the semi-classical regime where $f_{p}\gg1$, etc., for the other momenta appearing.
Hence, only the terms quadratic (cubic) in $f$ can be assumed to contribute to $I_{3}$ ($I_{4}$).
As a result, power counting of \eq{I_3} and \eq{I_4} in this semi-classical regime, taking into account that the scaling 
\begin{equation}
\label{eq:T_scalingstat}
  |T_l(\mathbf{k}_{1},\dots,\mathbf{k}_{l};t_\mathrm{ref})| 
  = s^{-m_{l}} |T_l(s \mathbf{k}_{1},\dots,s \mathbf{k}_{l};t_\mathrm{ref})|\,, 
\end{equation}
of the $T$-matrices \eq{T_3} and \eq{T_4}, respectively, is valid also at a fixed reference time, gives
\begin{align}
&l = 3: \qquad &\mu_{\kappa,3} =&\ 4+d+\gamma-2z-2\kappa\,,
  \label{eq:mukappa3}
\\
&l = 4: \qquad &\mu_{\kappa,4} =&\ 8+2d+2\gamma-5z-3\kappa.
  \label{eq:mukappa4}
\end{align}
These scaling exponents, together with \Eq{kappafrommukappa}, yield the momentum exponent of the quasiparticle scaling function as
\begin{align}
&l = 3: \qquad &\kappa=\kappa_{3} =&\ 4+d+\gamma-2z\,,
  \label{eq:kappa3}
\\
&l = 4: \qquad &\kappa=\kappa_{4} =&\ 4+d+\gamma-5z/2\,,
  \label{eq:kappa4}
\end{align}
given that either of the scattering integrals with $l=3,4$ dominates in the region of momenta considered.

For a given $\kappa_{l}$, and with the large-$N$ exponents $z=2$ and $\gamma=2$ inserted, one finds that
\begin{align}
  \mu_{\kappa,3}-\mu_{\kappa,4} = \kappa_{l}- d\geq1.
  \label{eq:mukappa3m4}
\end{align}
As a result, the four-wave scattering integral is expected to dominate at small momenta, $k\to0$, over the three-wave term, such that
\begin{align}
  \kappa=\kappa_{4} = d+1\,\qquad(z=2)
  \label{eq:kappa4result}
\end{align}
results as the momentum scaling exponent of the quasiparticle distribution $f(\mathbf{k},t)\sim k^{-\kappa}$ at the non-thermal fixed point.

This  appears to contradict the previous analysis of spatio-temporal scaling [cf.~Eqs.~\eq{Iltimemomentumscaling} and \eq{mu3QP} and \eq{mu4QP}], which, for $z=2$ showed equal importance of $I_{3}$ and $I_{4}$ while, for $z=1$, the integral $I_{3}$ dominated the late-time scaling in the IR.
We emphasize, however, that the spatio-temporal scaling exponents \eq{betaQ3} and \eq{betaQ4} were obtained from scaling relations which are independent of the precise form of the scaling function $f(\mathbf{k},t)$ but only require the scaling relation \eq{fScalingForm}. 
Hence, the question as to which of the vertices dominates the transport can be independently answered from the question of which vertex is responsible for the shape of the scaling function.

In contrast, in this section, we argue that the scaling function $f(\mathbf{k},t_\mathrm{ref})$ at a fixed reference time, in the IR scaling limit, i.e., for $k_{\Lambda}(t_\mathrm{ref})\to0$, is the solution of a fixed-point equation dominated by $I_{4}(\mathbf{k},t_\mathrm{ref})$.
We note that, at a given early time of the evolution, when $k_{\Lambda}$ is still large and this spatial scaling limit has not been reached yet, our estimate may not apply to the evolution starting from a given initial condition.
We will discuss this in more detail when comparing with numerical results in the following section.

\subsubsection{Relation between phase and phase-angle scaling forms}
\label{sec:ScalingFunctionBosePhase}
The result \eqref{eq:kappa4result}, as well as the temporal scaling exponents \eqref{eq:betalargeN} and \eqref{eq:alphalargeN} derived above appear, remarkably, in full agreement with earlier findings from non-perturbatively resummed kinetic theory (see \cite{Orioli:2015dxa,Chantesana:2018qsb.PhysRevA.99.043620}). To show this rigorously, however, still requires a translation between the quasiparticle distribution $f_{a}(\mathbf{k},t)$ characterizing the phase-angle excitations considered here and the scaling of the particle number distribution $n_{a}(\mathbf{k},t)$ encoded in the fundamental Bose field,
\begin{equation}
  n_{a}(\mathbf{k},t)
  =\langle\Phi_{a}^{\dag}(\mathbf{k},t)\Phi_{a}(\mathbf{k},t)\rangle
  \simeq \rho^{(0)}_{a}\mathcal{F}[C_{a}(\mathbf{x}-\mathbf{y},t)](\mathbf{k})\,,
  \label{eq:napFTofPhaseCorr}
\end{equation}
where the Fourier transform $\mathcal{F}$ (cf.~Appendix~\ref{app:Not}) is taken with respect to $\mathbf{r}=\mathbf{x}-\mathbf{y}$ of the on average translation-invariant phase correlator
\begin{align}
  C_{a}(\mathbf{r},t)
  &=\langle e^{-\i\theta_{a}(\mathbf{x}+\mathbf{r},t)}e^{\i\theta_{a}(\mathbf{x},t)}\rangle,
  \label{eq:PhaseCorrCa}
\end{align}
independent of $\mathbf{x}$.

Defining the (single-source) Schwinger-Keldysh generating functional of correlation functions of the phase angle field $\theta_{a}$,
\begin{equation}
\label{eq:ZgenF}
Z[J] =  \int D \theta \, e^{\i S_{\mathrm{eff}}[\theta] + \i\int_{x,\mathcal{C}} J_{a}(x)\theta_{a}(x)},
\end{equation}  
we can express the phase correlator as
\begin{align}
 C_{a}(\mathbf{r},t)
  &= Z\left[ \left\{J_{a}=\left[-\delta \left(\mathbf{x}-\mathbf{r}\right)+\delta \left(\mathbf{x}\right)\right]\delta_{\mathcal{C}}(x_{0}-t),0\right\}\right],
  \label{eq:PhaseCorrZgen}
\end{align}
where we take into account that the generating functional, for $J=0$, evaluates to unity on the closed time path, $Z[0]=1$, and that all other $J_{b}$, $b\not=a$, vanish. Using the standard representation of the generating functional in terms of connected Greens functions~\footnote{Note that the summation goes from $n=1$, as opposed to the standard zero-$T$ expansion, which reflects that $W[0] \equiv 0$ on the Schwinger-Keldysh contour.},
\begin{equation}
Z[J] 
= 
\exp{\Bigg\lbrace \sum_{n=1}^{\infty} \int_{\lbrace x_i \rbrace_{i=1}^n,\mathcal{C}} \frac{\i^n}{n!}\, G_{\theta}^{(n)}(x_1,\ldots,x_n) J(x_1) \cdots J(x_n) \Bigg\rbrace},
\end{equation}
with
\begin{align}
G_{\theta,a}^{(n)}(\mathbf{x}_{1},\dots,\mathbf{x}_{n},t)
  &= \langle\theta_{a}(\mathbf{x}_{1},t)\cdots\theta_{a}(\mathbf{x}_{n},t)\rangle_{c}\,
  \label{eq:PhaseCumulants}
\end{align}
being an equal-time $n$-point connected correlator of the phase-angle field $\theta_{a}$, we can express the two-point phase correlator as
\begin{equation}
 C_{a}(\mathbf{r},t)
  = \exp\Bigg\{\sum_{n=1}^{\infty}\sum_{\mathbf{x}_{i}=\mathbf{0},\mathbf{r}}
        c_{n}G_{\theta,a}^{(n)}(\mathbf{x}_{1},\dots,\mathbf{x}_{n},t)\Bigg\}\,,
  \label{eq:PhaseCorrCumSum}
\end{equation}
where 
\begin{align}
 c_{n}= \frac{\i^n}{n!}\exp\Bigg\{\sum_{i=1}^{n}\i\pi\delta_{r,|\mathbf{x_{i}}|}\Bigg\}
  \label{eq:cn}
\end{align}
contains a combinatorial factor $\i^n/n!$ as well as a sign which depends on the coordinates of the cumulants.

The above result provides a strong constraint on the scaling of the cumulants of the phase angle given that the phase correlator $C_{a}$ satisfies the scaling relation
\begin{equation}
 C_{a}(\mathbf{r},t)
  = s^{\alpha/\beta-d}C_{a}(s^{-1}\mathbf{r},s^{-1/\beta}t)\,,
  \label{eq:PhaseCorrScalingRel}
\end{equation}
which follows directly from the scaling of $n_a(\mathbf{k},t)$. It should be emphasized that the above scaling exponents $\alpha$ and $\beta$ are those associated with the scaling of fundamental bosonic fields $\Phi$, for which the particle number conservation is expected due to $\mathrm{U}(1)$ symmetry. Then, assuming the absence of accidental cancellations, the relation \eqref{eq:PhaseCorrCumSum} between the two-point correlator and the phase-angle connected correlation functions thus implies 
\begin{align}
\exp \Big\lbrace G^{(n)}_{\theta,a}&(\lbrace\mathbf{x}_i\rbrace,t)\Big\rbrace \Big|_{\mathbf{x}_{i}\in\{\mathbf{0},\mathbf{r}\}}\nonumber\\
&=
s^{-\alpha/\beta + d} \left.\exp{\left\lbrace G^{(n)}_{\theta,a}(\lbrace s^{-1} \mathbf{x}_i\rbrace,s^{-1/\beta} t) \right\rbrace}\right|_{\mathbf{x}_{i}\in\{\mathbf{0},\mathbf{r}\}},
\end{align} 
which, after taking the logarithm and recalling that $\alpha = d \beta$ is satisfied, yields
\begin{equation}
 \left.G_{\theta,a}^{(n)}(\{\mathbf{x}_{i}\},t)\right|_{\mathbf{x}_{i}\in\{\mathbf{0},\mathbf{r}\}}
  = \left.G_{\theta,a}^{(n)}(\{s^{-1}\mathbf{x}_{i}\},s^{-1/\beta}t)\right|_{\mathbf{x}_{i}\in\{\mathbf{0},\mathbf{r}\}},
  \label{eq:PhaseCumulants2n}
\end{equation}
or, after taking $s = t^{\beta}$,
\begin{equation}
\left.G_{\theta,a}^{(n)}(\{\mathbf{x}_{i}\},t)\right|_{\mathbf{x}_{i}\in\{\mathbf{0},\mathbf{r}\}}
  = G_{\mathrm{S},\theta,a}^{(n)}([t/t_\mathrm{ref}]^{-\beta}\mathbf{r})\,
\end{equation}
with $\beta$ independent of the correlator order $n$.

To summarize, we have shown that if density fluctuations are suppressed and hence \eqref{eq:napFTofPhaseCorr} and \eqref{eq:PhaseCorrCa} are applicable, the two-point equal-time correlator of fundamental Bose fields can be expressed in terms of connected correlation functions of the phase field as \eqref{eq:PhaseCorrCumSum}. This furthermore implies, recalling that the particle number associated with the fundamental field $\Phi_a$ is conserved, a strong constraint on the scaling of the phase field correlators, \eqref{eq:PhaseCumulants2n}. This proves that, under the above assumptions, the scaling exponents derived from the Luttinger-liquid-like low-energy EFT coincide with those of the fundamental bosonic particles.  

Finally, let us consider the leading-order term, $n=2$~\footnote{Recall that the expectation value of the phase field is zero, so that the $n=1$ term does not contribute.}, in \eqref{eq:PhaseCorrCumSum} which corresponds to the Gaussian approximation, 
\begin{equation}
 C_{a}(\mathbf{r},t)
  \simeq \exp\left\{G^{(2)}_{\theta,a}(\mathbf{r},t) - G^{(2)}_{\theta,a}(\mathbf{0},t)\right\}.
  \label{eq:PhaseCorrCumGauss}
\end{equation}
If the Gaussian approximation is sufficient, which is going to be discussed below, we can infer the scaling of the phase correlator $C_{a}$ from the scaling of the 2-point equal-time cumulant $G^{(2)}_{\theta,a}$ of the angle. According to \eqref{eq:fScalingForm}, \eqref{eq:fScalingFunctionForm}, and \eqref{eq:kappa4result}, we can take the momentum-space correlator \eqref{eq:DefPhaseAngleCorr} to have the form
\begin{equation}
  f_{a}(\mathbf{k},t) 
  \sim k_{\Lambda}(t)\,|\mathbf{k}|^{-(d+1)}\,,
 \label{eq:PhaseAngleCorrofp}
\end{equation}
for momenta $k_{\mathrm{IR}} < |\mathbf{k}| < k_{\mathrm{UV}}$ in the power-law region. Here, the IR scale is expected to be associated with $k_{\Lambda}(t)$, while the UV scale reflects the presence of a thermal tail that is typically developed in quenched ultracold atomic systems. To see \eqref{eq:PhaseAngleCorrofp} explicitly, we note that \eqref{eq:fScalingFunctionForm} and \eqref{eq:kappa4result} require $f(\mathbf{k},t_{\mathrm{ref}}) \sim |\mathbf{k}|^{-(d+1)}$. On the other hand, \eqref{eq:fScalingForm} implies $f(\mathbf{k},t) \sim t^{\alpha} (t^{\beta} |\mathbf{k}|)^{-(d+1)}$, which combined with the aforementioned constraint $\alpha = d \beta$ results in \eqref{eq:PhaseAngleCorrofp}, where $k_{\Lambda}(t) \sim t^{-\beta}$.

A simple scaling analysis of the Fourier transform of \eqref{eq:PhaseAngleCorrofp} in $d$ dimensions gives the spatial two-point function $G^{(2)}_{\theta,a}(\mathbf{r},t)=\mathcal{F}[f_{a}](\mathbf{r},t)$ as 
\begin{equation}
  G^{(2)}_{\theta,a}(\mathbf{r},t)-C_{\theta,a}(\mathbf{0},t)
  =-\mathrm{const.}\times k_{\Lambda}(t)\,|\mathbf{r}|\,,
 \label{eq:app:ScalingPhaseAngleCorr}
\end{equation}
which is normalized to some constant at $r=0$, and we added a minus sign reflecting that the correlations are expected to fall off at larger distances. Combining these results, we find that the two-point correlation function of the fundamental field should have a pure exponential decay at intermediate distances, which is consistent with our numerical results (cf.~Fig.~\ref{fig:NTFPEvolution}c),
\begin{equation}
 C_{a}(\mathbf{r},t)
  \simeq \exp\left\{
        -k_{\Lambda}(t) \,|\mathbf{r}|\right\},
        \quad (k_{\mathrm{UV}}^{-1} \lesssim 
        |\mathbf{r}| \lesssim k_\mathrm{IR}^{-1}).
  \label{eq:PhaseCorrPureExp}
\end{equation}
This corresponds to the angle-averaged spatial first-order coherence function $g^{(1)}(\mathbf{x}-\mathbf{y},t)=\langle\Phi^{\dag}(\mathbf{x},t)\Phi(\mathbf{y},t)\rangle$ of the Bose field $\Phi(\mathbf{x},t)$
\begin{equation}
g^{(1)}_{a}(\mathbf{r},t) = \rho_{a}^{(0)}\exp\Big\{-k_{\Lambda}(t) \, |\mathbf{r}|\Big\}\,,\quad (k_{\mathrm{UV}}^{-1} \lesssim 
   |\mathbf{r}| \lesssim k_\mathrm{IR}^{-1}),
 \label{eq:app:NTFPscalingg1}
\end{equation}
with uniform particle density $\rho^{(0)}_{a}$ and background phase $\theta^{(0)}_{a}$.
Its Fourier transform yields the occupation number distribution of the bosons,
\begin{equation}
n_{a}^{\mathrm{G}}(\mathbf{k},t)=\frac{C_{d}k_{\Lambda}(t)}{\left[k_{\Lambda}(t)^{2}+|\mathbf{k}|^{2}\right]^{\zeta/2}}\,, \quad (k_{\mathrm{IR}} \lesssim 
|\mathbf{k}| \lesssim k_\mathrm{UV})
\label{eq:app:NTFPscalingnkG}
\end{equation}
with normalization constant $C_{d}=\rho^{(0)}_{a}\Gamma([d+1]/2)/\pi^{(d+1)/2}$ and momentum exponent $\zeta=d+1$.

Precisely this scaling form has been assumed in the previous non-perturbative analysis   \cite{Chantesana:2018qsb.PhysRevA.99.043620}, where the scaling at large momenta $|\mathbf{k}|\gg k_{\Lambda}(t)$ was predicted to be given by $\zeta=d+1$ for $z=2$ as recovered in the Gaussian approximation assumed here. Furthermore, $n_{a}(\mathbf{k},t)$ was assumed to satisfy a scaling relation of the form \eqref{eq:fScalingForm}, 
\begin{equation}
\label{eq:nakScaling}
n_{a}(\mathbf{k},t) = s^{\alpha/\beta} n_{a}(s \mathbf{k},s^{-1/\beta}t)\,,
\end{equation}
such that the time-dependent constant scales as $k_{\Lambda}(t)\sim t^{-\beta}$.
The  exponents were predicted to be $\beta=1/z=1/2$ and $\alpha=\beta\,d$, assuming $z=2$ and taking into account the conservation of the momentum integral over $n_{a}(\mathbf{k},t)$, cf.~\cite{Orioli:2015dxa,Chantesana:2018qsb.PhysRevA.99.043620}.

Hence, provided the validity of the Gaussian approximation \eqref{eq:PhaseCorrCumGauss}, the scaling properties derived here in the large-$N$ limit, defined by the exponents $z=2$, $\alpha=d/2$, $\beta=1/2$, and $\kappa=d+1$, as well as by the scaling form \eqref{eq:app:ScalingPhaseAngleCorr}, are consistent with the scaling properties derived within the non-perturbative approach of \cite{Orioli:2015dxa,Chantesana:2018qsb.PhysRevA.99.043620}, for the case of a dynamical exponent $z=2$, and a vanishing anomalous dimension $\eta=0$.

As we will discuss in more detail in the following, the Gaussian approximation, for the situations studied here, is supported by the scaling analysis of the effective action. One expects this limit to be reached, ideally, for infinite system size in position as well as momentum space, at $t\to\infty$. In this limit, the interaction part of the action vanishes as compared to the part quadratic in the phase angle fields, such that also any correlations are expected to become Gaussian. This will, however, not be the general case. 
As we will argue, for sufficiently small $\beta$, the respective fixed point is expected to have a non-Gaussian character.

In summary, we observe that the values of the exponents obtained above are in a agreement with the results of \cite{Chantesana:2018qsb.PhysRevA.99.043620}, under the assumption of a vanishing anomalous scaling, $\eta = 0$. 
With this, we conclude that, if 
(a) the kinetic description is adequate;
(b) the $T$-matrix has the form \eqref{eq:T_3} and \eqref{eq:T_4} for the three- and four-point interactions, respectively;
(c) the quasiparticle number or energy of the system are conserved in the scaling regime;
then our low-energy EFT predicts the universal self-similar dynamics close to the non-thermal fixed point in agreement with the previously developed non-perturbative approach in the special limit of infinite $N$.

\subsubsection{The case $N=1$ of a single field component}
\label{sec:SingleCompScaling}
One can also apply our scaling analysis to the limiting case $N=1$ in which the action, in the next-to-leading order of a derivative expansion, reads as
\begin{align}
S_{\mathrm{eff}}[\theta] 
=
\frac{1}{2 g} &\int_{x,\mathcal{C}} \Big\lbrace \theta(x) \left(-\partial^2_t + c_s^2 \nabla^2 \right) \theta(x) \nonumber\\
&+
\frac{1}{m} \partial_t \theta(x) \left(\nabla \theta (x) \right)^2 + \frac{1}{4 m^2} \left(\nabla \theta(x)\right)^4 \Big\rbrace\,,
\end{align}
with $c_s^2 = \rho^{(0)} g/m$ being the speed of sound. 
It describes, in the IR limit, the scattering of modes with linear Bogoliubov dispersion, with $z=1$ and $\gamma=0$.
Equivalently, the tensor couplings in Eqs.~\eq{SeffnG3thetaa} and \eq{SeffnG4thetaa} reduce to
${g^{-1}_{\mathrm{1/N}}(\mathbf{k})} [1 - ({1 + \mathbf{k}^2/2 k^2_{\Xi}})^{-1}]=g^{-1}({1 + \mathbf{k}^2/2 k^2_{\Xi}})^{-1}$, i.e., at low energies $k\ll k_{\Xi}$, the coupling $g_{1/N}(\mathbf{k})$ in \eq{Seff4} is replaced by the constant bare coupling $g$. 

Given the scaling behavior of the dispersion and the coupling, the scaling analysis of the kinetic equation proceeds, for $N=1$, as described above in the large-$N$ limit. 
In particular, the scaling relations between the different exponents, for general $z$ and $\gamma$, are equally valid in both cases.
Inserting $\gamma=0$ and $z = 1$ into \eqref{eq:betaQ3} and \eqref{eq:betaQ4}, one finds for quasiparticle transport the exponents 
\begin{align}
&l = 3: \quad \beta = 1/2\,,
\label{eq:beta3N1}
\\
&l = 4: \quad \beta = 1/3\,,
\label{eq:beta4N1}
\end{align}
and, from \eqref{eq:mu3QP} and \eqref{eq:mu4QP}, that $-\mu_{3}=d-2>-\mu_{4}=d-3$. Hence, with increasing time, the scattering integral $I_{4}$, for $z<2$, loses out against $I_{3}$ such that the transport is predicted to occur, in the scaling limit, with $\beta=1/2$ according to \eqref{eq:beta3N1}. As a result, despite $z=1$, the same values $\alpha=d/2$ and $\beta=1/2$ describe transport of quasiparticles towards the IR, dominated by one-to-two scattering.

Determining the spatial exponent $\kappa$ by using the same arguments as for the case $z=2$, which imply that  the 4-wave interactions dominate, one finds $\kappa=\kappa_{4}=d+3/2$. Note, however, that according to the arguments leading to the dominance of $I_{3}$ in the long-time limit, one rather expects, in the case $N=1$, the three-wave interaction to also dominate the stationary fixed-point equation. Under these circumstances, we rather predict
\begin{align}
  \kappa=\kappa_{3} = d+2\, \qquad(z=1)
  \label{eq:kappa3resultBog}
\end{align}
to represent the momentum scaling exponent in the large-time scaling limit $t \to \infty$. It should be remarked that this exponent describes the scaling of a vortex-dominated state, for $d\ge2$, as found in many simulations for the case $N=1$ (cf., e.g., \cite{Nowak:2010tm,Nowak:2011sk,Schole:2012kt,Nowak:2012gd,Karl2017b.NJP19.093014}).
This is rather unexpected as the Luttinger-liquid-based kinetic approach chosen here does not take into account the compact property of the phase and thus the excitations of vortices. The question as to whether the above result is just a coincidence or consequence of some physics thus remains open.

\subsubsection{Upper ``critical'' dimension}
\label{sec:UpperCritDim}
Given the above predictions for the scaling behavior at non-thermal fixed points we can use standard power-counting arguments to obtain the respective upper ``critical'' dimension. With this, we refer to the spatial dimension above which non-Gaussian terms in the action scale to zero at the non-thermal fixed point.

Note, however, that there is no critical point reached in the conventional sense: the scaling limit is expected to be attained, in the thermodynamic limit, at infinite evolution times. In practice, however, scaling already shows up at intermediate times, while the approach of the precise scaling function defining the scaling limit takes asymptotically long times. This can be conjectured from the results of our simulations shown in Fig.~\ref{fig:NTFPEvolution}, in particular also from the observation that the higher-order powers in $k_{\Lambda}(t)r$ contributing to the scaling form of $g^{(1)}(r,t)$ take much longer to form than the low-order ones (see \cite{Schmied:2018upn.PhysRevLett.122.170404}).

To obtain an estimate for the upper ``critical'' dimension, we analyze the relative scaling of the quadratic and higher-order contributions to the effective action \eqref{eq:Seff4}. Considering the Schwinger-Keldysh action $S(t)=\int_{\mathcal{C},t_\mathrm{ref}}^{t}dt'L(t')$ integrating from a reference time $t_\mathrm{ref}$ to the present evolution time $t$ and back, its scaling is defined as
\begin{equation}
\label{eq:dSdefinition}
S(s^{-1/\beta} t) = s^{d_{S}} S(t),
\end{equation}
with canonical scaling dimension $d_{S}=[S]$ of the action. Using the dynamical canonical scaling dimension of the field $\theta_{a}$ at the non-thermal fixed point,
\begin{equation}
  [\theta_{a}] = -\alpha/2\beta=-d/2\,,
  \label{eq:canScDimtheta}
\end{equation}
we obtain (see Appendix~\ref{app:ActionScaling} for details) the canonical scaling of the quadratic, cubic and quartic parts of the action \eqref{eq:Seff4},
\begin{subequations}
\begin{align}
  [S^{(2)}] &= 2z-\gamma-1/\beta &=&\ 0\,,
  \label{eq:canScDimS2}
  \\
  2[S^{(3)}] &= d+4+2z-2\gamma-2/\beta&=&\ d+4-2z\,,
  \label{eq:canScDimS3}
  \\
  [S^{(4)}] &= d+4-\gamma-1/\beta&=&\ d+4-2z\,,
  \label{eq:canScDimS4}
\end{align}
\end{subequations}
where we considered twice the scaling dimension of the 3-vertex as it occurs in even multiples in any diagram contributing to the two-particle irreducible (2PI) effective action, and inserted, in the respective second equations, $\gamma=2z-2$ and $1/\beta=2$.

For the non-thermal fixed point to be a Gaussian fixed point in the IR scaling limit, the relative scaling conditions
\begin{align}
  2[S^{(3)}]>[S^{(2)}]\,,\qquad
  [S^{(4)}]>[S^{(2)}]
  \label{eq:GaussianFPcond}
\end{align}
need to be fulfilled.
This is the case in any dimension $d>0$, as $d+4-2z>0$ for $z<(d+4)/2$.

A remark is in order. The above analysis is to be understood, in general, as giving a first estimate as, strictly speaking, time-dependent correlation functions  need to be analyzed, e.g., within a functional renormalization-group approach \cite{Pawlowski:2005xe,Gasenzer:2008zz,Berges:2008sr} by studying the flows of $\Gamma^{(n)}$. Nonetheless, for the fixed point considered here, the above arguments are unlikely to fail for our fixed point in the large-$N$ limit since $1/\beta=z$ and, thus, any scaling in time is governed by the same exponent. The Gaussianity of the fixed point is further corroborated by the fact that also the scattering described by the integrals $I_{l}[f]$ scales to zero in the infinite-time limit. This can be inferred from their scaling exponents $\mu_{l}$ defined in \eqref{eq:mu3QP} and \eqref{eq:mu4QP}, since, for $z=2$, $\mu_{3}=\mu_{4}=2-d$, such that $-\mu_{l}-\alpha/\beta=-2<0$ and thus $I_{l}[f](\mathbf{k},t)=t^{-2}I_{l}[f]([t/t_0]^{1/2}\mathbf{k},t_0)$. One should, however, be careful since that may rather indicate a \emph{critical slow down}.  
Furthermore, we emphasize that Gaussianity of the non-thermal fixed point here refers to phase quasiparticles only, while in terms of the fundamental fields the fixed point can easily appear to be non-Gaussian.

In this context, we furthermore point out that the scaling of the $T$-matrix with $m_{l}=2$ is crucial for the validity of the perturbative Boltzmann approach.
As laid out in detail in Ref.~\cite{Chantesana:2018qsb.PhysRevA.99.043620}, the overoccupied momentum distributions would lead to a divergence of the scattering integral in the low-momentum limit and thus to a breakdown of the perturbative kinetic Boltzmann description of the transport process. 
In the context of wave turbulence, this problem is well known to lead to what is called strong wave turbulence \cite{Nazarenko2011a}, requiring a modified kinetic equation.
Within the formulation in fundamental fields, a non-perturbative approximation of the kinetic equation is required to remedy this problem for the present case of a non-thermal fixed point, implying a $T$-matrix which scales to zero at low-momentum transfers \cite{Orioli:2015dxa,Berges:2015kfa,Chantesana:2018qsb.PhysRevA.99.043620}.
Here, however, the $T$-matrix scales to zero at low-momentum transfers already within the perturbative Boltzmann approximation, implying a Gaussian non-thermal fixed point and thus validity of the approach also in the scaling limit. 

Given the applicability of the above arguments, comparing \eqref{eq:canScDimS2} and \eqref{eq:canScDimS4}, we find that the 4-vertex is irrelevant as compared to the free action for $z<(d+4)/2$, independent of the value of $\beta$. Leaving $\beta$ arbitrary, we find that the condition for the non-thermal fixed point to be non-Gaussian in $d\leq d_\mathrm{up}$ gives the upper critical dimension,   
\begin{align}
 d_\mathrm{up}=2z-6+\beta^{-1}\,,
  \label{eq:dup}
\end{align}
which is determined solely by the 3-vertex. 
To obtain an upper critical dimension $d_\mathrm{up}\geq1$ requires
\begin{align}
 \beta^{-1}\geq d_\mathrm{up}+6-2z.
  \label{eq:condupdgt1}
\end{align}
For example, considering the scaling exponent $\beta^{-1}\simeq5$ found in \cite{Karl2017b.NJP19.093014,Johnstone2018a.arXiv180106952} in $d=2$ dimensions and assuming the above general arguments to be applicable there, the value of $\beta$  implies $d_\mathrm{up}=2z-1\geq1$ for $z\geq1$.
Hence, for the fixed point to be a non-Gaussian one in $d=2$ dimensions, the relevant $z$ is required to be equal to or exceed $z=3/2$.  

We finally remark that one can analyze, in an analogous manner, the stationary spatial scaling of the cubic and quartic parts of the Lagrangian as compared to the Gaussian part.
This shows as well that, for the cases considered here, the upper critical dimension vanishes or is negative.
Hence, we expect that in the scaling limit, the Gaussian approximation considered in the previous sections is valid. 
Nevertheless, the interaction part of the action is crucial for the universal scaling transport to occur at all. 
In the course of the rescaling, leading closer to the fixed point, these interactions become increasingly suppressed, vanishing asymptotically at the fixed point.

\subsection{Concluding remarks on the scaling dynamics}
To close the topic, let us briefly summarize the results obtained above and compare them with already known ones. 

For the model \eq{ONGPH}, the scaling exponent $\beta$ has previously been predicted, within a large-$N$ approximation, by means of a scaling analysis of non-perturbative wave-Boltzmann-type kinetic equations governing the time evolution of $n_{a}(\mathbf{k},t)$ \cite{Orioli:2015dxa,Berges:2015kfa,Chantesana:2018qsb.PhysRevA.99.043620}. As the scaling evolution is subject to a conservation law, $\alpha$ and $\beta$ are not independent. In the large-$N$ limit studied there, the scaling dynamics corresponds to transport in momentum space which conserves the density, $\rho_{a}=\int_{\mathbf k}n_{a}(\mathbf{k},t)\equiv$ const. \footnote{At finite times, on the way towards the fixed point, the integral only includes the long-wave-length modes up to a scale $k_{\Lambda}(t)$ evolving with the distribution as $k_{\Lambda}(t)\sim t^{-\beta}$ and is expected to be approximately conserved. Furthermore, as we showed, the transport shifts relative-phase excitations to the IR which only in the $N\to\infty$ limit decouple from each other and correspond to single-particle excitations in each component.}, i.e., one finds $\alpha=\beta\,d$ in $d$ dimensions. As $\beta$ is positive, the transport is directed toward the infrared.

Analyzing this transport to lower $k$ by means of kinetic equations describing the four-wave scattering between free quasiparticle modes, it was found that $\beta=1/z$ \cite{Chantesana:2018qsb.PhysRevA.99.043620}. This result is independent of  $d$ and $N$ and related to the dynamical scaling exponent $z$ governing the dispersion $\omega_{\mathrm{qp}}(\mathbf{k})\sim k^{z}$ of the quasiparticles entering the kinetic description of the transport.

Our low-energy effective theory derived above provides an alternative, perturbative way of predicting scaling evolution of the type \eq{NTFPscaling}, which is complementary to the non-perturbative approach chosen in \cite{Orioli:2015dxa,Berges:2015kfa,Chantesana:2018qsb.PhysRevA.99.043620}.
According to our theory, the occupation numbers at  $k\lesssim k_{\Xi}$ are  dominated by the phase excitations $n_{a}(\mathbf{k},t)\simeq \rho^{(0)}_{a}\mathcal{F}[\langle e^{-\i\theta_{a}(\mathbf{x},t)}e^{\i\theta_{a}(\mathbf{y},t)}\rangle](\mathbf{k})$, where $\mathcal{F}$ denotes the Fourier transform with respect to $\mathbf{x}-\mathbf{y}$ of the on average translation invariant phase correlator (cf.~Appendix~\ref{app:Not}). We find analytically that the dynamical exponent $z$, in the large-$N$ limit, is $z=2$, corresponding to the Goldstone dispersion \eq{GoldstoneFreq} and confirming the numerical results of, e.g., \cite{Schachner:2016frd,Walz:2017ffj.PhysRevD.97.116011}.

The effective action \eqref{eq:Seff4} includes non-linear interaction terms $\sim\theta^{l}$, $l=3,4$, between the phase modes. These interactions lead to the non-linear transport describing scaling evolution of the type \eq{NTFPscaling}. This can be shown in terms of kinetic wave-Boltzmann equations governing the phase excitation spectra $\langle\theta_a(\mathbf{k},t) \theta_a(-\mathbf{k},t)\rangle$, in analogy to \cite{Orioli:2015dxa,Chantesana:2018qsb.PhysRevA.99.043620}, cf.~\Sect{Kinetic}. As we outline in more detail in \Sect{ScalingLEEFT}, for the four-wave interaction term, which dominates the spatial scaling in the IR limit, the scaling analysis of our kinetic integrals in a large-$N$ approximation yields $\beta=1/2$ and $\zeta=d+1$.
Hence, the infrared scale is predicted to algebraically decrease in time, $k_{\Lambda}(t)\sim t^{-\beta}$.

We have, moreover, studied the case of $N=1$. We find, in particular, the same value for the exponent $\beta=1/2$ is predicted despite the different dynamical exponent $z=1$ of the sound-wave dispersion, linearly scaling at low momenta, see \Sect{SingleCompScaling}. The cases $1\leq N<\infty$ are less straightforward and will be the subject of a forthcoming publication. 

Our analytical results are, within errors, consistent with the truncated Wigner simulations shown in \Fig{NTFPEvolution}. This is seen when taking into account that in leading approximation, the phase and phase-angle correlators of the translationally invariant system are related by $\langle e^{-\i\theta_{a}(\mathbf{x},t)}e^{\i\theta_{a}(\mathbf{y},t)}\rangle\simeq\exp\{\langle\theta_a(\mathbf{x},t) \theta_a(\mathbf{y},t)\rangle\}$.
From this relation, the phase-angle correlator is approximately given by
\begin{equation}
  \langle\theta_a(\mathbf{r}+\mathbf{x},t) \theta_a(\mathbf{x},t)\rangle 
  \approx -k_{\Lambda}(t)\,|\mathbf{r}| + \mathcal{O}(|\mathbf{r}|^{2}).
  \label{eq:PhaseCorr}
\end{equation}
If this is valid in $d$ dimensions, up to additive terms involving cutoffs, the phase correlator scales as $\langle\theta_a(\mathbf{k},t) \theta_a(-\mathbf{k},t)\rangle\sim |\mathbf{k}|^{-(d+1)}$, as analytically predicted within our low-energy effective theory ($\zeta=d+1$).

We find that the consistency between our perturbatively obtained scaling exponent and the non-perturbative results of \cite{Orioli:2015dxa,Walz:2017ffj.PhysRevD.97.116011,Chantesana:2018qsb.PhysRevA.99.043620} crucially relies on the momentum scaling of the effective coupling \eq{app:g1N} which is shown here to be related to the scaling ${\sim}\mathbf{k}^{2}$ of the hydrodynamic kinetic energy.
In fact, $g_\mathrm{1/N}$ depends only on momentum, density, and particle mass, but not on the microscopic coupling $g$. These same properties were found for the universal effective coupling function which enters the non-perturbative kinetic scattering integrals derived previously \cite{Chantesana:2018qsb.PhysRevA.99.043620}.

We stress that, whereas in \cite{Maraga2015a.PhysRevE.92.042151, Maraga2016b.PhysRevB.94.245122, Chiocchetta2015a.PhysRevB.91.220302, Chiocchetta2016a.PhysRevB.94.134311,
Chiocchetta:2016waa.PhysRevB.94.174301,
Chiocchetta2016b.161202419C.PhysRevLett.118.135701} 
scaling evolution after quenches of the gap and interaction parameters from a thermal initial state have been discussed in the context of a large-$N$ approximation applied to a scalar $\mathrm{O}(N)$ model, as well as perturbative, $\varepsilon$-expansion, and functional renormalization group, the resulting scaling forms do not allow for the scaling to be discussed here.
While the cited work accounts for scaling in the context of critical coarsening, initial-slip dynamics and ageing phenomena \cite{Janssen1989a,Janssen1992a,Calabrese2002a.PhysRevE.65.066120,Calabrese2005a.JPA38.05.R133,Gambassi2006a.JPAConfSer.40.2006.13}, for the type of quenches considered there, the same (initial-slip) scaling exponent results for the correlation and response functions \cite{Gambassi2006a.JPAConfSer.40.2006.13}.
It is an interesting question beyond the scope of this work as to how initial-slip scaling manifests in the context of a non-thermal fixed point.

We finally emphasize that the non-thermal fixed point identified within the low-energy effective theory derived here turns out to have a Gaussian character with respect to the relevant degrees of freedom. As we discussed in more detail in \Sect{UpperCritDim}, the contributions to the action which are cubic and quartic in $\theta$, exhibit subdominant scaling as compared to the Gaussian part and thus rescale to zero in the scaling limit.
One thus obtains, by standard scaling arguments for the bare action,  an upper `critical' dimension which is negative for the non-thermal fixed point with $\beta=1/2$, dominated by the sound and relative-phase excitations.
We anticipate that fixed points with smaller $\beta$, as observed in \cite{Karl2017b.NJP19.093014,Johnstone2018a.arXiv180106952}, involving the interaction of non-linear excitations and topological defects, may have a positive upper `critical' dimension and thus have a non-Gaussian character
\footnote{The scaling limit of the non-thermal fixed point is reached at late evolution times provided that the initial condition is tuned such that no competing processes can prevent this.
For a given system size this is typically ensured by a sufficiently extreme out-of-equilibrium initial state, such as the box distribution used here.}.

\begin{figure}
\includegraphics[width=0.85\columnwidth]{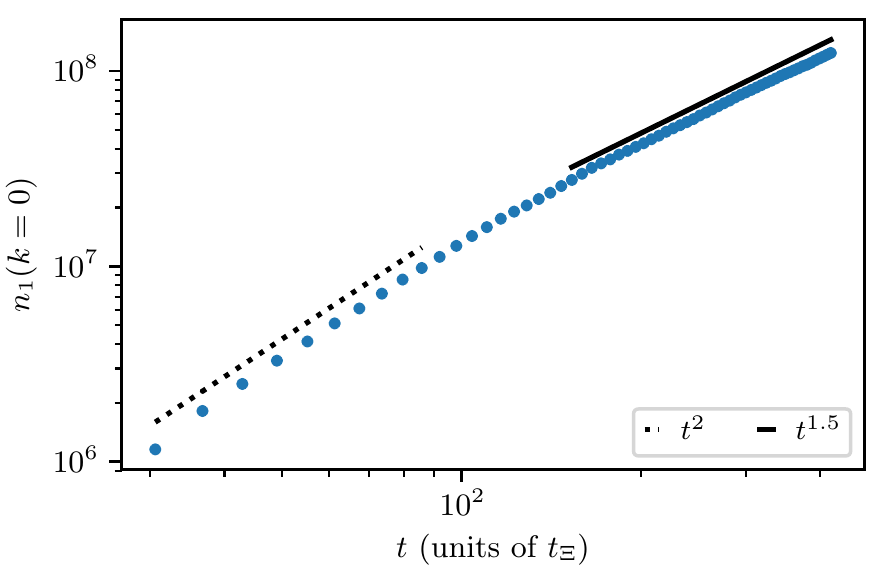}
\caption{
Zero-momentum mode of the single-component occupation number $n_1(k=0,t)$ (blue dots). According to \eq{NTFPscaling} the universal time evolution is given by $n_1(k=0,t) = (t/t_{\mathrm{ref}})^{\alpha} f_{\mathrm{S},1}(k=0)$. 
At late times, $t \gtrsim 200\, t_{\Xi}$, we find $\alpha \simeq d/2 = 3/2$ (black solid line) consistent with the analytical prediction within our low-energy EFT.
\label{fig:n1k0}
}
\end{figure}

\begin{figure}
\includegraphics[width=0.96\columnwidth]{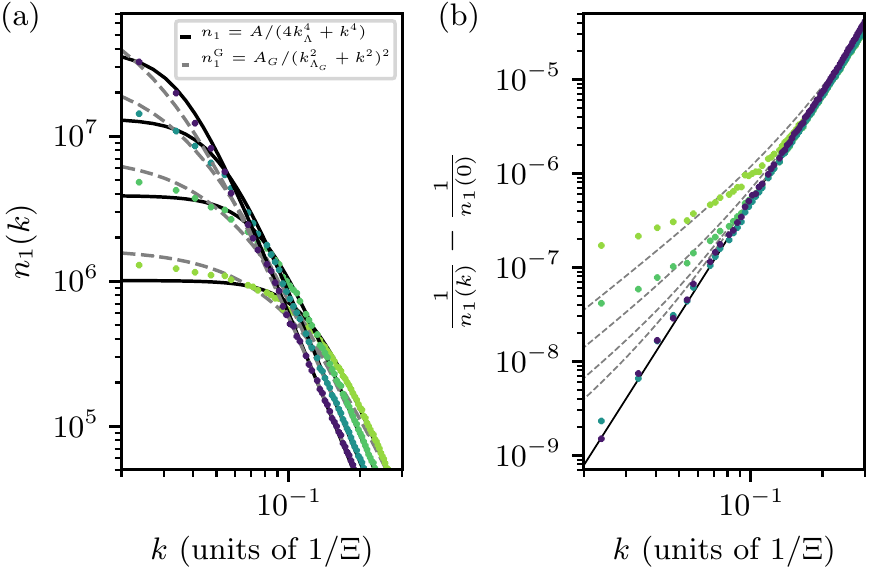}
\caption{(a) Enlarged representation of the infrared scaling evolution of the single-component occupation number $n_{1}(k)\equiv n_1(\mathbf{k},t)$, for the same evolution times as shown in \Fig{NTFPEvolution} (same color coding). 
The solid black and dashed grey lines show the results obtained by fitting the corresponding scaling functions to the IR part of the distribution.
At late times (cyan and purple), the data are close to the scaling function $n_{1}(\mathbf{k},t)$, defined in \eq{NTFPscalingnk}, 
which corresponds to a first-order coherence function with exponential times cardinal-sine form, \eq{NTFPscalingg1}.
For all evolution times the data differs from the scaling function $n_{1}^{\mathrm{G}}(\mathbf{k},t)$ defined in \eq{app:NTFPscalingnkG} which corresponds to the purely exponential first-order coherence function \eq{app:NTFPscalingg1}.
This shows that, although we observe an exponential decay at short distances in the first-order coherence function, an additional oscillatory contribution is present (see \cite{Schmied:2018upn.PhysRevLett.122.170404} for more details). 
Note that we do not claim \eq{NTFPscalingnk} to be the precise scaling form.
(b) $n_{1}(k)^{-1}-n_{1}(0)^{-1}$, with the respective extrapolated fit value inserted for $n_{1}(k=0)$ in order to be independent of possible deviations due to the build-up of a condensate in the zero mode. This representation clearly shows that the data are better described by the scaling function $n_{1}$ at late times. The solid black line corresponds to the fit of  \Eq{NTFPscalingnk} to the data for the latest time shown (purple dots).
\label{fig:InverseOcc}
}
\end{figure}

\begin{figure*}
\includegraphics[width=0.96\textwidth]{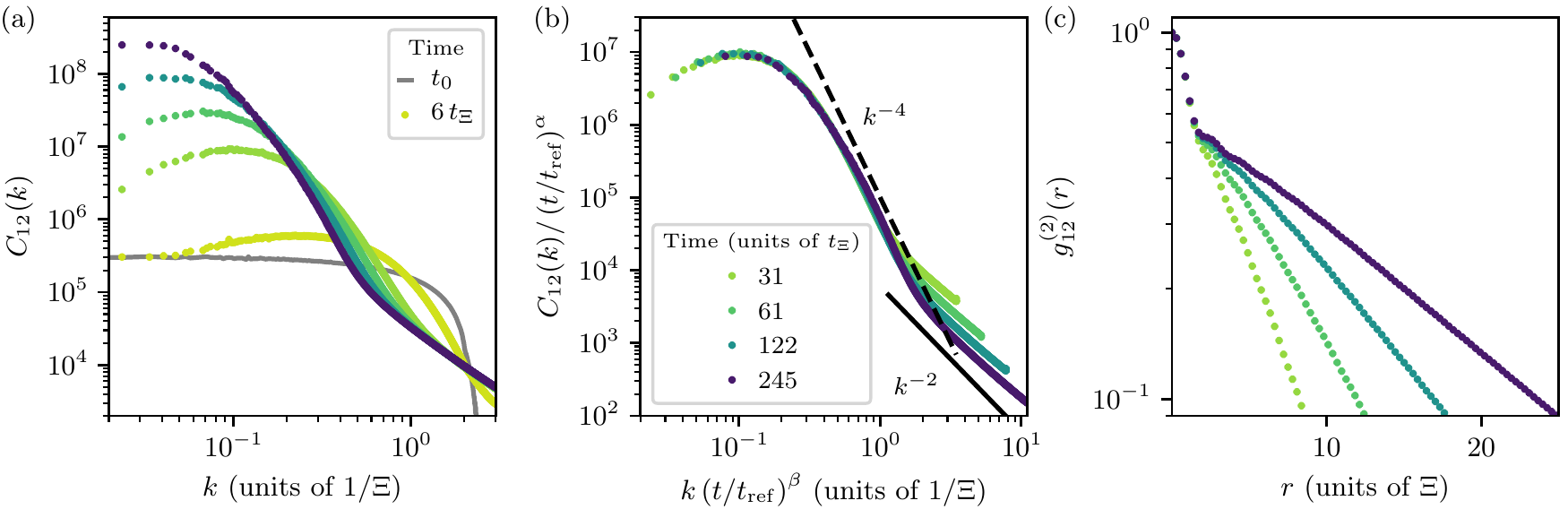}
\caption{(a) Universal scaling dynamics of the relative phases $C_{12}(k,t) = \langle\lvert (\Phi_1^{\dagger} \Phi_2)(\mathbf{k},t)\rvert ^2\rangle $ for a ($d=3$)-dimensional gas with $N=3$ components. The time evolution is starting from the initial box distribution. Colored dots show $C_{12}$ at five different times. (b) The collapse of the data to the universal scaling function $f_{S,C} (k) = C_{12}(k,t_{\mathrm{ref}})$, with reference time $t_{\mathrm{ref}} / t_{\Xi} = 31$ (units of $t_{\Xi}^{-1} = g \rho^{(0)} / 2\pi$), shows the scaling \eq{fScalingForm} in space and time.
The universal scaling of $C_{12}(k,t)$ confirms our hypothesis that the scaling behavior also affects the relative phases of the components. 
Note that $C_{12}(k,t)$ does not show a plateau in the IR but goes over to follow a scaling function with similar fall-off at higher momenta as the one for $n_{a}(k,t)$.
The power law $C_{12}\sim k^{-4}$ in the scaling regime is only slightly modified as compared to that of $n_{a}$. 
(c) Corresponding first-order coherence function of the relative phases $g_{12}^{(1)} (r,t)$ (colored dots), for the same system and the same color encoding of time as in (a) and (b). 
Similar to the single-component first-order coherence functions we an exponential form of $g_{12}^{(1)} (r,t)$ at short distances $r \gtrsim \Xi$.
Note the semi-log scale.
\label{fig:RelativePhase12}}
\end{figure*}

\section{Numerical simulations for $N=3$ and $d=3$}
\label{sec:TWA}
In the remainder of this article we complement our numerical results presented in Ref.~\cite{Schmied:2018upn.PhysRevLett.122.170404} and briefly in \Sect{Intro} with further details and observables.
We have solved the coupled Gross-Pitaevskii equations resulting from the Hamiltonian \eq{ONGPH} for $N=3$, in $d=3$ spatial dimensions, by means of a spectral split-step algorithm.
We compute the time evolution of the correlation functions within the semi-classical truncated Wigner approximation \cite{Blakie2008a}.

We start with a far-from-equilibrium initial condition given by a ``box'' momentum distribution $n_{a}(\mathbf{k},t_0) = n_0\, \Theta(k_{\mathrm{q}} - \lvert \mathbf{k}\rvert)$, where, in the thermodynamic limit one can express the occupation number as  $n_{0}=(4\pi k_{q}^{3})^{-1}\rho^{(0)}$, and all phases $\theta_{a}(\mathbf{k},t_0)$ are chosen randomly on the circle.
As before, $a = 1,2,3$ enumerates the three components. 
We choose the momentum cutoff for the initial condition to be $k_{\mathrm{q}} = 1.4 k_{\Xi}$ leading to an initial occupation number of $n_{0} \simeq 2350$.
All simulations are performed on a grid with $N_{\mathrm{g}} = 256^3$ grid points using periodic boundary conditions. 
The corresponding physical volume of our system is $V = N_{\mathrm{g}} \Xi^3$. 
The total particle number is $\mathcal{N} =\rho^{(0)}V= 6.7 \times 10^9$, i.e., we have $\mathcal{N}_a = 2.23 \times 10^9$ particles in each of the three components. 
The correlation functions are averaged over 144 trajectories.
Varying the UV cutoff of the grid does not change our numerical results.

The initial condition in our simulations is chosen to take an extreme out-of-equilibrium form.
The modes are strongly occupied, on the order of $n_{0}\sim\zeta_\mathrm{d}^{-3/2}$. 
Here, $\zeta_\mathrm{d}=a{\rho^{(0)}}^{1/3}$ is the diluteness parameter which measures the scattering length $a$ in units of the inter-particle spacing.
The occupancy distribution is cut off at a maximum scale $k_\mathrm{q}$ which is on the order of the relevant (healing) length scale set by the interactions, i.e., of $k_{\Xi}=(8\pi a\rho^{(0)})^{1/2}$.
In physical terms, this means that all particles are placed, predominantly, at a momentum scale with kinetic energy  equal to the interaction energy or chemical potential $\mu=g\rho^{(0)}$ the gas has stored if it was fully Bose condensed.
The phases of the excited modes are random, i.e., no significant coherence prevails.

In practice, one may prepare such an initial state, e.g., by applying a strong cooling quench to an equilibrated system or by allowing the system to pass through an instability after a parameter quench \cite{Berges:2008wm,Chantesana:2018qsb.PhysRevA.99.043620}.

During the intermediate to long-time evolution, the system can approach a non-thermal fixed point and obey the scaling evolution \eq{NTFPscaling} within a range of modes $0<k\lesssim k_{\Xi}$.

In \Fig{NTFPEvolution}a, we depict the evolution of the occupation numbers of our system, starting from the above-described  initial condition (gray box).
Figure~\ref{fig:NTFPEvolution}b shows that, for times $t\gtrsim t_\mathrm{ref}=31\,t_{\Xi}$, the momentum distribution, to a good approximation, undergoes a scaling evolution according to \eq{NTFPscaling}. 
Rescaling the distributions at different times during this period they collapse to a single universal scaling function. 

We have determined the respective scaling exponents $\alpha$ and $\beta$ resulting from least-square rescaling fits of the occupation spectra within a time window $[t_\mathrm{ref}, t_\mathrm{ref} + \Delta t]$ with $\Delta t = 146\,t_{\Xi}$. 
The resulting $t_\mathrm{ref}$-dependent exponents are shown in \Fig{WindowFit1}, confirming, within errors, the relation $\alpha=\beta\,d$ in $d=3$ dimensions.
We find similar results (not shown) for the other components.
For all three components, the exponent $\beta$ is slightly larger than the value $\beta=1/2$  analytically predicted in the large-$N$ limit and for $N=1$.

Analyzing the time evolution of the zero-momentum mode occupation $n_1 (k=0,t)$ gives direct access to the scaling exponent $\alpha$ as the universal dynamics according to \eq{NTFPscaling} reduces to $n_1(k=0,t) = (t/t_{\mathrm{ref}})^{\alpha} f_{S,1}(k=0)$.
Figure~\ref{fig:n1k0} shows that at late times, $t \gtrsim 200\,t_{\Xi}$, the evolution is  governed by $\alpha \simeq d/2 = 3/2$ consistent with our analytic prediction in the case of a conserved quasiparticle number.

In \Fig{NTFPEvolution}c, we show the first-order spatial coherence function $g^{(1)}_{1}(\mathbf{r},t)=\langle\Phi_{1}^{\dag}(\mathbf{x}+\mathbf{r},t)\Phi_{1}(\mathbf{x},t)\rangle$ for different times.
They take, to a good approximation, an exponential form at short distances $r \gtrsim \Xi$.
Similar results are found (not shown) for the components $a=2,3$. 
Such an exponential form of the coherence function is reminiscent of that of a quasicondensate in an equilibrium gas in one spatial dimension \cite{Cazalilla2011a}.
Hence, the coherence function signals the buildup of a non-equilibrium quasicondensate in three spatial dimensions, rescaling in time and space towards a longer-range coherence.
Further studies of the correlation properties of this state concerning its relation to an equilibrium quasicondensate are beyond the scope of this work and are to be presented elsewhere.

In the following, we want to briefly comment on the form of the scaling function of the momentum distribution $n_a(k)$.
Therefore, we compare the momentum distribution, at different times, with two different scaling functions (see \Fig{InverseOcc}).
The first one, quoted in \eq{app:NTFPscalingnkG}, is obtained within the Gaussian approximation \eq{PhaseCorrCumGauss} of the relation between the phase-angle and the phase correlators.
It corresponds to the purely exponential first-order coherence function  \eq{app:NTFPscalingg1}.
The second scaling function used takes the form 
\begin{equation}
n_{a}(\mathbf{k},t)=C_{d}k_{\Lambda}(t)\left[4k_{\Lambda}^{d+1}(t)+|\mathbf{k}|^{d+1}\right]^{-1}. 
\label{eq:NTFPscalingnk}
\end{equation}
Taking the Fourier transform of \eq{NTFPscalingnk}, for $d=3$, gives 
\begin{equation}
  g^{(1)}(\mathbf{r},t) \approx \rho^{(0)}e^{-k_{\Lambda}(t)\,|\mathbf{r}|}
  \,{\mathrm{sinc}\big(k_{\Lambda}(t)\,|\mathbf{r}|\big)}\,,
\label{eq:NTFPscalingg1}
\end{equation}
which has an oscillatory contribution due to the cardinal-sine function.
We find that at late times, our data are approximatively described by the scaling function \eq{NTFPscalingnk}.
The scaling form \eq{app:NTFPscalingnkG}, however, differs from the data for all evolution times.
Hence, the corresponding first-order coherence function has an additional oscillatory contribution, which becomes visible at larger distances.
See \cite{Schmied:2018upn.PhysRevLett.122.170404} for more details.
Accounting for the oscillatory part within our analytical treatment requires a more refined analysis of non-linear phase excitations which is beyond the scope of this work.

We  remark that the scaling form \eq{NTFPscalingnk} is only one exemplary choice which takes into account the most striking features of $n_a(k)$.
To capture all details of the data, the scaling form might involve additional terms. 
To determine the exact scaling function is beyond the scope of this work. 

\begin{figure}
\includegraphics[width=0.7\columnwidth]{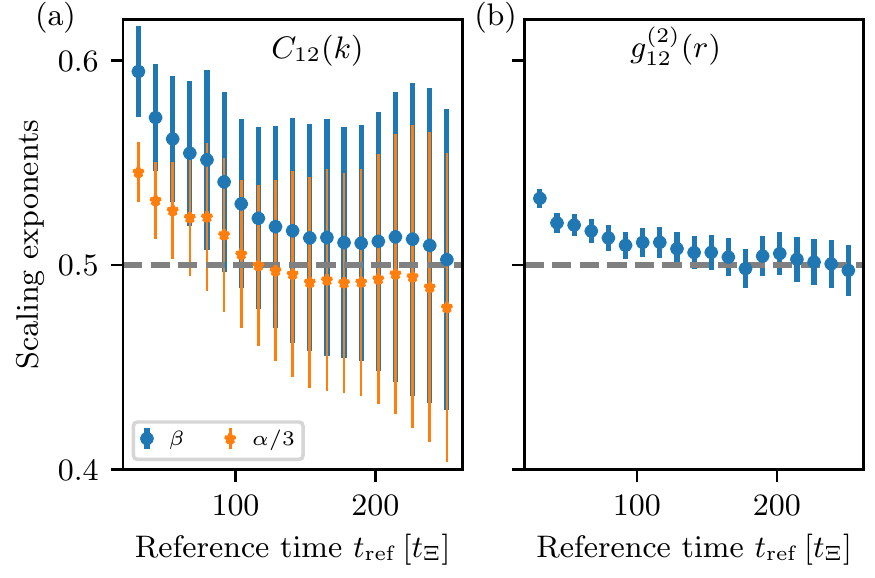}
\caption{Scaling exponents $\alpha/3$ (orange stars) and $\beta$ (blue dots) obtained from least-square rescaling fits of the occupancy spectra $C_{12}(k)\equiv C_{12}(\mathbf{k},t)$ shown in \Fig{RelativePhase12}b. 
The exponents correspond to the mean required to collapse the spectra within the time window $[t_\mathrm{ref},t_\mathrm{ref}+\Delta t]$ with $\Delta t = 146\,t_{\Xi}$.
Error bars denote the least-square fit error.
\label{fig:WindowFit12}
}
\end{figure}

\begin{figure*}
\includegraphics[width=2\columnwidth]{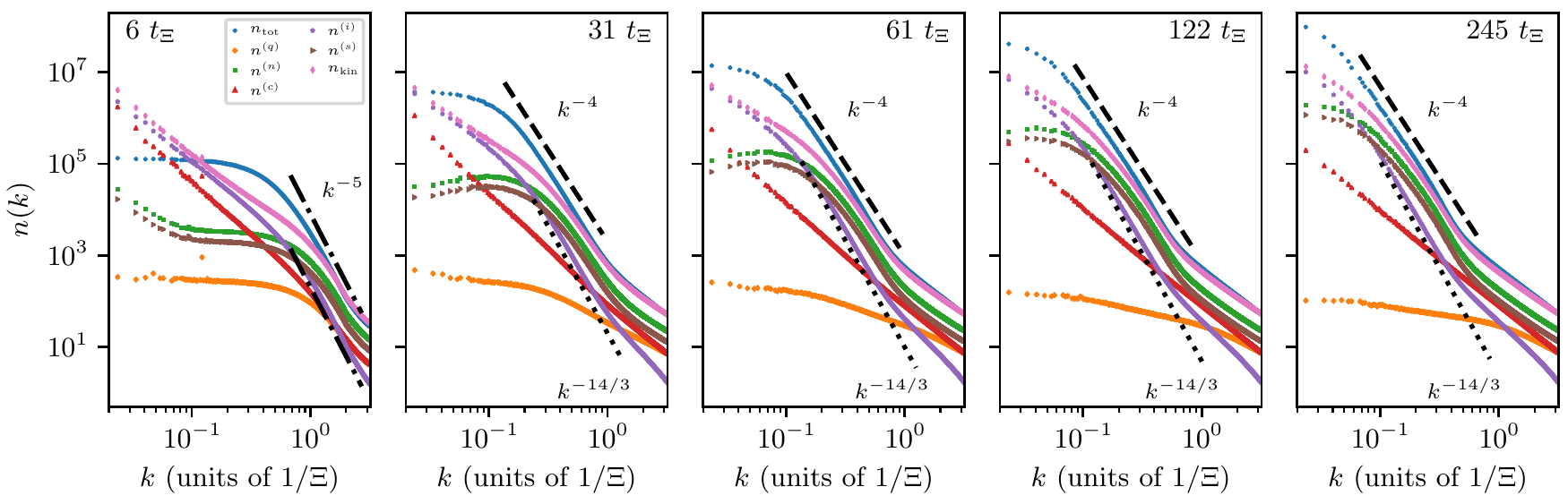}
\caption{Hydrodynamic decomposition of the flow pattern encoded in the phase-angle field $\theta_{a}$, at an early time as well as for the four evolution times in the universal scaling regime.
Shown are momentum distributions $n^{(\delta)}(k)$, derived from the decomposition of the kinetic energy density into $\varepsilon^{(\delta)}(k)=k^{2}\,n^{(\delta)}(k)$. 
The total occupation number $n_{\mathrm{tot}}$ (blue dots) is compared to the distributions representing the quantum pressure part $n^{(q)}$ (orange diamonds), the nematic $n^{(n)}$ (green squares), compressible $n^{(c)}$ (red triangles), incompressible $n^{(i)}$ (purple pentagons) and spin $n^{(s)}$ (brown arrows) parts, as well as their sum $n_\mathrm{kin}$ (pink thin diamonds). 
The incompressible part arising from vorticity in the system exhibits a power law decay with $n^{(i)}(k) \sim k^{-14/3}$ (see dotted line). 
The total occupation number is  dominated by the nematic and the spin parts except for momenta deep in the IR regime where the decomposition is expected to fail. 
The interplay of the dominant parts with the steep power-law in the incompressible part leads to an overall power-law decay of $n_{\mathrm{tot}}(k) \sim k^{-4}$ (see dashed line) at intermediate momenta. 
\label{fig:HydroDecompEvo}
}
\end{figure*}

\begin{figure*}
\includegraphics[width=0.96\textwidth]{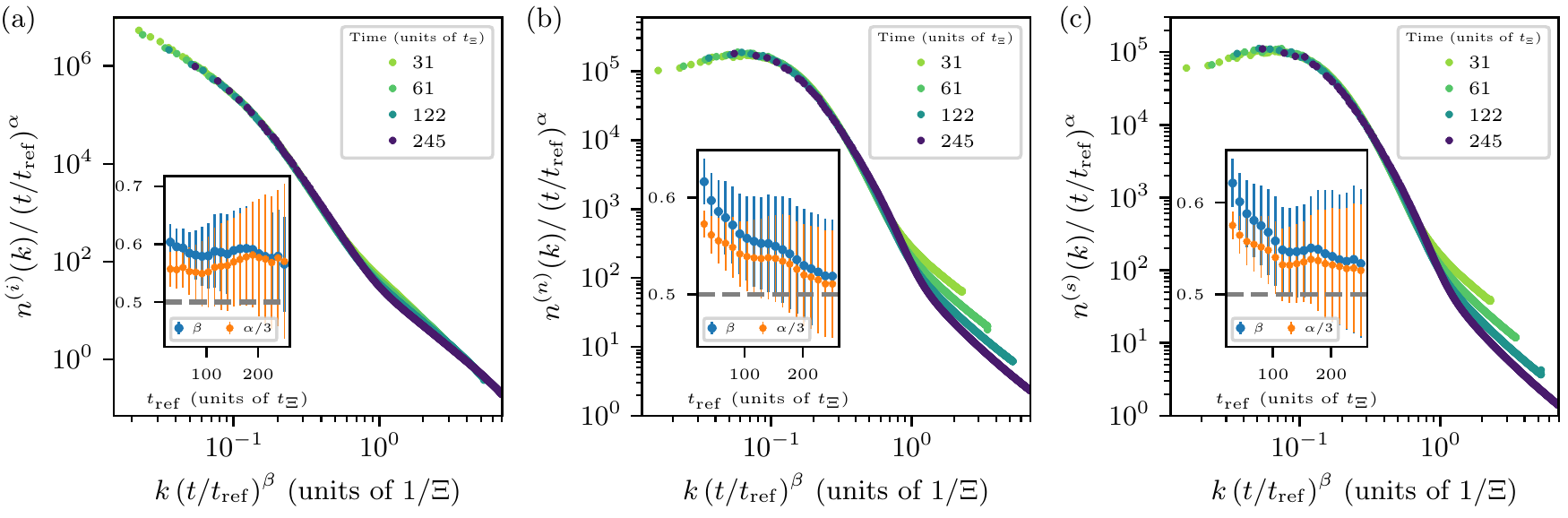}
\caption{Universal scaling dynamics of the occupation numbers representing the (a) incompressible, $n^{(i)}$, (b) nematic, $n^{(n)}$, and (c) spin, $n^{(s)}$, parts of the hydrodynamic decomposition (see \Fig{HydroDecompEvo}) according to \eq{NTFPscaling}.  The collapse of the data to the universal scaling functions $f_{\mathrm{S}}^{(\delta)}(\mathbf{k})=n^{(\delta)}(\mathbf{k},t_\mathrm{ref})$, with reference time $t_\mathrm{ref}=31\,t_{\Xi}$, shows, in each case $\delta = i,n,s$, the scaling \eq{NTFPscaling} in space and time.
The universal scaling exponents $\alpha/3$ (orange stars) and $\beta$ (blue dots) obtained by means of a least-square fit to the data within the time window $[t_{\mathrm{ref}}, t_{\mathrm{ref}} + \Delta t]$ with $\Delta t = 146\,t_{\Xi}$ are depicted in the insets of each panel. 
For late times we find that the scaling exponent $\beta$ extracted for the nematic and spin part approaches $\beta \simeq 0.5$ while for the incompressible part it settles in to $\beta \simeq 0.58$.
The scaling exponents corresponding to the time evolution of the nematic and spin part corroborate our hypothesis that the scaling behavior is dominated by the relative phases of the components. 
\label{fig:HydroDecompRescaled}
}
\end{figure*}

We furthermore analyze the scaling evolution of the momentum distribution $C_{12}(k,t) = \langle\lvert \Phi_1^{\dagger} \Phi_2(k,t)\rvert ^2\rangle$ and the corresponding spatial coherence function $g_{12}(r,t)$ of the relative phase operator which we show in \Fig{RelativePhase12}.
The rescaled data collapse in a similar manner as the single-component occupation numbers, with extracted exponents presented in \Fig{WindowFit12}.
The scaling function $g^{(1)}_{12}$ in position space (\Fig{RelativePhase12}c) shows a form reminiscent of an exponential decay comparable to the first-order coherence function.

We finally analyze, in \Fig{HydroDecompEvo}, the time evolution of the total momentum distribution $n_\mathrm{tot}(k,t)=\sum_{a=1}^{3}n_{a}(k,t)$ in hydrodynamic decomposition.
This is obtained by decomposing the kinetic energy distribution in momentum space into quantum-pressure (q), spin (s), nematic (n), incompressible (i) and compressible (c) parts, as defined in detail in \App{HydroDecomp}.
The decomposition provides additional information on the character of the hydrodynamic flows corresponding to the phase field $\theta_{a}$.

We point to the observation that the incompressible flow arising from vortical motion is subdominant as compared to the nematic and spin parts which determine the leading scaling of the occupation number as $n_\mathrm{tot}(k)\sim k^{-4}$.
The vortex flow contributes significantly only at very small momenta, where the plateau appears in the total number distribution at very low momenta. 
In this regime, as is seen at times $t/t_{\Xi}=31$, $61$, the nematic and spin parts fall off towards zero momentum but this is compensated for by the incompressible part.
Note that the distributions shown in \Fig{HydroDecompEvo} result from the tensor decomposition of the current which is a four-point function of the Bose fields. 
Hence,  $n_\mathrm{kin}(k,t)$ which is the sum of the hydrodynamic parts, deviates from $n_\mathrm{tot}(k,t)$ in the infrared \cite{Nowak:2011sk}. 

We emphasize that the spin and nematic fluctuations are determined by the fluctuations of the relative phases between the field components $a=1,2,3$.
The corresponding momentum distributions take the same form as the scaling function shown in \Fig{RelativePhase12}b. The universal scaling of $C_{12}(k,t)$ confirms our hypothesis that the scaling behavior is dominated, for non-zero $N$, by the relative phases of the components. 

We additionally find that the incompressible as well as the spin and nematic parts of the hydrodynamic decomposition also exhibit universal scaling according to \eq{NTFPscaling} (see \Fig{HydroDecompRescaled}). 
The late-time value of $\beta \simeq 0.5$ extracted for the universal scaling of the spin and nematic parts (see insets \Fig{HydroDecompRescaled}b and c) demonstrates the crucial role of the relative phases to the scaling behavior of the system.
Interestingly the scaling behavior of the subdominant vortical flow in the system seems to be characterized by a scaling exponent $\beta \simeq 0.58$ nearly unaffected by variations of the reference time (see inset \Fig{HydroDecompRescaled}a).
For smaller $N$, in particular $N=1$, vortices are expected to play a more prominent role \cite{Nowak:2010tm,Nowak:2011sk,Schole:2012kt,Nowak:2012gd,Karl:2013mn,Karl:2013kua,Karl2017b.NJP19.093014}  in the evolution following a quench of the type considered here
\footnote{The transport associated with a non-thermal fixed point in the $N=1$ case has recently been associated with a `transport peak' occurring in the spectral decomposition of the statistical correlation function, independent of the thermal Bogoliubov excitations in the system \cite{PineiroOrioli:2018hst}.}.

We finally mention that during the approach of the scaling limit and thus of the non-thermal fixed point, the system shows \emph{prescaling}  \cite{Schmied:2018upn.PhysRevLett.122.170404} (see also \cite{Mazeliauskas:2018yef} in a perturbative high-energy context).
Our analysis demonstrates that (cf.~in particular \Fig{HydroDecompEvo}) effects of non-linear excitations such as vortices, contributing to the incompressible flow, can induce scaling violations of the single-component occupation numbers and coherence functions.
On the other hand, also other contributions such as $1/N$ corrections to the scaling analysis could explain certain systematic deviations of the exponents found at the largest simulation times from the predictions presented in this work.
A more detailed finite-$N$ analysis is the subject of future work.

\section{Conclusions}
\label{sec:Conclusions}
We present a low-energy effective theory for the interacting phase-angle excitations of a $\mathrm{U}(N)$-symmetric Gross-Pitaevskii model of $N$ Bose fields with local quartic self-coupling of the total particle density. The theory provides a perturbative formulation of far-from-equilibrium low-energy universal scaling dynamics at a non-thermal fixed point. This is complementary to the non-perturbative approach on the basis of fundamental Bose fields chosen previously, while being technically easier to evaluate in the scaling regime.  

Our approach provides the leading-order dynamical scaling exponent $z=2$ of the quasiparticle eigenfrequencies in the large-$N$ limit.
This result for $z$ closes a longstanding gap in the non-perturbative kinetic-theory formulation of non-thermal fixed points \cite{Berges:2008wm,Berges:2008sr,Scheppach:2009wu,Berges:2010ez,Orioli:2015dxa,Berges:2015kfa,Chantesana:2018qsb.PhysRevA.99.043620}.
Applying the theory in the large-$N$ limit, we recover the universal scaling exponents at a non-thermal fixed point predicted previously within the non-perturbative approach (cf.~\cite{Orioli:2015dxa,Chantesana:2018qsb.PhysRevA.99.043620}).

We find analytically that, in leading order, the first-order coherence function, close to a non-thermal fixed point, falls off exponentially in space, $g^{(1)}(\mathbf{r},t) \sim e^{-k_{\Lambda}(t)\,|\mathbf{r}|}$, with a coherence-length scale rescaling as $k_{\Lambda}(t)\sim t^{-\beta}$ in time, with $\beta=1/2$ both, for the case $N=1$ and $N\to\infty$.
This is reminiscent of equilibrium systems in $d<3$ dimensions where such an exponentially reduced long-range phase order indicates the presence of a quasicondensate, with the there static coherence length depending on the coupling and/or temperature.
Also in these situations, without spontaneous symmetry breaking and a field expectation value singling out a zero-momentum condensate mode, the states are characterized by strong occupancies of the low-energy momentum modes of the system.

Our analytical predictions are corroborated by the results of Truncated-Wigner simulations for $N=3$ in $d=3$ dimensions.
Considering $g^{(1)}(r,t)$ at short distances $r=|\mathbf{r}|$ where it falls off exponentially, we are able to confirm, with high accuracy, the analytic predictions $\beta=1/2$ and $\alpha=d/2$ for the spatio-temporal scaling exponents, effectively leaving little space for anomalous deviations.

This finding is consistent with our analytical result that the corresponding non-thermal fixed point, which is approached in the scaling limit of infinite evolution times, has a Gaussian character as we argue by standard scaling arguments applied to the bare action of the low-energy effective theory.
In contrast, to obtain a positive upper ``critical'' dimension a much smaller exponent $\beta$ is required as, e.g., was found numerically for an anomalous fixed point dominated by vortex interactions in two spatial dimensions \cite{Karl2017b.NJP19.093014,Johnstone2018a.arXiv180106952}.

We emphasize that the numerically determined coherence function, however, shows oscillations at larger distances which are not covered by the analytical approach which rests on a homogeneous background phase $\theta^{(0)}_{a}$.
An extension of the theory to a non-uniform $\theta^{(0)}_{a}(\mathbf{r},t)$ seems viable but is beyond the scope of this work.

When evaluating our low-energy effective description for the case of a single-component Bose gas, $N=1$, we find the same spatio-temporal scaling exponent $\beta=1/2$, despite the fact that the dynamical exponent is $z=1$ in this case.
This analytical result corroborates earlier numerical findings presented in \cite{Orioli:2015dxa,Schachner:2016frd,Walz:2017ffj.PhysRevD.97.116011}.

Our results support the conjecture, that related universal scaling in $\mathrm{O}(N)$-symmetric relativistic models (cf.~\cite{Orioli:2015dxa,Moore:2015adu}), while the $\mathrm{O}(N)$ symmetry is broken by the evolution, is connected with the appearance of an approximately conserved charge due to a remaining $\mathrm{U}(1)$ symmetry not broken by the flow \cite{Schmied:2018upn.PhysRevLett.122.170404}.
This was also seen in numerical simulations~\cite{Gasenzer:2011by}.
In consequence, we presume that $\beta=1/2$ would apply also for these systems, independent of $N$, reflecting  relative phase fluctuations between the field components and their universal transport toward low momenta. 
As the non-thermal fixed-point scaling relation $\alpha=d\beta$, to a good approximation, holds also for the relative-phase correlators $C_{ab}(k,t)$ shown in \Fig{RelativePhase12} as well as for the spin-contributions to the energy spectrum seen in \Fig{HydroDecompRescaled}, a further emerging symmetry is expected to play an important role.
This symmetry has been conjectured to be related to the suppression of density fluctuations in the system \cite{Schmied:2018upn.PhysRevLett.122.170404} and will be further discussed elsewhere.

From our analytical results in the large-$N$ limit and for the case $N=1$ together with our numerical results for $N=3$, we propose that the universality class corresponding to the scaling exponent $\beta$ is related to the dynamical breaking of a $\mathrm{U}(1)$ symmetry and therefore independent of $N$.
Our approach offers itself for a refinement using field-theoretic renormalization-group techniques and for applications to small-$N$ spin systems available in experiment.

In the following Appendices, we provide further details of
the calculations leading to our results presented in the main
text.

\section*{Acknowledgements} 
The authors thank I.~Aliaga Sirvent, J.~Berges, K.~Boguslavski, R.~B\"uck\-er, I.~Chantesana, S.~Diehl, S.~Erne, F.~Essler, S.~Heupts, M.~Karl, P.~Kunkel, S.~Lannig, D.~Linnemann, A.~Mazeliauskas, M.~K.~Oberthaler, J.~M.~Pawlowski, A.~Pi{\~n}eiro Orioli, M.~Pr\"ufer, R.~F.~Rosa-Medina Pimentel, J.~Schmiedmayer, T.~Schr\"oder, H.~Strobel, and C. Wetterich for discussions and collaboration on the topics described here. 
This work was supported by the Horizon-2020 framework programme of the European Union (FET-Proactive, AQuS, No. 640800, ERC Advanced Grant EntangleGen, Project-ID 694561),  by Deutsche Forschungsgemeinschaft (SFB 1225 ISOQUANT), by Deutscher Akademischer Austauschdienst (No.~57381316), and by Center for Quantum Dynamics, Heidelberg University.
C.-M.S. thanks the Dodd-Walls Centre for Photonic and Quantum Technologies, University of Otago, Dunedin, New Zealand, for hospitality and support. 
T.G. thanks the Erwin Schr\"odinger International Institute for Mathematics and Physics, Wien, Austria, for hospitality and support within their topical program \emph{Quantum Paths}.\\


\vspace{\columnsep}

\begin{appendix}
\setcounter{equation}{0}
\setcounter{table}{0}
\makeatletter

\section{Notation}
\label{app:Not}
In this paper, we adopt the $(-,-,-,+)$ signature convention for the metric so that the scalar product of two $(d + 1)$-vectors $x = (x_1,\ldots,x_d,x_0) = (\mathbf{x},x_0)$ and $y = (y_1,\ldots,y_d,y_0)$ is given by $x y = x_0 y_0 - \mathbf{x} \mathbf{y}$. Define the Fourier transform as:
\begin{align}
\mathcal{F}[f(x)](p) &\equiv f(p) = \int_x e^{-\i p x} f(x)\,, \\
\mathcal{F}^{-1}[f(p)](x) &\equiv f(x) = \int_p e^{\i p x} f(p)\,,
\end{align}
with the short-hand notations $\int_x \equiv \int \mathrm d^{d+1} x$ and ${\int_p \equiv\int {\mathrm d^{d+1} p}/(2 \pi)^{d+1}}$.
We also introduce the Fourier transform over only space (time) variables:
\begin{align}
f(\mathbf{k},t) &= \int_\mathbf{x} e^{\i \mathbf{k x}} f(\mathbf{x},t) = \int_{\omega} e^{\i \omega t} f(\mathbf{k},\omega)\,,\\
f(\mathbf{x},\omega) &= \int_t e^{-\i \omega t} f(\mathbf{x},t) = \int_\mathbf{k} e^{-\i \mathbf{k x}} f(\mathbf{k},\omega).
\end{align}

\section{Low-energy effective action out of equilibrium}
\label{app:EFT_general}
In many cases, the description of a system of interest can be significantly simplified if one only considers the physics above some length scale $1/\Lambda$, or, equivalently, below a momentum and thus energy scale $\Lambda$. 
This typically is the case if there is a separation of scales and the dynamics distinguishes fields with different masses (``heavy'' vs.~``light'' field modes) or if some degrees of freedom are suppressed compared to the others (``fast'' vs.~``slow'' modes). 
A low-energy effective field theory is then obtained by integrating out the fast (i.e., heavy) fluctuations. 
The matching condition reads as
\begin{equation}
\label{eq:cond}
\Tr{\rho (t) \mathcal{O}(t)} = \Tr{\rho_{\mathrm{eff}} (t) \mathcal{O}(t)},
\end{equation}
where $\mathcal{O}(t)$ is an operator measuring only slow (light) degrees of freedom.

In the following, we will discuss the evaluation of the above expectation value within the Schwinger-Keldysh approach and the Feynman-Vernon influence-functional approach \cite{Feynman:1963fq} (for details, see also \cite{Calzetta:2008iqa}).
Assuming that the initial density matrix $\hat\rho(t_0)$ at a distant past time $t_0$ splits into phase and density fluctuations, and that the density fluctuation part can be taken in Gaussian approximation, it becomes clear that the computation of the low-energy effective action can be done as in a standard zero-temperature approach.
The reason behind this is that, when moving towards a non-thermal fixed point in a closed system, the system loses its memory about the details of the initial state, similarly as driven-dissipative \cite{Sieberer_2016} and equilibrium systems. 
As a consequence the above assumptions are expected to be well-justified.

Let us consider the example where $\varphi (x)$ describes the slow modes and $\psi (x)$ the fast ones with the following action:
\begin{equation}
S[\varphi,\psi] = S_{\varphi}[\varphi] + S_{\psi}[\psi] + S_{\mathrm{int}}[\varphi,\psi]\,,
\end{equation}
where $S_{\varphi}[\varphi]$ and $S_{\psi}[\psi]$ are the (semi-)classical actions that do not contain any mixing terms and thus describe independent dynamics of the fields, while $S_{\mathrm{int}}[\varphi,\psi]$ corresponds to interaction between two fields. 

The total density matrix of the whole system is given by
\begin{equation}
\rho[\varphi,\psi;t] = \left\langle \varphi^{+},\psi^{+}|\,\hat{\rho}(t) \,| \varphi^{-},\psi^{-}\right\rangle,
\end{equation}
where $\pm$ denotes that the field is defined on the positive or negative branch of the closed time path, respectively. According to the above, the reduced (or effective) density matrix, which is the object of interest, is then defined by tracing out the fast modes at the time $t$ considered,
\begin{equation}
\rho_{\mathrm{eff}}[\varphi;t] = \int d \psi \, \left\langle \varphi^{+},\psi \,|\, \hat{\rho}(t) \,|\, \varphi^{-},\psi \right\rangle .
\end{equation}
Quite often, the initial state can be chosen to be described by a Gaussian density matrix. Higher-order correlations in this case are built up during the evolution. Similarly, hereafter we assume that the initial density matrix is factorized, i.e.,
\begin{equation}
\label{eq:initalfact}
\hat{\rho}(t_0) = \hat{\rho}_{\varphi}(t_0) \times \hat{\rho}_{\psi}(t_0)\,,
\end{equation}
where $\hat{\rho}_{\varphi}(t_0)$ and $\hat{\rho}_{\psi}(t_0)$ are the initial density operators of the fields $\varphi$ and $\psi$, respectively. The condition \eqref{eq:initalfact} implies the absence of correlation between $\varphi$ and $\psi$ at $t = t_0$. Nevertheless, due to the presence of the interaction term $S_{\mathrm{int}}[\varphi,\psi]$, the latter will be generated at later times $t > t_0$. In our case, $\varphi \to \theta$ and $\psi \to \delta \rho$ so that \eqref{eq:initalfact} means that there is no correlation between phase and density fields in the initial configuration. In the absence of topological defects, this is typically an adequate assumption.

\begin{figure}[h]
\centering
\includegraphics[scale=0.7]{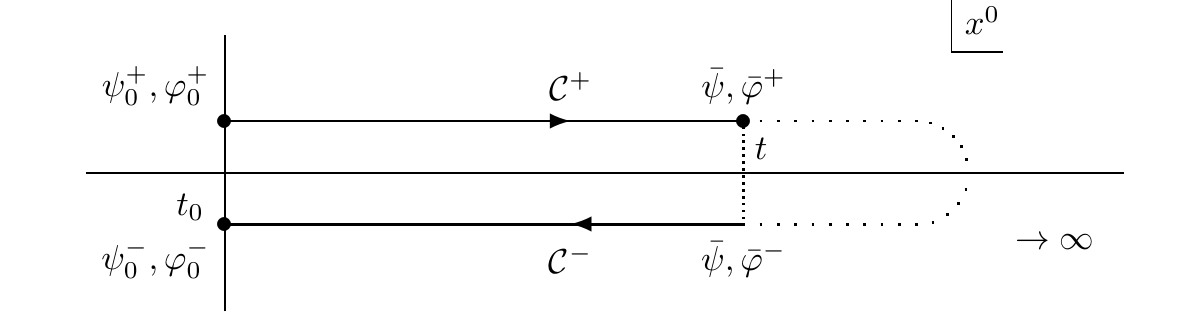}
\caption{Contour summarizing \eqref{eq:infpathint} and \eqref{eq:infavg}.}
\label{fig:contour_inf}
\end{figure}

The evolution of the reduced matrix $\rho_{\mathrm{eff}}[\varphi,t]$ then reads as
\begin{equation}
\rho_{\mathrm{eff}}[\bar{\varphi};t] = \int [d \varphi_0^{+}] [d \varphi_0^{-}] \, \left(\bar{\varphi};t \,|\, \varphi_0;t_0\right) \rho_{\mathrm{eff}}[\varphi_0;t_0]\,,
\end{equation}
where we use the bar to label the value at the time $t$. The transition amplitude can be expressed in terms of a closed-time path functional integral
\begin{equation}
\left(\bar{\varphi};t \,|\, \varphi_0;t_0\right) = \int_{\varphi_0^{+}}^{\bar{\varphi}^{+}} \mathcal{D}' \varphi^{+} \int_{\varphi_0^{-}}^{\bar{\varphi}^{-}} \mathcal{D}' \varphi^{-} \, e^{\i S_{\mathrm{eff}}[\varphi]}\,,
\end{equation}
with
\begin{equation}
\label{eq:seff_fv}
S_{\mathrm{eff}}[\varphi] = S_{\varphi}[\varphi^{+}] - S_{\varphi}[\varphi^{-}] + S_{\mathrm{IF}}[\varphi^{+},\varphi^{-}]
\end{equation}
being the full effective action while $S_{\mathrm{IF}}$ is the \emph{influence action}. The latter is given by
\begin{align}
\label{eq:infpathint}
\mathfrak{F}[\varphi^{+},\varphi^{-}] \equiv e^{\i S_{\mathrm{IF}}[\varphi^{+},\varphi^{-}]} = \int [d \bar{\psi}] \int [d \psi_0^{+}] [d \psi_0^{-}] \, \rho_{\psi} [\psi_0;t_0] \nonumber\\ 
\times\int_{\psi_0^{+}}^{\bar{\psi}} \mathcal{D}' \psi^{+} \int_{\psi_0^{-}}^{\bar{\psi}} \mathcal{D}' \psi^{-} \, e^{\i \left\lbrace S_{\psi}[\psi^{+}] + S_{\mathrm{int}}[\varphi^{+},\psi^{+}] - S_{\psi}[\psi^{-}] - S_{\mathrm{int}}[\varphi^{-},\psi^{-}] \right\rbrace}\,,
\end{align}
where we have introduced the so-called \emph{Feynman-Vernon influence functional} $\mathfrak{F}$ \cite{Feynman:1963fq}, which can be conveniently rewritten as
\begin{equation}
\label{eq:infavg}
\mathfrak{F}[\varphi^{+},\varphi^{-}] = \left\langle \mathcal{T_C} e^{\i S_{\mathrm{int}}[\varphi,\psi]} \right\rangle_{\psi}\,,
\end{equation}
with average being defined with respect to the $\psi$-field density matrix. Note that the Schwinger-Keldysh contour can be depicted as shown in Fig.~\ref{fig:contour_inf}. 

Now, given the effective action $S_{\mathrm{eff}}$ [Eqs.~\eqref{eq:seff_fv}, \eqref{eq:infpathint}], we can define the effective generating functional $Z_{\mathrm{eff}}[J,R;\hat{\rho}_{\mathrm{eff},0}]$:
\begin{widetext}
\begin{align}
\label{eq:zeffpathint}
Z_{\mathrm{eff}}[J,R;\hat{\rho}_{\mathrm{eff},0}] 
&= 
\int [d \varphi_0^{+}] [d \varphi_0^{-}] \, \rho_{\mathrm{eff}}[\varphi_0;t_0] \int\limits_{\varphi_0^{+}}^{\varphi_0^{-}} \mathcal{D}' \varphi \, e^{\i \left\lbrace S_{\mathrm{eff}}[\varphi] +  \int_{x,\mathcal{C}} \varphi (x) J(x) + \frac{1}{2} \int_{xy,\mathcal{C}} \varphi(x) R(x,y) \varphi(y) \right\rbrace}\nonumber\\
&=
\int [d \varphi_0^{+}] [d \varphi_0^{-}] \, \rho_{\mathrm{eff}}[\varphi_0;t_0] \int\limits_{\varphi_0^{+}}^{\varphi_0^{-}} \mathcal{D}' \varphi \, e^{\i \left\lbrace S_{\varphi}[\varphi^{+}] - S_{\varphi}[\varphi^{-}] + S_{\mathrm{IF}}[\varphi^{+},\varphi^{-}] +  \int_{x,\mathcal{C}} \varphi (x) J(x) + \frac{1}{2} \int_{xy,\mathcal{C}} \varphi(x) R(x,y) \varphi(y) \right\rbrace}\nonumber\\
&=
\int [d \varphi_0 d \psi_0] \, \rho[\varphi_0,\psi_0;t_0] \int\limits_{\varphi_0^{+},\psi_0^{+}}^{\varphi_0^{-},\psi_0^{-}} \mathcal{D}' \varphi \mathcal{D}' \psi \, e^{\i \left\lbrace S_{\mathrm{full}}[\varphi,\psi] +  \int_{x,\mathcal{C}} \varphi (x) J(x) + \frac{1}{2} \int_{xy,\mathcal{C}} \varphi(x) R(x,y) \varphi(y) \right\rbrace}\,,
\end{align}
where we have introduced 
\begin{equation}
S_{\mathrm{full}}[\varphi,\psi] 
\equiv 
\left(S_{\varphi}[\varphi^{+}] + S_{\psi}[\psi^{+}] + S_{\mathrm{int}}[\varphi^{+},\psi^{+}]\right) -
\left(S_{\varphi}[\varphi^{-}] + S_{\psi}[\psi^{-}] + S_{\mathrm{int}}[\varphi^{-},\psi^{-}]\right)\,,
\end{equation}
and the condition $\bar{\psi}^{+} = \bar{\psi}^{-} = \bar{\psi}$ is implied. In the last line, we also used the following property:
\begin{equation}
\rho_{\mathrm{eff}}[\varphi_0;t_0] 
= 
\int d \psi \, \left\langle \varphi_0^{+},\psi \,|\, \hat{\rho}(t_0) \,|\, \varphi_0^{-},\psi \right\rangle 
= 
\rho_{\varphi}[\varphi_0;t_0] \int d \psi \, \langle \psi \, | \, \hat{\rho}_{\psi}(t_0) \, | \, \psi \rangle 
=
\rho_{\varphi}[\varphi_0;t_0] \Tr{\hat{\rho}_{\psi}(t_0)} = \rho_{\varphi}[\varphi_0;t_0].
\end{equation}
By comparing the first line of \eqref{eq:zeffpathint} with the last one we get the final expression for the effective action:
\begin{align}
\label{eq:seff_final}
\exp{\left\lbrace \i S_{\mathrm{eff}}[\varphi]\right \rbrace} 
&=
\exp{\left\lbrace \i \left(S_{\varphi}[\varphi^{+}] - S_{\varphi}[\varphi^{-}]\right) \right\rbrace} \int [d\psi_0^{+}] [d\psi_0^{-}] \,\rho_{\psi}[\psi_0;t_0] \nonumber\\
&\times
\int_{\psi_0^{+}}^{\psi_0^{-}} \mathcal{D}' \psi \, \exp{\left\lbrace \i \left(S_{\psi}[\psi^{+}] + S_{\mathrm{int}}[\varphi^{+},\psi^{+}] - S_{\psi}[\psi^{-}] - S_{\mathrm{int}}[\varphi^{-},\psi^{-}]\right) \right\rbrace}\,,
\end{align}
\end{widetext}
with condition $\bar{\psi}^{+} = \bar{\psi}^{-} = \bar{\psi}$ being, again, implied.
As a result, assuming that the initial density matrix of the heavy modes $\rho_{\psi}[\psi_0;t_0]$ takes the form of a Gaussian ground states, the integral in \eq{seff_final} falls into two independent factors, the positive and negative closed-time path zero-temperature contributions. 
It follows that the influence functional $\mathfrak{F}$ is formally equivalent to the Keldysh vacuum generating functional, and the influence action \eq{infpathint} is the vacuum effective action. 
Since in this work we concentrate on dynamics at the non-thermal fixed point, which is formally reached only at $t \to \infty$, we can use the memory loss property of non-thermal fixed points, as well as the fact that density fluctuations are suppressed, to choose the reference time $t_0$ such that the density matrix of the corresponding density fluctuations can be taken in Gaussian approximation in the ground state, thus reducing the computation of the effective action to the zero-temperature case, as pointed out above.

\section{Eigenvalues of the mass matrix}
\label{app:Mab}
In this Appendix, we determine the eigenvalues of the mass matrix
\begin{equation}
M^{ab} = \left(\frac{\mathbf{k}^2}{2 m}\right)^2 \delta^{ab} + 2 g \rho^{(0)}_n \frac{\mathbf{k}^2}{2m} \delta^{nb}, 
\end{equation}   
where no assumptions on the values of $\rho^{(0)}_b$ were made.
We define $a = (\mathbf{k}^2/2 m)^2$, $b_n = 2 g \rho^{(0)}_n (\mathbf{k}^2/2m)$ such that this matrix can be rewritten as
\begin{equation}
M = a \, I_{N \times N} + J_{N \times N} \, \mathcal{P} = \begin{pmatrix}
  a + b_1 & b_2 & \cdots & b_N \\
  b_1 & a + b_2 & \cdots & b_N \\
  \vdots  & \vdots  & \ddots & \vdots  \\
  b_1 & b_2 & \cdots & a + b_N 
 \end{pmatrix}.
\end{equation}
Here, $I_{N \times N}$ is the unit matrix, $J_{N \times N}$ is the matrix of ones, and $\mathcal{P} = \diag \left(b_1,\ldots,b_N \right)$.

We now take a look at the second term. This matrix contains $N$ identical rows, and hence it has the eigenvalue $\lambda = 0$ with an algebraic multiplicity $N - 1$. 
It is easy to check that the corresponding eigenvectors are given by 
\begin{align}
  \mathbf{e}_{1} &= ( -1, 1, 0,...,0 )^{T}\,, \quad
  \mathbf{e}_{2} = ( -1, 0, 1,...,0 )^{T}\,, \ldots,\nonumber\\
  \mathbf{e}_{N-1} &= ( -1, 0, 0,...,1 )^{T}\,,
  \label{eq:app:GoldstoneEV}
\end{align}
meaning that the geometric multiplicity is also $N - 1$. 
The remaining eigenvalue,
\begin{equation}
 J_{N \times N} \, \mathcal{P}\,\mathbf{e}_{N} = \left(\sum_i b_i \right)\mathbf{e}_{N}\,,
\end{equation} 
corresponds to the eigenvector
\begin{equation}
\mathbf{e}_{N} = ( 1, 1,\ldots, 1 )^{T}.
\label{eq:app:GoldstoneBog}
\end{equation}
Hence, we conclude that $M$ possesses the eigenvalue $\lambda = a$ with multiplicity $N - 1$ and $\lambda = a + \sum_i b_i$ with multiplicity $1$. 
We thus have two types of excitation modes, with 
eigenfrequencies being the square roots of these eigenvalues,
\begin{align}
  \omega_1(\mathbf{k}) 
  &= \omega_2 (\mathbf{k}) = \ldots = \omega_{N-1} (\mathbf{k}) = \frac{\mathbf{k}^2}{2m}\,,
  \label{eq:app:GoldstoneFreq}
  \\ 
  \omega_N (\mathbf{k}) 
  &= \sqrt{\frac{\mathbf{k}^2}{2m} \left(\frac{\mathbf{k}^2}{2m} + 2 g \rho^{(0)} \right)}.
  \label{eq:app:BogHiggsFreq}
\end{align}

\section{Spontaneous symmetry breaking}
\label{app:SSB}
In this appendix we discuss the breaking of the $\mathrm{U}(N)$ symmetry by the particular choice of initial conditions we apply. 
In the respective ground state, this corresponds to spontaneous symmetry breaking. 
Consider the $N$-component Gross-Pitaevskii Lagrangian in the absence of external potential:
\begin{align}
\label{eq:lagrangian}
\mathcal{L} 
\ = \ &
\frac{\i}{2}\big[\varphi^{*}_a(x) \partial_t \varphi_a(x) - \varphi_a(x) \partial_t \varphi^{*}_a(x)\big]\nonumber\\
&- \frac{1}{2m} \big[\nabla \varphi_a^{*}(x)\big] \cdot \big[\nabla \varphi_a(x)\big] - \frac{g}{2} \big[\varphi_a(x) \varphi_a(x)\big]^2\,,
\end{align}
On a classical level, a mean-field expectation value of $\varphi$ can be derived from the classical potential:
\begin{equation}
\mathcal{V}(|\varphi_a|) =  \frac{g}{2} \left(\sum_b |\varphi_a|^2\right)^2,
\end{equation}
which yields a zero-valued ground state. This is not surprising since \eqref{eq:lagrangian} corresponds to a vacuum QFT. In condensed matter physics, one is typically interested in field theories with non-zero background densities. To obtain a finite background density, one introduces Lagrange multipliers associated with the $\mathrm{U}(1)$ charges for each component,
\begin{equation}
  j_a^0 
  = \frac{\partial \mathcal{L}}{\partial(\partial_0 \varphi_a)} \frac{\delta \varphi_a}{\delta \alpha} + \mathrm{c.c.}
  = \varphi_{a}^{*}\varphi_{a}=\rho_{a}\, 
  \label{eq:NoetherCharge}
\end{equation}
where $\partial_{0}\equiv\partial_{t}$, and the aforementioned $\mathrm{U}(1)$ symmetries read 
\begin{equation}
\varphi_a \rightarrow e^{-\i \alpha} \varphi_a\,, \quad \varphi^{*}_a \rightarrow e^{\i \alpha} \varphi^{*}_a.
\end{equation}
Introducing the chemical potentials one replaces
\begin{equation}
H \rightarrow H - \mu_a N_a,
\end{equation}
with particle number $N_{a}=\rho_{a}V$ in volume $V$, and thus, in non-relativistic case,
\begin{equation}
\mathcal{L} \rightarrow \mathcal{L} + \mu_a \varphi_a^{*} \varphi_a.
\end{equation}
We note that this, in general, explicitly breaks the $\mathrm{U}(N)$ symmetry into a product $\mathrm{U}(1)^{\times N}$. The classical potential then takes the form
\begin{equation}
\mathcal{V}(|\varphi_a|) = -\mu_a |\varphi_a|^2 + \frac{g}{2} \left(\sum_b |\varphi_a|^2\right)^2.
\end{equation}
Varying with respect to, e.g., $\varphi_b^{*}$ one gets the conditions
\begin{equation}
\left(\mu_b -  g \sum_a |\varphi_a|^2 \right)\, \varphi_b = 0, \quad b = 1,...,N.
\end{equation}
It follows directly that, in order to have a non-vanishing amplitude $\varphi_{b}$ of component $b$, the associated chemical potential must be $\mu_{b}=g\rho$, with total density $\rho=\sum_{a}\rho_{a}$. Hence, the chemical potentials for all non-vanishing densities need to be equal and given by the total density. Therefore, in the ground state, the densities will be non-zero in the subspace having identical chemical potentials with the largest value, while the rest are zero. These non-vanishing ground-state densities $\rho_{a}^{(0)}$ add up to the total density $\rho^{(0)}=\mu/g$ set by the chemical potential $\mu$ for the respective components.

In order to have a non-zero total density without specifying densities of each component and thus remain a $\mathrm{U}(N)$ symmetry, a chemical-potential term is introduced as
\begin{equation}
H \rightarrow H - \mu N \longleftrightarrow \mathcal{L} \rightarrow \mathcal{L} + \mu\, \varphi_{a}^{*} \varphi_{a}\,,
\end{equation}
which is equivalent to fixing only the total density. We note that the above is equivalent to replacing, with $\mu = g \rho^{(0)}$, 
\begin{equation}
-\frac{g}{2}\left(\sum_b \varphi_{b}^{*} \varphi_{b}\right)^2 \rightarrow 
-\frac{g}{2}\left(\sum_b \varphi_{b}^{*} \varphi_{b} - \rho^{(0)}\right)^2.
\label{eq:backgroundshift}
\end{equation}

Let us close the discussion with the following remark. Above we tried to preserve a full $\mathrm{U}(N)$ symmetry. On the other hand, this symmetry might be explicitly broken in experimentally relevant scenarios. Indeed, in the experiment, particle number is typically specified for each component. One can take care of this by choosing an appropriate initial density matrix $\hat{\rho}(t_0)$ which fixes the expectation values of the particle number operators at $t = t_0$. In this paper, we disregard this subtlety and only take it into account by choosing the vacuum with $\rho_a^{(0)} \neq 0$ for each component. 

\section{Inverse of the kernel $\tilde{g}^{ab}$}
\label{app:Inversegab}
To derive the inverse of $\tilde{g}^{ab}$ defined in \Eq{app:G_inv}, we consider a general $N \times N$ matrix of the form $U(u,v) = u I + \mathbf{v} \otimes \mathbf{v}\,,$ where $\mathbf{v} = (v_1,v_2,\ldots,v_N)^{T}$ and $\otimes$ denotes an outer (tensor) product. To find the inverse of $U(u,v)$ we use the ansatz $U^{-1}(u,v) = x I - \mathbf{y} \otimes \mathbf{y}\,,$ which yields the condition
\begin{equation}
x u\, \delta_{ij} - u \,y_i y_j + x\, v_i v_j - (\mathbf{v} \cdot \mathbf{y})\, v_i y_j = \delta_{ij}\,,
\end{equation}
where we used $(\mathbf{v} \otimes \mathbf{v}) \cdot (\mathbf{y} \otimes \mathbf{y}) = (\mathbf{v} \cdot \mathbf{y}) \mathbf{v} \otimes \mathbf{y}$.
We note that, if $\mathbf{v} = \mathbf{0}$, then obviously also $\mathbf{y} = \mathbf{0}$ such that we get $x = 1/u$ and the above condition becomes
\begin{equation}
y_i y_j - \frac{v_i}{u} \frac{v_j}{u} + (\mathbf{v} \cdot \mathbf{y}) \frac{v_i}{u} y_j = 0.
\end{equation} 
This is a system of non-linear equations, which, in principle, cannot be solved analytically. However, we note that the above system is almost symmetric in $\mathbf{y} \leftrightarrow \mathbf{v}/u$, which gives a hint to use the following ansatz as a solution: $\mathbf{y} = c \mathbf{v}\,,$ where $c > 0$ is some positive constant. Using this we obtain
\begin{equation}
\left(c^2 - \frac{1}{u^2} + \frac{c^2 \sum_j v_j^2}{u} \right) v_i v_j = 0\,,
\end{equation}
such that
\begin{equation}
c^2 = \frac{1}{u \left(u +\sum_j v_j^2 \right)} \rightarrow y_i = \frac{v_i}{\sqrt{u \left(u + \sum_j v_j^2 \right)}}.
\end{equation}
According to \eq{app:G_mod} we can use these results with $\displaystyle u = ({N g /2})({\mathbf{k}^2/ k^2_{\Xi}})$ and $v_{a} = ({N g\rho_a^{(0)}/\rho^{(0)}})^{1/2}$, $a=1,\ldots,N$, to obtain the inverse of $\tilde{g}^{ab}$ as given in \Eq{app:G_inv}.

\begin{widetext}
\section{Rearrangement of $3$- and $4$-vertices}
\label{app:vertices}
The three- and four-wave coupling terms \eq{SeffnG3thetaa} and \eq{SeffnG4thetaa}, respectively, can be rewritten using
\begin{align}
\int_{\mathbf{k},\mathbf{k}'} \frac{\mathbf{k}' \cdot (\mathbf{k}'-\mathbf{k})}{g_{\mathrm{1/N}}(\mathbf{k})} \partial_t \theta_a(-\mathbf{k},t) \theta_a (\mathbf{k}',t) \theta_a (\mathbf{k} - \mathbf{k}',t) 
&=\phantom{-} 
\int_{\mathbf{k}_2,\mathbf{k}_3} 
\frac{\mathbf{k}_2 \cdot (\mathbf{k}_2 + \mathbf{k}_3)}{g_{\mathrm{1/N}} (-\mathbf{k}_3)}
\partial_t \theta_a (\mathbf{k}_3,t) \theta_a (\mathbf{k}_2,t) \theta_a (-\mathbf{k}_2 - \mathbf{k}_3,t) \nonumber\\
&=
-\int_{\mathbf{k}_1,\mathbf{k}_2,\mathbf{k}_3} \frac{\mathbf{k}_1 \cdot \mathbf{k}_2}{g_{\mathrm{1/N}}(\mathbf{k}_3)} \delta (\mathbf{k}_1 + \mathbf{k}_2 + \mathbf{k}_3) \theta_a (\mathbf{k}_1, t) \theta_a (\mathbf{k}_2,t) \partial_t \theta_a (\mathbf{k}_3,t) \,,
\end{align}
and
\begin{align}
\int_{\mathbf{k},\mathbf{k}',\mathbf{k}''}& {g_{\mathrm{1/N}}^{-1} (\mathbf{k})}
  {\,\mathbf{k}' \cdot (\mathbf{k} + \mathbf{k}') [\mathbf{k}'' \cdot (-\mathbf{k} + \mathbf{k}'')]}  
  \theta_a(\mathbf{k}') \theta_a(-\mathbf{k} - \mathbf{k}') \theta_a(\mathbf{k}'') \theta_a(\mathbf{k} -\mathbf{k}'') \nonumber \\
&=\
-\ \int_{\mathbf{k}_1, \mathbf{k}_2, \mathbf{k}_3} 
  {g_{\mathrm{1/N}}^{-1} (\mathbf{k}_1 - \mathbf{k}_2)}
  {(\mathbf{k}_1 \cdot \mathbf{k}_2) [\mathbf{k}_3 \cdot (\mathbf{k}_1 + \mathbf{k}_2 + \mathbf{k}_3))]} 
  \theta_a(\mathbf{k}_1) \theta_a(\mathbf{k}_2) \theta_a(\mathbf{k}_3) \theta_a(-\mathbf{k}_1 - \mathbf{k}_2 - \mathbf{k}_3) 
  \nonumber\\ 
 &=\ \phantom{-}\ 
  \int_{\mathbf{k}_1,\mathbf{k}_2, \mathbf{k}_3, \mathbf{k}_4} 
  {g_{\mathrm{1/N}}^{-1} (\mathbf{k}_1 - \mathbf{k}_2)}
  {(\mathbf{k}_1 \cdot \mathbf{k}_2) (\mathbf{k}_3 \cdot \mathbf{k}_4)} \theta_a(\mathbf{k}_1) \theta_a(\mathbf{k}_2) 
  \theta_a(\mathbf{k}_3) \theta_a(\mathbf{k}_4) \delta (\mathbf{k}_1 + \mathbf{k}_2 + \mathbf{k}_3 + \mathbf{k}_4)\,,
\end{align}
where we also used that $g_{\mathrm{1/N}} (\mathbf{k}) = g_{\mathrm{1/N}} (-\mathbf{k})$.
Taking the above terms together gives the effective action \eq{Seff4}.
\end{widetext}
\section{Non-linear sigma model}
\label{app:NLSM}
In Sect.~\ref{sec:LEEFT} in the main text, the derivation of the low-energy effective action required suppression of the density fluctuations for \emph{each} component. While possible, such an assumption is definitely not always satisfied. For that reason, we outline, in this appendix, following \cite{Watanabe2014a.PhysRevX.4.031057}, the derivation of the effective theory which is independent of the aforementioned assumption and only requires the suppression of the total density fluctuations.

\subsubsection{Field parametrization}
To begin with, let us recall the basic statement of the (Lorentz-invariant) Goldstone theorem. Suppose the Lagrangian is left invariant by a symmetry group $G$ with $n(G)$ generators, whereas the vacuum is only invariant under a subgroup $H$ with $n(H)$ generators. Then, there are $n(G) - n(H)$ Nambu-Goldstone gapless bosons. In other words, Goldstone excitations live in the coset space $G/H$. In our case, $G$ is $\mathrm{U}(N)$ and $H$ is $\mathrm{U}(N-1)$. Having the above in mind, we write down the following mathematical statement:
\begin{equation}
\mathrm{U}(N)/\mathrm{U}(N-1) = S^{2N-1}.
\end{equation}
In principle, we can now use the standard field parametrization procedure as described in, e.g., \cite{Pich:2018ltt}. However, since our theory is not Lorentz-invariant, the Goldstone theorem cannot be applied directly \cite{Watanabe2012a.PhysRevLett.108.251602,Watanabe2014a.PhysRevX.4.031057}. Furthermore, recalling that the Bogoliubov excitation is associated with a global phase, it is tempting to set it apart directly. Since, a phase can always be seen as a point on a circle $S^1$, we would like to, in some sense, divide $S^{2N-1}$ into an $S^1$ piece and the rest. Using the mathematical language, this can be done by a \emph{Hopf-like fibration}. One considers the $S^{2N-1}$ sphere as an $S^1$ bundle over the complex \emph{projective space} $\mathbb{CP}^{N-1}$. 

The elements of the projective space $\mathbb{CP}^{N-1}$ are the complex planes through the origin. For a complex plane $\text{span}_{\mathbb{C}}(w=(Z_1,\dots Z_N)^T)\subseteq \mathbb{R}^{2N}$ all points are associated with each other and represented by a \emph{representative} geometry. In principle, the choice of the representative is completely arbitrary, though usually it is possible to give a construction that relates this abstract space to geometric shapes. To do so, we normalize $w$: $w' = w/|w|$. We note that $w^\prime$ now lies on the $S^{2N-1}$ sphere. The intersection of the complex plane $\text{span}_{\mathbb{C}}(w)$ and $S^{2N-1}$ defines a circle of points $S^1$. Any point on that circle can now be chosen as the representative. Here, we just take all points of the circle to be associated with one point of the complex projective space. This defines the following projection:
\begin{equation}
  \pi: \quad S^{2N-1}\to S^{2N-1}/S^1\cong \mathbb{CP}^{N-1}, \quad w' \to w'/S^1.
\end{equation}
This projection cannot be inverted globally but only locally. $S^{2N-1}$ can be understood as the set of representatives called \emph{projective space} with a circle $S^1$ attached to each point. Mathematically, this, together with some requirements on the smoothness of the attachment, defines a \emph{fiber bundle} on the $\mathbb{CP}^{N-1}$ with fiber $S^1$. Compared to the product $\mathbb{CP}^{N-1}\times S^1$ the above fiber bundle has to be seen as a collection of circles $(S^1_z)_{z\in \mathbb{CP}^{N-1}}$, where copies are distinguished by the base point $z$. This type of decomposition of a sphere is also called Hopf fibration. 

The fiber bundle should locally look like a product. Therefore, around the ground state we are allowed to write the point of the sphere as a product,
\begin{equation}
\psi = e^{\i\theta} \chi, \quad \text{with } \ \psi \in S^{2N-1}, \ e^{i\theta} \in S^{1}, \ \chi \in \mathbb{CP}^{N-1}.
\end{equation}
Next, we parametrize the complex projective space using homogeneous coordinates. Choosing some point $(1,z)$, $z \in \mathbb{C}^{N-1}$, one can find a representative of the complex plane spanned by this point,
\begin{equation}
\chi(z) = \frac{1}{\sqrt{1+z ^{\dagger}z}} \begin{pmatrix}
1 \\
z
\end{pmatrix}\,, 
\quad \text{with } z \in \mathbb{C}^{N-1}\,,
\end{equation}
which lies on a unit sphere. We are now able to assemble our field parametrization:
\begin{equation}
\label{eq:parametrization}
\varphi = \sqrt{\rho} \psi = \sqrt{\rho} e ^{\i\theta} \frac{1}{\sqrt{1+z ^{\dagger}z}} \begin{pmatrix}
1 \\
z
\end{pmatrix}\,, \quad z \in \mathbb{C}^{N-1}\,,
\end{equation}
with the ground state chosen $\varphi^{(0)} = (\sqrt{\mu/g},0,\ldots,0)^{T}$, for the sake of simplicity. 
%

\subsubsection{Classical picture}
Plugging the parametrization \eqref{eq:parametrization} into the Gross-Pitaevskii Lagrangian, one can then find that the Lagrangian now takes the following form:
\begin{align}
\label{eq:actionsigma}
\mathcal{L}
=
&- 
\rho \partial_t \theta + \frac{\i}{2} \frac{\rho}{1 + z^{\dagger} z} \left(z^{\dagger} \partial_t z - \text{h.c.} \right) \nonumber\\
&- 
\frac{1}{2m}\Bigg\{(\nabla\sqrt{\rho})^2 + 
\rho\,\Bigg[ \left(\nabla \theta +\frac{\i}{2} \frac{\left(\nabla z^\dagger\right)z- \text{h.c}}{(1+z^\dagger z)^2}\right)^2 \nonumber\\
&+
\frac{(\nabla z^\dagger)(\nabla z)}{1+z^\dagger z} - \frac{(\nabla z^\dagger)z\cdot z^\dagger (\nabla z)}{(1+z^\dagger z)^2}\Bigg]\Bigg\} - \frac{g}{2}\rho^2 + \mu\rho.
\end{align}
We first analyze the classical picture behind \eqref{eq:actionsigma}. To that end, we expand the Lagrangian up to the second order in fields and use $\mu = g \rho^{(0)}$. The result reads:
\begin{align}
\mathcal{L}^{(2)} 
= 
&-\delta\rho \partial_t \theta  +\frac{i}{2}\rho^{(0)} \left(z^\dagger \partial_t z - \text{h.c.} \right)- \frac{g}{2}\delta\rho^2 \nonumber\\
&-
\frac{1}{2m}\bigg[\frac{1}{4\rho^{(0)}}(\nabla \delta\rho)^2 +\rho^{(0)}(\nabla\theta)^2 +\rho^{(0)}\nabla z \nabla z^{\dagger}  \bigg].
\end{align}
The corresponding Euler-Lagrange equations take the following form:
\begin{subequations}
\begin{align}
\label{eq:eqrho}
\partial_t \delta \rho 
&= 
- \frac{\rho^{(0)}}{m} \nabla^2 \theta \,,\\
\label{eq:eqtheta}
\partial_t \theta 
&=
\frac{1}{4 m \rho^{(0)}} \nabla^2 \delta \rho - g \delta \rho \,,\\
\label{eq:eqz}
\i\partial_t z 
&=
-\frac{\rho^{(0)}}{2m} \nabla^2 z\,,
\end{align}
\end{subequations}
and the same (complex conjugate) equation for $z^{\dagger}$. Combining \eqref{eq:eqrho} with \eqref{eq:eqtheta}, we get:
\begin{subequations}
\begin{align}
\partial_t^2 \theta &= \frac{\nabla^2}{2m} \left(-\frac{\nabla^2}{2m}  + 2 g \rho^{(0)} \right) \theta\,,\\
\i\partial_t z &= -\frac{\rho^{(0)}}{2m}\nabla^2 z.
\end{align}
\end{subequations}
With this, one obtains a global phase excitation with Bogoliubov dispersion and $N-1$ coupled (particle-hole) $z$-excitations with free-particle dispersion. 

\subsubsection{Effective action}
Following the same procedure as in \Sect{luttingerleeft} we now want to integrate out heavy degrees of freedom. In this case, those are the longitudinal density fluctuations $\delta\rho$ since they correspond to positive eigenmodes of the potential's curvature. By integrating them out, one can find, to the leading order in derivative expansion, the following effective Lagrangian \cite{Watanabe2014a.PhysRevX.4.031057}:
\begin{align}
\label{EffectiveActionWatanabe}
\mathcal{L}_{\mathrm{eff}} 
=
&\rho^{(0)}\left(-\partial_t\theta + \frac{\i}{2} \frac{z^{\dagger} \partial_tz - \partial_t z^{\dagger} z}{1 + z^\dagger z}\right) \nonumber\\
&-
\frac{\rho^{(0)}}{2m}\left(\nabla\theta + \frac{\i}{2} \frac{z^{\dagger} \nabla z - \nabla z^{\dagger} z}{1 + z^{\dagger} z}\right)^2 \nonumber\\
&-
\frac{\rho^{(0)}}{2m}\left( \frac{\nabla z^{\dagger} \nabla z}{1 + z^{\dagger} z} - \frac{(\nabla z^{\dagger} z)(z^{\dagger} \nabla z)}{(1+z^{\dagger} z)^2}\right) \nonumber\\
&-
\frac{1}{2g}\left(-\partial_t\theta + \frac{\i}{2} \frac{z^{\dagger} \partial_t z - \partial_t z^{\dagger} z}{1 + z^{\dagger} z}\right)^2 + \mathrm{h.o.t}.
\end{align}
We note that the resulting effective Lagrangian has a form of a (non-relativistic) non-linear sigma model (NLSM) with the Fubini-Study metric. As it was shown in \cite{Watanabe2012a.PhysRevLett.108.251602}, different types of NLSM also appear in the experimentally relevant system of an ultracold Bose gas with spin-spin interactions. Furthermore, such models are known to enjoy stable topologically non-trivial configurations, and it was observed that dynamical properties of a non-thermal fixed point can change dramatically in presence of topological defects (vortices) in a Bose gas \cite{Nowak:2010tm,Nowak:2011sk,Schole:2012kt,Nowak:2012gd,Karl2017b.NJP19.093014}. 
Finally, to close the discussion, let us mention that out-of-equilibrium dynamics of non-linear sigma models has been considered in the context of disordered metals and superconductors \cite{Kamenev:2009}. The discussion of far-from-equilibrium dynamics within such non-linear sigma models is beyond the scope of this work and of interest for future research.

\section{Kinetic equation}
\label{app:Boltzmann}
In this Appendix, we discuss the derivation of the quantum Boltzmann equation for the Luttinger-liquid-like effective action \eqref{eq:Seff4} following the standard routine (see, e.g., \cite{Berges:2004yj,Berges:2015kfa,Lindner:2005kv,Lindner:2007am}). We start with the Dyson equation written in the following form:
\begin{align}
\label{eq:dyson}
\int_{z,\mathcal{C}}& G_0^{-1}(x,z) G(z,y) - \int_{z,\mathcal{C}} \left[\Sigma(x,z) + \i R(x,z) \right] G(z,y)\nonumber\\
= \ &
\delta_{\mathcal{C}}(x-y).
\end{align}
In the absence of explicit external sources, the term $R(x,y)$ has a support only for $x^0 > t_0$ reflecting the initial density matrix contribution. The initial condition for $G(x,y)$ has to be specified at $x^0,y^0 = t_0$, while the subsequent evolution for times $t > t_0$ follows from \eqref{eq:dyson} with $R = 0$. Therefore, we can set $R(x,y)$ to zero at $t>0$.
See \App{EFT_general} for a more detailed discussion of the relevance of the initial state within our low-energy effective field theory description. 

\onecolumngrid
Recalling that $\theta(x)$ has a zero macroscopic value, we note the classical inverse propagator $G_0(x,y)$ coincides with the free propagator \eqref{eq:freeinverse}, which in real space reads as~\footnote{We changed the notation of the mean density to $n_a^{(0)}$ for this section to avoid a possible confusion with the spectral function.}
\begin{equation}
\label{eq:realspaceeq}
\i G^{-1}_0(x,y; \langle \theta \rangle = 0) = -\mathcal{K}(\mathbf{x} - \mathbf{y}) \partial_{x^0}^2 \delta_{\mathcal{C}}(x^0 - y^0) + \frac{n_a^{(0)}}{m} \nabla^2_{\mathbf{x}} \delta_{\mathcal{C}} (x - y)\,,
\end{equation}
with 
\begin{equation}
\label{eq:fouriergeff}
\mathcal{K}(\mathbf{x} - \mathbf{y}) = \int_{\mathbf{k}} \frac{e^{-\i \mathbf{k} (\mathbf{x} - \mathbf{y})}}{g_G(\mathbf{k})}.
\end{equation}
Equation \eqref{eq:realspaceeq} then takes the form
\begin{equation}
\label{eq:prekadanoffbaym}
\partial_{x^0}^2 \int_{\mathbf{z}} \mathcal{K}(\mathbf{x} - \mathbf{z}) G(\mathbf{z},x^0;y) - \frac{n_a^{(0)}}{m} \nabla^2_{\mathbf{x}} G(x,y) + 
\i \int_{z,\mathcal{C}} \Sigma(x,z) G(z,y) = -\i \delta_{\mathcal{C}}(x-y).
\end{equation}
Next, we decompose the propagator $G(x,y)$ and the proper self-energy $\Sigma(x,y)$ into spectral and statistical parts:
\begin{subequations}
\begin{align}
\label{eq:decompos}
G(x,y) &= F(x,y) - \frac{\i}{2} \rho(x,y) \sgn_{\mathcal{C}}(x^0 - y^0),\\
\label{eq:decomposen}
\Sigma(x,y) &= -\i \Sigma^{(0)}(x) \delta(x-y) + \Sigma^{F}(x,y) - \frac{\i}{2} \Sigma^{\rho}(x,y) \sgn_{\mathcal{C}}(x^0 - y^0),
\end{align}
\end{subequations}
where we also introduced the local part of the self-energy $\Sigma^{(0)}(x)$. Let us analyze each term in \eqref{eq:prekadanoffbaym} separately. The first term takes the form
\begin{align}
\int_{\mathbf{z}}& \mathcal{K}(\mathbf{x} - \mathbf{z}) \partial^2_{x^0} \left[F(\mathbf{z},x^0;y) - \frac{\i}{2} \rho(\mathbf{z},x^0;y) \sgn_{\mathcal{C}}(x^0 - y^0) \right] \nonumber\\
= &\int_{\mathbf{z}} \mathcal{K}(\mathbf{x} - \mathbf{z}) \bigg[\partial_{x^0}^2 F(\mathbf{z},x^0;y)  
- \frac{\i}{2} \sgn_{\mathcal{C}}(x^0 - y^0) \partial_{x^0}^2 \rho(\mathbf{z},x^0;y) - 2 \i \delta_{\mathcal{C}}(x^0 - y^0) \partial_{x^0} \rho(\mathbf{z},x^0;y) - \i \rho(\mathbf{z},x^0;y) \partial_{x^0} \delta_{\mathcal{C}}(x^0 - y^0) \bigg]  \nonumber\\
= &\int_{\mathbf{z}} \mathcal{K}(\mathbf{x} - \mathbf{z}) \bigg[\partial_{x^0}^2 F(\mathbf{z},x^0;y) - \frac{\i}{2} \partial_{x^0}^2 \rho(\mathbf{z},x^0;y) \sgn_{\mathcal{C}}(x^0 - y^0) \bigg] - \i \delta_{\mathcal{C}}(x - y),
\end{align}
where we have used that~\footnote{The following statements should, of course, be understood in terms of distributional derivatives}:
\begin{equation}
\partial_{x^0} \sgn_{\mathcal{C}}(x^0 - y^0) = 2 \delta_{\mathcal{C}}(x^0 - y^0), \quad \rho(\mathbf{z},x^0;y) \partial_{x^0} \delta_{\mathcal{C}}(x^0 - y^0) = -\delta_{\mathcal{C}}(x^0 - y^0) \rho(\mathbf{z},x^0;y),
\end{equation}
and
\begin{align}
\delta_{\mathcal{C}}(x^0 &- y^0) \int_{\mathbf{z}} \mathcal{K}(\mathbf{x} - \mathbf{z}) \partial_{x^0} \rho(\mathbf{z},x^0;y) = \i \delta_{\mathcal{C}}(x^0 - y^0) \left\langle \left[\int_{\mathbf{z}} \mathcal{K}(\mathbf{x} - \mathbf{z}) \partial_{x^0} \theta(\mathbf{z},x^0), \theta(\mathbf{y},y^0)\right] \right\rangle \nonumber\\
&=
- \i \delta_{\mathcal{C}}(x^0 - y^0) \left\langle\left.\left[\theta(\mathbf{y},y^0),\pi_{\theta}(\mathbf{x},x^0) \right] \right\rvert_{x^0=y^0}\right\rangle = \delta_{\mathcal{C}} (x^0 - y^0) \delta(\mathbf{x} - \mathbf{y}) = \delta_{\mathcal{C}}(x - y).
\end{align}
Since the second term and the local part of the third one in \eqref{eq:prekadanoffbaym} are trivial, it is only left to consider the non-local part of the latter:
\begin{align}
\i &\int_{z,\mathcal{C}} \left(\Sigma^F(x,z) - \frac{\i}{2} \Sigma^{\rho}(x,z) \sgn_{\mathcal{C}}(x^0 - z^0) \right) \left(F(z,y) - \frac{\i}{2} \rho(z,y) \sgn_{\mathcal{C}}(z^0 - y^0) \right) \nonumber\\
&=\ \int_{\mathbf{z}} \bigg(\int_{t_0}^{x^0} d z^0 \, \Sigma^{\rho}(x,z) F(z,y)
- \frac{\i}{2}\sgn_{\mathcal{C}}(x^0 - y^0) \int_{y^0}^{x^0} d z^0 \, \Sigma^{\rho}(x,z) \rho(z,y)  
- \int_{t_0}^{y^0} d z^0 \, \Sigma^{F}(x,z)\rho(z,y) \bigg),
\end{align}
where we have used $\sgn_{\mathcal{C}}(x) = \theta_{\mathcal{C}}(x) - \theta_{\mathcal{C}}(-x)$ and properties of the Schwinger-Keldysh contour. 

Collecting the results obtained above and comparing the real and imaginary parts of the resulting equation, we find the following set of equations for statistical and spectral functions:
\begin{subequations}
\label{eq:kadanoffbaym}
\begin{align}
\label{eq:kbstat}
(\hat{\mathcal{D}}_x F)(x,y) &= -\int_{t_0}^{x^0} dz \, \Sigma^{\rho}(x,z) F(z,y) + \int_{t_0}^{y^0} dz \, \Sigma^{F}(x,z) \rho(z,y), \\
\label{eq:kbspect}
(\hat{\mathcal{D}}_x \rho)(x,y) &= -\int_{y^0}^{x^0} dz \, \Sigma^{\rho}(x,z) \rho(z,y),
\end{align}
\end{subequations}
where we have adopted the short-hand notation $\int_{t_1}^{t_2} d z = \int_{t_1}^{t_2} d z^0 \int d^d z$. Above, we also introduced an integro-differential operator $\hat{\mathcal{D}}_x$ related to the real-space Green's function and defined as
\begin{equation}
(\hat{\mathcal{D}}_x h)(x,y) = \int_{\mathbf{z}}\mathcal{K}(\mathbf{x} - \mathbf{z}) \partial_{x^0}^2 h(\mathbf{z},x^0;y) - \frac{n_a^{(0)}}{m} \nabla_{\mathbf{x}}^2 h(x,y) + \Sigma^{(0)}(x)h(x,y).
\end{equation}  
The system of coupled equations \eqref{eq:kadanoffbaym} describing the evolution of a two-point Green's function is commonly referred as the \emph{Kadanoff-Baym equations}. Note that in the absence of approximations on the self-energy, these equations are exact. These equations, however, are not feasible to solve analytically. Therefore, some further approximations are required, leading to Boltzmann-type kinetic equations. 
Note that, in this work, we intend to perform a scaling analysis of the system at non-thermal fixed points.
This can be achieved using the Kadanoff-Baym equations directly \cite{Berges:2008wm,Berges:2008sr,Scheppach:2009wu}, while, however, the scaling analysis of the kinetic equations \cite{Berges:2010ez,Orioli:2015dxa,Chantesana:2018qsb.PhysRevA.99.043620} is more straightforward and employs standard procedures \cite{Svistunov1991a,Zakharov1992a}.

To proceed, we perform the so-called \emph{gradient expansion} \cite{Branschadel:2008sk,Berges:2005md} of the Kadanoff-Baym equations. To do so, we first interchange $x$ and $y$ in the equations and subtract the result from \eqref{eq:kbstat} and \eqref{eq:kbspect}, respectively. Using the relation between retarded/advanced and spectral components one then gets:
\begin{subequations}
\label{eq:pretransport}
\begin{align}
\left\lbrace\left(\hat{\mathcal{D}}_x - \hat{\mathcal{D}}_y \right) F \right\rbrace(x,y) &= \int_{z} \theta(z^0) \bigg(F(x,z) \Sigma^{A}(z,y) + G^{R} (x,z) \Sigma^{F}(z,y) - \Sigma^{R}(x,z)F(z,y) - \Sigma^{F}(x,z)G^{A}(z,y)\bigg),\\
\left\lbrace\left(\hat{\mathcal{D}}_x - \hat{\mathcal{D}}_y \right) \rho \right\rbrace(x,y) &= \int_{z} \bigg(G^{R}(x,z) \Sigma^{\rho}(z,y) + \rho(x,z) \Sigma^{A}(z,y) - \Sigma^{\rho}(x,z)G^{A}(z,y) - \Sigma^{R}(x,z)\rho(z,y)\bigg).
\end{align}
\end{subequations}
Next, we introduce \emph{center and relative coordinates}:
\begin{equation}
\label{eq:centrelcoord}
X_{xy}^{\mu} = \frac{x^{\mu} + y^{\mu}}{2}, \qquad s_{xy}^{\mu} = x^{\mu} - y^{\mu}.
\end{equation}
The idea is to approximate a finite time description of the system by employing the above equations with $t_0 \to -\infty$. This then allows for a systematic expansion with respect to derivatives of $X^{\mu}$ and powers of $s^{\mu}$. The analysis of the right-hand side of \eqref{eq:pretransport} is standard and can be found in, e.g., \cite{Berges:2015kfa}, and we therefore concentrate on the left-hand side. From \eqref{eq:centrelcoord} it immediately follows that
\begin{subequations}
\begin{align}
\partial_{x^0}^2 &= \partial_{s^0_{xy}}^2 + \partial_{s^0_{xy} X^0_{xy}}^2 + \partial_{X^0_{xy}}^2,\\
\partial_{y^0}^2 &= \partial_{s^0_{xy}}^2 - \partial_{s^0_{xy} X^0_{xy}}^2 + \partial_{X^0_{xy}}^2,
\end{align}
\end{subequations}
which yields
\begin{align}
\label{eq:fpart}
\int_{\mathbf{z}}\bigg\lbrace \mathcal{K}(\mathbf{x} - \mathbf{z}) \partial_{x^0}^2 F(\mathbf{z},x^0;y) - \mathcal{K}(\mathbf{z} - \mathbf{y}) \partial_{y^0}^2 F(x;\mathbf{z},y^0) \bigg\rbrace 
=
\int_{\mathbf{z}}& \bigg\lbrace \partial^2_{s_{xy}^0 X_{xy}^0} \big[\mathcal{K}(\mathbf{x} - \mathbf{z}) F(\mathbf{z},x^0;y) + \mathcal{K}(\mathbf{z} - \mathbf{y}) F(x;\mathbf{z},y^0)\big] \nonumber\\
&+ 
\left(\partial_{s_{xy}^0}^2 + \partial_{X_{xy}^0}^2 \right) \big[\mathcal{K}(\mathbf{x} - \mathbf{z}) F(\mathbf{z},x^0;y) 
- \mathcal{K}(\mathbf{z} - \mathbf{y}) F(x;\mathbf{z},y^0)\big] \bigg\rbrace .
\end{align}
Next, following the idea of the \emph{Wigner distribution function}, we perform a Fourier transform with respect to relative coordinates, i.e., we define:
\begin{align}
\label{eq:wigner}
F(\mathbf{k},\omega,X) &= \int_{-2 X^0}^{2 X^0} d s^0 \, e^{-\i \omega s^0} \int_{-\infty}^{\infty} d^ds \, e^{\i \mathbf{k} \mathbf{s}} F \left(X + \frac{s}{2},X - \frac{s}{2} \right),\\
\tilde{\rho}(\mathbf{k},\omega,X) &= -\i \int_{-2 X^0}^{2 X^0} d s^0 \, e^{-\i \omega s^0} \int_{-\infty}^{\infty} d^ds \, e^{\i \mathbf{k} \mathbf{s}} \rho \left(X + \frac{s}{2},X - \frac{s}{2} \right), 
\end{align}
and similarly for $\Sigma^{F}$ and $\Sigma^{\rho}$. Here, the imaginary unit $\i$ factor was added to have $\tilde{\rho}(X,\mathbf{k},\omega)$ real~\footnote{Recall that $F(x,y)$ is even, whereas $\rho(x,y)$ is odd with respect to relative coordinates, which implies that the Fourier transform of the former is real, while the latter one is purely imaginary.}. Finally, while, for given finite $X^0$, relative coordinate $s^0$ is limited by $\pm X^0$, it is standard to extend the integration limits in \eqref{eq:wigner} to $\pm \infty$, which is related to the assumption $t \to -\infty$ that has been done above. With all that in mind, we perform a Fourier transform of both sides of \eqref{eq:pretransport} with respect to the relative coordinate $s_{xy}$. Again, we focus on \eqref{eq:fpart} since the rest is either similar or standard for any (scalar) theory. To that end, we first perform a spatial transformation, e.g.,
\begin{align}
\int_{\mathbf{z}}\int_{\mathbf{s}_{xy}}& e^{\i \mathbf{k} \mathbf{s}_{xy}} \mathcal{K}(\mathbf{s}_{xz}) F \left(\mathbf{X}_{zy} + \frac{\mathbf{s}_{zy}}{2}, X_{xy}^0 + \frac{s^0_{xy}}{2}; \mathbf{X}_{zy} - \frac{\mathbf{s}_{zy}}{2}, X_{xy}^0 - \frac{s^0_{xy}}{2}\right) \nonumber\\
=
&\int_{\mathbf{z}}\int_{\mathbf{s}_{xy}}\int_{\mathbf{p}} e^{\i \mathbf{k} \mathbf{s}_{xy} - \i \mathbf{p} \mathbf{s}_{xz}} \mathcal{K}(\mathbf{X}_{zy} + \frac{\mathbf{s}_{zy}}{2}) F \left(\mathbf{p},X_{xy}^0 + \frac{s^0_{xy}}{2}; \mathbf{X}_{zy} - \frac{\mathbf{s}_{zy}}{2}, X_{xy}^0 - \frac{s^0_{xy}}{2}\right) \nonumber\\
=
&\int_{\mathbf{z}}\int_{\mathbf{s}_{xy}}\int_{\mathbf{p}}\int_{\mathbf{q}} e^{\i \mathbf{k} \mathbf{s}_{xy} - \i \mathbf{p} \mathbf{x} + \i \mathbf{q} \mathbf{y} + \i(\mathbf{p} - \mathbf{q})\mathbf{z}} \mathcal{K}(\mathbf{p}) F(\mathbf{q},x^0,y^0) = \int_{\mathbf{s}_{xy}}\int_{\mathbf{p}} e^{\i (\mathbf{k} - \mathbf{p}) \mathbf{s}_{xy}} \mathcal{K}(\mathbf{p}) F(\mathbf{p},x^0,y^0) = \mathcal{K}(\mathbf{k}) F(\mathbf{k},x^0,y^0), 
\end{align}
where $\mathcal{K}(\mathbf{k}) = (g_G(\mathbf{k}))^{-1}$, which immediately follows from \eqref{eq:fouriergeff}. Here, we have already assumed no dependence on spatial central coordinates due to homogeneity of the system. We thus conclude that after the spatial Wigner transform the RHS of \eqref{eq:fpart} becomes:
\begin{align}
\eqref{eq:fpart} 
\to 
\partial_{s_{xy}^0 X_{xy}^0}^2 \left[\mathcal{K}(\mathbf{k}) F(\mathbf{k},x^0,y^0) + \mathcal{K}(\mathbf{k}) F(\mathbf{k},x^0,y^0)\right] + \left(\partial_{s_{xy}^0}^2 + \partial_{X_{xy}^0}^2 \right) \left[\mathcal{K}(\mathbf{k}) F(\mathbf{k},x^0,y^0) - \mathcal{K}(\mathbf{k}) F(\mathbf{k},x^0,y^0) \right] \nonumber\\
=
2 \mathcal{K}(\mathbf{k}) \partial_{s_{xy}^0 X_{xy}^0}^2 F(\mathbf{k},x^0,y^0) .
\end{align}
As it was already mentioned, the remaining part of the analysis is standard and can be found in \cite{Berges:2015kfa}. In short, we finally end up with the set of the so-called \emph{transport equations}:
\begin{subequations}
\label{eq:transport}
\begin{align}
\label{eq:transportstat}
\frac{2 k^0}{g_G(\mathbf{k})} \frac{\partial F(k,X^0)}{\partial X^0} &= \tilde{\Sigma}^{\rho}(k,X^0) F(k,X^0) - \Sigma^{F}(k,X^0) \tilde{\rho}(k,X^0),\\
\label{eq:transportspect}
\frac{2 k^0}{g_G(\mathbf{k})} \frac{\partial \tilde{\rho} (k,X^0)}{\partial X^0} &= 0 ,
\end{align}
\end{subequations}
which are the leading order equations with respect to $\partial_{X^{\mu}}$ and $k^{\mu}$ expansion. Note that \eqref{eq:transportspect} implies that the spectral remains unchanged during the evolution.

\begin{figure}
\centering
\includegraphics[scale=0.8]{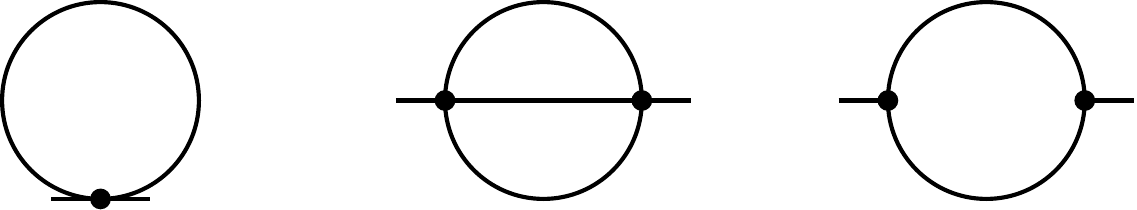}
\caption{First non-trivial diagrams that contribute to the proper self-energy $\Sigma(x,y)$. The lines represent full propagators $G(x,y)$, the vertices are introduced in Fig.~\ref{fig:verts}}.
\label{fig:diagrams}
\end{figure}

It remains to compute the self-energies $\tilde{\Sigma}^{\rho}$ and $\Sigma^{F}$ in some chosen approximation. In this paper, we restrict ourselves to only first non-trivial contributions to the self-energy, see Fig.~\ref{fig:diagrams} for a diagrammatic representation. We discuss the adequacy of such an approximation {\it a posteriori} when discussing the properties of the scattering matrix, see \Sect{ScalingLEEFT}.

Before we proceed, let us define the real-space couplings:
\begin{subequations}
\begin{align}
\gamma(\mathbf{x}_1,\ldots,\mathbf{x}_3) 
&=
\int_{\mathbf{p}_1,\ldots,\mathbf{p}_3} e^{\i \left(\mathbf{p}_1\mathbf{x}_1 + \ldots + \mathbf{p}_3 \mathbf{x}_3 \right)} \gamma(\mathbf{p}_1,\ldots,\mathbf{p}_3) (2 \pi)^d \, \delta\left(\sum_{i=1}^3 \mathbf{p}_i \right),\\
\lambda(\mathbf{x}_1,\ldots,\mathbf{x}_4) &=
\int_{\mathbf{p}_1,\ldots,\mathbf{p}_4} e^{\i\left(\mathbf{p}_1 \mathbf{x}_1 + \ldots + \mathbf{p}_4 \mathbf{x}_4 \right)} \lambda(\mathbf{p}_1,\ldots,\mathbf{p}_4) (2 \pi)^d \delta \left(\sum_{i=1}^{4} \mathbf{p}_i \right).
\end{align}
\end{subequations}
Note that the sign might seem to be off (c.f.~Appendix~\ref{app:Not}), but it reflects the definition
\begin{equation}
\int_{\lbrace \mathbf{k}_i \rbrace} g_l(\mathbf{k}_1, \ldots, \mathbf{k}_l) \theta(\mathbf{k}_1,t) \ldots \theta(\mathbf{k}_l,t) \equiv \int_{\lbrace \mathbf{x}_i \rbrace} g_l(\mathbf{x}_1, \ldots, \mathbf{x}_l) \theta(\mathbf{x}_1,t) \ldots \theta(\mathbf{x}_l,t). 
\end{equation}

We are now in the position to analyze the diagrams. First, we note that the left, so-called \emph{tadpole}, diagram is typically local and only gives a shift of a mass. However, since the interactions of our low-energy effective action are non-local, one has to be a bit more careful. Because the interaction term is still local in time, so is the tadpole. Therefore, it only contributes to $\Sigma^{F}$ and has only a central coordinate $X^0 = t$. This implies that in Wigner space, i.e., after taking a Fourier transform with respect to relative coordinates, it is proportional to $\delta (k^0)$, as one expects from a time-local tadpole. Lastly, we notice that it only enters the transport equations \eqref{eq:transport} together with $\tilde{\rho}(X^0,k)$, which is zero at $k^0 = 0$. Hence, we conclude that the tadpole diagram does not contribute to the kinetic equation.

Next, we consider the central (\emph{sunset}) diagram. Its symbolic expression reads as
\begin{equation}
\Sigma_{2-\mathrm{loop}}(x,y) 
=
-\int_{\substack{\mathbf{x}_1,\mathbf{x}_2,\mathbf{x}_3 \\ \mathbf{y}_1,\mathbf{y}_2,\mathbf{y}_3}} \lambda(\mathbf{x},\mathbf{x}_1,\mathbf{x}_2,\mathbf{x}_3) \lambda(\mathbf{y}, \mathbf{y}_1,\mathbf{y}_2,\mathbf{y}_3) G(\mathbf{x}_1, t;\mathbf{y}_1, t') G(\mathbf{x}_2, t; \mathbf{y}_2, t') G(\mathbf{x}_3, t;\mathbf{y}_3, t').
\end{equation}
Introducing the short-hand notation, $G(\mathbf{x}_i, t;\mathbf{y}_i, t') \equiv G_i$, and employing the definition \eqref{eq:decompos}, it is straightforward to show that
\begin{equation}
G_1 G_2 G_3 = \Big[F_1 F_2 F_3 - \frac{1}{4} \left(F_1 \rho_2 \rho_3 + \rho_1 F_2 \rho_3 + \rho_1 \rho_2 F_3 \right)\Big] + \Big[F_1 F_2 \rho_3 + F_1 \rho_2 F_3 + 
 \rho_1 F_2 F_3 - \frac{1}{4} \rho_1 \rho_2 \rho_3 \Big] \left(-\frac{\i}{2} \sgn_{\mathcal{C}}(t - t') \right),
\end{equation}
which combined with \eqref{eq:decomposen} yields
\begin{subequations}
\begin{align}
\label{eq:2loopse}
\Sigma_{2-\mathrm{loop}}^{F}(x,y) 
&= 
-\int_{\substack{\mathbf{x}_1,\mathbf{x}_2,\mathbf{x}_3 \\ \mathbf{y}_1,\mathbf{y}_2,\mathbf{y}_3}} \lambda(\mathbf{x},\mathbf{x}_1,\mathbf{x}_2,\mathbf{x}_3) \lambda(\mathbf{y}, \mathbf{y}_1,\mathbf{y}_2,\mathbf{y}_3) \Big[F_1 F_2 F_3 - \frac{1}{4} \big(F_1 \rho_2 \rho_3 + \rho_1 F_2 \rho_3 + \rho_1 \rho_2 F_3 \big)\Big],\\
\Sigma^{\rho}_{2-\mathrm{loop}}(x,y) 
&=
-\int_{\substack{\mathbf{x}_1,\mathbf{x}_2,\mathbf{x}_3 \\ \mathbf{y}_1,\mathbf{y}_2,\mathbf{y}_3}} \lambda(\mathbf{x},\mathbf{x}_1,\mathbf{x}_2,\mathbf{x}_3) \lambda(\mathbf{y}, \mathbf{y}_1,\mathbf{y}_2,\mathbf{y}_3) \Big[F_1 F_2 \rho_3 + F_1 \rho_2 F_3 + \rho_1 F_2 F_3 - \frac{1}{4} \rho_1 \rho_2 \rho_3 \Big].
\end{align}
\end{subequations}
Now we perform a Wigner transformation. Since all the terms have a similar structure, we only consider a single of them:
\begin{align}
\label{eq:interstep}
\int_{\substack{\lbrace\mathbf{x}_i \rbrace, \lbrace\mathbf{y}_i \rbrace\\ s_{xy}}} \lambda(\mathbf{x},\mathbf{x}_1,\mathbf{x}_2,\mathbf{x}_3) \lambda(\mathbf{y}, \mathbf{y}_1,\mathbf{y}_2,\mathbf{y}_3) \rho_1 F_2 \rho_3 e^{-\i k s_{xy}} 
=
&\int \lambda(\mathbf{p},\mathbf{p}_1,\mathbf{p}_2,\mathbf{p}_3) \lambda(\mathbf{q},\mathbf{q}_1,\mathbf{q}_2,\mathbf{q}_3) \rho_1 F_2 \rho_3 \, e^{-\i k s_{xy} + \i (\mathbf{p} \mathbf{x}  + \ldots + \mathbf{q}_3 \mathbf{y}_3)} \nonumber\\
&\times
(2 \pi)^{2d} \, \delta\left( \mathbf{p} + \mathbf{p}_1 + \mathbf{p}_2 + \mathbf{p}_3 \right) \delta \left(\mathbf{q} + \mathbf{q}_1 + \mathbf{q}_2 + \mathbf{q}_3\right).
\end{align}
Here, we suppressed the integration variables to ease the notation. Next, we notice that the integration over $\lbrace \mathbf{x}_i \rbrace$ and $\lbrace \mathbf{y}_i \rbrace$ can be transformed into the integration over $\lbrace \mathbf{s}_{xy,i} \rbrace$ and $\lbrace \mathbf{X}_{xy,i} \rbrace$ via a unitary transformation. Furthermore, we use that
\begin{equation}
e^{\i \mathbf{p}_i\mathbf{x}_i + \i \mathbf{p}_i\mathbf{x}_i} = e^{\i (\mathbf{p}_i + \mathbf{q}_i) \mathbf{X}_{xy,i} + \i(\mathbf{p}_i - \mathbf{q}_i) \mathbf{s}_{xy,i}/2},
\end{equation}
which, recalling that, for a spatially homogeneous system, $F_i$ and $\rho_i$ do not depend on spatial central coordinates, yields a series of delta-functions $(2\pi)^{d}\, \delta(\mathbf{p}_i + \mathbf{q}_i)$. We thus obtain:
\begin{align}
\eqref{eq:interstep} 
&=
\int\lambda(\mathbf{p},\mathbf{p}_1,\mathbf{p}_2,\mathbf{p}_3) \lambda(\mathbf{q},-\mathbf{p}_1,-\mathbf{p}_2,-\mathbf{p}_3) \, e^{\i (-k s_{xy} + \mathbf{p} \mathbf{x} + \mathbf{q} \mathbf{y} + \mathbf{p}_1 \mathbf{s}_{xy,1} + \mathbf{p}_2 \mathbf{s}_{xy,2} + \mathbf{p}_3 \mathbf{s}_{xy,3})} \rho_1 F_2 \rho_3 (2 \pi)^{2d} \, \delta\left( \mathbf{p} + \mathbf{p}_1 + \mathbf{p}_2 + \mathbf{p}_3 \right) \nonumber\\
&\phantom{=}\times
 \delta \left(\mathbf{q} - \mathbf{p}_1 - \mathbf{p}_2 - \mathbf{p}_3\right) 
=
\int |\lambda(\mathbf{p},\mathbf{p}_1,\mathbf{p}_2,\mathbf{p}_3)|^2 
e^{\i (- k s_{xy} + \mathbf{p} \mathbf{s}_{xy} + \mathbf{p}_1 \mathbf{s}_{xy,1} + \mathbf{p}_2 \mathbf{s}_{xy,2} + \mathbf{p}_3 \mathbf{s}_{xy,3})} \rho_1 F_2 \rho_3 (2\pi)^d\, \delta (\mathbf{p} + \mathbf{p}_1 + \mathbf{p}_2 + \mathbf{p}_3) \nonumber\\
&=
\i^2 \int |\lambda(-\mathbf{k},\mathbf{p}_1,\mathbf{p}_2,\mathbf{p}_3)|^2 e^{\i(-k^0 + p_1^0 + p_2^0 + p_3^0) s_{xy}^0} \rho(\mathbf{p}_1,p_1^0,X^0) F(\mathbf{p}_2,p_2^0,X^0) 
\rho(\mathbf{p}_3,p_3^0,X^0) (2 \pi)^d \, \delta (-\mathbf{k} + \mathbf{p}_1 + \mathbf{p}_2 + \mathbf{p}_3) \nonumber\\
&=
-\int_{p,q,r} |\lambda(-\mathbf{k},\mathbf{p},\mathbf{q},\mathbf{r})|^2 
\rho(p,X^0) F(q,X^0) \rho(r,X^0) (2 \pi)^{d+1} \, \delta(k - p - q - r),
\end{align} 
where in the last line we have changed the notation of dummy variables. Here, we adopted the notation $|\lambda(\mathbf{k},\mathbf{p},\mathbf{q},\mathbf{r})|^2 \equiv \lambda(\mathbf{k},\mathbf{p},\mathbf{q},\mathbf{r}) \lambda(-\mathbf{k},-\mathbf{p},-\mathbf{q},-\mathbf{r})$. Repeating the same procedure for each term in \eqref{eq:2loopse} and recalling that for $\Sigma^{\rho}$ an additional factor $-\i$ has to be added, we conclude that up to $2$ loops the transport equations take the form:
\begin{subequations}
\begin{align}
\label{eq:preboltzmann}
\frac{2 k^0}{g_{G}(\mathbf{k})}\frac{\partial{F_k}}{\partial X^0} 
=&
\int_{p,q,r} |\lambda(-\mathbf{k},\mathbf{p},\mathbf{q},\mathbf{r})|^2 \Big[\tilde{\rho}_k  F_p F_q F_r  + \frac{1}{4} \tilde{\rho}_k \left(F_p \tilde{\rho}_q \tilde{\rho}_r + \tilde{\rho}_p F_q \tilde{\rho}_r + \tilde{\rho}_p \tilde{\rho}_q F_r \right) \nonumber\\
&-
F_k F_p F_q \tilde{\rho}_r  - F_k F_p \tilde{\rho}_q F_r - F_k \tilde{\rho}_p F_q F_r - \frac{1}{4} F_k \tilde{\rho}_p\tilde{\rho}_q \tilde{\rho}_r \Big] (2 \pi)^{d+1}\, \delta (k - p - q - r),\\
\frac{2 k^0}{g_G(\mathbf{k})} \frac{\partial \tilde{\rho}_k}{\partial X^0} 
=&\ 0,
\end{align}
\end{subequations}
where $F_k \equiv F(k,X^0)$, $\tilde{\rho}_k \equiv \tilde{\rho}(k,X^0)$. We now write, without any loss of generality, the generalized relation
\begin{equation}
\label{eq:callenwelton}
F(k,X^0) = \left(f(k,X^0) + \frac{1}{2} \right) \tilde{\rho}(k,X^0)\,,
\end{equation}
which in the equilibrium case reduces to the \emph{Callen-Welton fluctuation-dissipation theorem}. Changing in \eqref{eq:preboltzmann} the integration variable $p \to -p$ and employing that for a real scalar theory one has 
\begin{equation}
F(-p,X) = F(p,X), \quad \tilde{\rho}(-p,X) = -\tilde{\rho}(p,X),
\end{equation}
we obtain, after some simple algebra,
\begin{equation}
\label{eq:almostboltzmann}
\frac{2 k^0}{g_{G}(\mathbf{k})} \tilde{\rho}_k  \partial_t f_k = \int_{p,q,r} \left|\lambda (\mathbf{k},\mathbf{p},-\mathbf{q},-\mathbf{r}) \right|^2 (2 \pi)^{d+1}\, \delta(k + p - q - r) \big[(f_k + 1)(f_p + 1) f_q f_r - f_k f_p (f_q + 1) (f_r + 1)  \big]\, \tilde{\rho}_{p} \tilde{\rho}_{k} \tilde{\rho}_{q} \tilde{\rho}_{r},
\end{equation}
where we have relabeled the central coordinate $X^0$ to $t$, reflecting the fact that it plays the role of time in our limit, and used $|\lambda(-\mathbf{k},-\mathbf{p},\mathbf{q},\mathbf{r})|^2 \equiv |\lambda(\mathbf{k},\mathbf{p},-\mathbf{q},-\mathbf{r})|^2$.

At last, we adopt the \emph{quasiparticle} (or \emph{on-shell}) \emph{approximation}, i.e., assume that the spectral function takes the form of a free one, $\tilde{\rho}(p) \approx \tilde{\rho}^{(0)}(p)$~\footnote{Note that, for a non-local theory as given by the vertices depicted in Fig.~\ref{fig:verts}, the validity of the on-shell approximation needs to be verified
after all. This question is beyond the scope of the present work and is the subject of further study.}. From \eqref{eq:freeinverse} one can derive:
\begin{equation}
G^{R,(0)}(k) = \frac{g_G(\mathbf{k})}{\left(\omega + \i 0^{+}\right)^2 - \omega_{\mathbf{k}}^2} = \frac{g_G(\mathbf{k})}{2 \omega_{\mathbf{k}}} \left(\frac{1}{\omega - \omega_{\mathbf{k}} + \i 0^{+}} - \frac{1}{\omega + \omega_{\mathbf{k}} + \i 0^{+}} \right),
\end{equation}
with $k = (\mathbf{k}, \omega)$ and $\omega_{\mathbf{k}} = \mathbf{k}^2/2m$. Recalling that
\begin{equation}
\lim_{\epsilon \to 0} \frac{1}{x + \i \epsilon} = -\lim_{\epsilon \to 0} \frac{\epsilon}{x^2 + \epsilon^2} = -\pi \delta(x)
\end{equation}
and using the relation between spectral and retarded Green functions, one gets
\begin{equation}
\label{eq:spectralfree}
\tilde{\rho}(\mathbf{k},\omega) = \frac{\pi g_{G}(\mathbf{k})}{\omega_{\mathbf{k}}} \left[\delta (\omega - \omega_{\mathbf{k}}) - \delta (\omega + \omega_{\mathbf{k}})\right].
\end{equation}
It should be emphasized that naively it might seem that \eqref{eq:spectralfree} violates the sum rule. We note, however, that the prefactor suggests that the sum rule has to be enhanced with a prefactor $2 k^0/g_{G}(\mathbf{k})$ prefactor, similar to relativistic systems. This is further corroborated by \eqref{eq:LHS} and the commutation relation of the phase field. 

Finally, we substitute this into \eqref{eq:almostboltzmann} and integrate over $\int_0^{\infty} \frac{d k^0}{2 \pi}$. The left-hand side then reads as
\begin{equation}
\label{eq:LHS}
\text{LHS} = \partial_t \int_0^{\infty} \frac{d k^0}{2 \pi} \frac{2 k^0}{g_G(\mathbf{k})} \frac{\pi g_{G}(\mathbf{k})}{\omega_{\mathbf{k}}} \left[\delta (\omega - \omega_{\mathbf{k}}) - \delta (\omega + \omega_{\mathbf{k}})\right] f(k,t) = \partial_t f(\mathbf{k},t),
\end{equation}
where we have defined the occupation number distribution function as
\begin{equation}
f(\mathbf{k},t) \equiv f(\mathbf{k},\omega_{\mathbf{k}},t),
\end{equation}
This definition is motivated by the equilibrium limit of \eqref{eq:callenwelton}, for which $f(X^0,k)$ becomes a thermal distribution function. The right-hand side is a bit more elaborate, yet very straightforward as well. First, we notice that, since integration over $p$, $q$, and $r$ goes from $-\infty$ to $+\infty$, both $\delta$-functions in \eqref{eq:spectralfree} contribute, and we thus have to deal with $2^3 = 8$ different terms. In order to analyze them, it is convenient to split the frequency integrals, e.g., $\int_{-\infty}^{\infty} d p^0 \ldots =  \int_{-\infty}^0 d p^0 \ldots + \int_0^{\infty} d p^0 \ldots$ with $p \to -p$ variable change for the negative part and then use $f(\mathbf{p},-p^0,t) = -\left[f(\mathbf{p},p^0,t) + 1 \right]$. For completeness, let us consider one term explicitly:
\begin{align}
\label{eq:oneterm}
-&\int_{\mathbf{p},\mathbf{q},\mathbf{r}}\int_0^{\infty} d k^0 \int_0^{\infty} \frac{d p^0}{2\pi} 
\int_{-\infty}^0 \frac{d q^0}{2\pi} \int_0^{\infty} \frac{d r^0}{2\pi} |\lambda(\mathbf{k},\mathbf{p},-\mathbf{q},-\mathbf{r})|^2 \frac{\pi^4 g_{G}(\mathbf{k}) g_{G}(\mathbf{p}) g_{G}(\mathbf{q}) g_{G}(\mathbf{r})}{\omega_{\mathbf{k}} \omega_{\mathbf{p}} \omega_{\mathbf{q}} \omega_{\mathbf{r}}} \delta(k^0 - \omega_{\mathbf{k}}) \delta(p^0 - \omega_{\mathbf{p}}) \nonumber\\
&\hspace{65pt}\times
\delta(q^0 + \omega_{\mathbf{q}}) \delta(r^0 - \omega_{\mathbf{r}}) 
(2 \pi)^{d+1}\,\delta (k + p - q - r) \big[(f_k + 1)(f_p + 1) f_q f_r - f_k f_p (f_q + 1) (f_r + 1) \big] \nonumber\\
&= 
\int_{\mathbf{p},\mathbf{q},\mathbf{r}} 
|\lambda(\mathbf{k},\mathbf{p},-\mathbf{q},-\mathbf{r})|^2 
\frac{g_{G}(\mathbf{k}) g_{G}(\mathbf{p}) g_{G}(\mathbf{q}) g_{G}(\mathbf{r})}{2 \omega_{\mathbf{k}} 2\omega_{\mathbf{p}} 2 \omega_{\mathbf{q}} 2\omega_{\mathbf{r}}} 
(2 \pi)^{d+1}\,\delta (\omega_{\mathbf{k}} + \omega_{\mathbf{p}} + \omega_{\mathbf{q}} - \omega_{\mathbf{r}}) \delta(\mathbf{k} + \mathbf{p} - \mathbf{q} - \mathbf{r}) \nonumber\\
&\hspace{110pt}\times
\big[(f(\mathbf{k},t) + 1)(f(\mathbf{p},t) + 1) (f(\mathbf{q},t) + 1) f(\mathbf{r},t) - f(\mathbf{k},t) f(\mathbf{p},t) f(\mathbf{q},t) (f(\mathbf{r},t) + 1)  \big],
\end{align}
where the minus sign in the first line comes from the one in front of $\delta(q^0 + \omega_{\mathbf{q}})$. Finally, we assume that $f(t,\mathbf{p})$ has no explicit momentum dependence, i.e., it only depends on $\mathbf{p}$ via $\omega_{\mathbf{p}}$  (cf. \cite{Chantesana:2018qsb.PhysRevA.99.043620}), and change the integration variable $\mathbf{p} \to -\mathbf{p}$:
\begin{align}
\eqref{eq:oneterm} 
=
\int_{\mathbf{p},\mathbf{q},\mathbf{r}} &
|\lambda(\mathbf{k},\mathbf{p},\mathbf{q},-\mathbf{r})|^2 \frac{g_{G}(\mathbf{k}) g_{G}(\mathbf{p}) g_{G}(\mathbf{q}) g_{G}(\mathbf{r})}{2 \omega_{\mathbf{k}} 2\omega_{\mathbf{p}} 2 \omega_{\mathbf{q}} 2\omega_{\mathbf{r}}} (2 \pi)^{d+1}\,\delta (\omega_{\mathbf{k}} + \omega_{\mathbf{p}} + \omega_{\mathbf{q}} - \omega_{\mathbf{r}}) \delta(\mathbf{k} + \mathbf{p} + \mathbf{q} - \mathbf{r}) \nonumber\\
&\times
\big[(f_{\mathbf{k}} + 1)(f_{\mathbf{p}} + 1) (f_{\mathbf{q}} + 1) f_{\mathbf{r}} - f_{\mathbf{k}} f_{\mathbf{p}} f_{\mathbf{q}} (f_{\mathbf{r}} + 1)  \big],
\end{align}
where we have used that $g_{G}(\mathbf{p})$ and $\omega_{\mathbf{p}}$ are even functions and adopted the short-hand notation $f(\mathbf{k},t) \equiv f_{\mathbf{k}}$. 

The remaining terms can be computed following exactly the same procedure. With that we conclude:
\begin{equation}
\label{eq:boltzmann}
\partial_t f_{\mathbf{k}} = I[f](\mathbf{k},t) = I_{3}[f](\mathbf{k},t) + I_{4}[f](\mathbf{k},t),
\end{equation}
with
\begin{equation}
\label{eq:4vertex1}
I_4[f](\mathbf{k},t) = 
I^{2 \leftrightarrow 2}[f](\mathbf{k},t) + 
I^{3 \leftrightarrow 1}[f](\mathbf{k},t) + 
I^{1 \leftrightarrow 3}[f](\mathbf{k},t) +
I^{4 \leftrightarrow 0}[f](\mathbf{k},t),
\end{equation}
where
\begin{subequations}
\label{eq:4vertex2}
\begin{align}
I^{2 \leftrightarrow 2}[f](\mathbf{k},t) 
=
\int_{\mathbf{p},\mathbf{q},\mathbf{r}}& 
|\lambda(\mathbf{k},\mathbf{p},-\mathbf{q},-\mathbf{r})|^2 \frac{g_{G}(\mathbf{k}) g_{G}(\mathbf{p}) g_{G}(\mathbf{q}) g_{G}(\mathbf{r})}{2 \omega_{\mathbf{k}} 2\omega_{\mathbf{p}} 2 \omega_{\mathbf{q}} 2\omega_{\mathbf{r}}} (2 \pi)^{d+1}\,\delta (\omega_{\mathbf{k}} + \omega_{\mathbf{p}} - \omega_{\mathbf{q}} - \omega_{\mathbf{r}}) \delta(\mathbf{k} + \mathbf{p} - \mathbf{q} - \mathbf{r}) \nonumber\\
&\times
\big[(f_{\mathbf{k}} + 1)(f_{\mathbf{p}} + 1) f_{\mathbf{q}} f_{\mathbf{r}} - f_{\mathbf{k}} f_{\mathbf{p}} (f_{\mathbf{q}} + 1) (f_{\mathbf{r}} + 1)  \big] +
(\mathbf{p} \to -\mathbf{p}, -\mathbf{q} \to\mathbf{q}) + (\mathbf{p} \to -\mathbf{p}, -\mathbf{r} \to\mathbf{r}),
\end{align}
\begin{align}
I^{3 \leftrightarrow 1}[f](\mathbf{k},t) 
=
\int_{\mathbf{p},\mathbf{q},\mathbf{r}}&
|\lambda(\mathbf{k},\mathbf{p},\mathbf{q},-\mathbf{r})|^2 \frac{g_{G}(\mathbf{k}) g_{G}(\mathbf{p}) g_{G}(\mathbf{q}) g_{G}(\mathbf{r})}{2 \omega_{\mathbf{k}} 2\omega_{\mathbf{p}} 2 \omega_{\mathbf{q}} 2\omega_{\mathbf{r}}}  (2 \pi)^{d+1}\,\delta (\omega_{\mathbf{k}} + \omega_{\mathbf{p}} + \omega_{\mathbf{q}} - \omega_{\mathbf{r}}) \delta(\mathbf{k} + \mathbf{p} + \mathbf{q} - \mathbf{r}) \nonumber\\
&\times
\big[(f_{\mathbf{k}} + 1)(f_{\mathbf{p}} + 1) (f_{\mathbf{q}} + 1) f_{\mathbf{r}} - f_{\mathbf{k}} f_{\mathbf{p}} f_{\mathbf{q}} (f_{\mathbf{r}} + 1)  \big] +
(\mathbf{p} \to -\mathbf{p}, -\mathbf{r} \to \mathbf{r}) + (\mathbf{q} \to -\mathbf{q}, -\mathbf{r} \to\mathbf{r}),
\end{align}
\begin{align}
I^{1 \leftrightarrow 3}[f](\mathbf{k},t) 
=
\int_{\mathbf{p},\mathbf{q},\mathbf{r}} &
|\lambda(\mathbf{k},-\mathbf{p},-\mathbf{q},-\mathbf{r})|^2 \frac{g_{G}(\mathbf{k}) g_{G}(\mathbf{p}) g_{G}(\mathbf{q}) g_{G}(\mathbf{r})}{2 \omega_{\mathbf{k}} 2\omega_{\mathbf{p}} 2 \omega_{\mathbf{q}} 2\omega_{\mathbf{r}}} (2 \pi)^{d+1}\,\delta (\omega_{\mathbf{k}} - \omega_{\mathbf{p}} - \omega_{\mathbf{q}} - \omega_{\mathbf{r}}) \delta(\mathbf{k} - \mathbf{p} - \mathbf{q} - \mathbf{r}) \nonumber\\
&\times
\big[(f_{\mathbf{k}} + 1)f_{\mathbf{p}} f_{\mathbf{q}} f_{\mathbf{r}} - f_{\mathbf{k}} (f_{\mathbf{p}} + 1) (f_{\mathbf{q}} + 1) (f_{\mathbf{r}} + 1)  \big],
\end{align}
\begin{align}
I^{4 \leftrightarrow 0}[f](\mathbf{k},t)
=
\int_{\mathbf{p},\mathbf{q},\mathbf{r}}& 
|\lambda(\mathbf{k},\mathbf{p},\mathbf{q},\mathbf{r})|^2 \frac{g_{G}(\mathbf{k}) g_{G}(\mathbf{p}) g_{G}(\mathbf{q}) g_{G}(\mathbf{r})}{2 \omega_{\mathbf{k}} 2\omega_{\mathbf{p}} 2 \omega_{\mathbf{q}} 2\omega_{\mathbf{r}}} (2 \pi)^{d+1}\,\delta (\omega_{\mathbf{k}} + \omega_{\mathbf{p}} + \omega_{\mathbf{q}} + \omega_{\mathbf{r}}) \delta(\mathbf{k} + \mathbf{p} + \mathbf{q} + \mathbf{r}) \nonumber\\
&\times
\big[(f_{\mathbf{k}} + 1)(f_{\mathbf{p}} + 1) (f_{\mathbf{q}} + 1) (f_{\mathbf{r}} + 1) - f_{\mathbf{k}} f_{\mathbf{p}} f_{\mathbf{q}} f_{\mathbf{r}}  \big].
\end{align}
\end{subequations}
Above, $\mathbf{p} \leftrightarrow -\mathbf{p}$ means that the same term has to be taken but with the change of sign for $\mathbf{p}$ and $f_{\mathbf{p}} \leftrightarrow f_{\mathbf{p}} + 1$ and $\lambda(\mathbf{k},\mathbf{p},\mathbf{q},\mathbf{r})$ is the momentum-dependent $4$-coupling constant of the action \eqref{eq:Seff4}:  
\begin{equation}
\lambda(\mathbf{k},\mathbf{p},\mathbf{q},\mathbf{r}) = \frac{(\mathbf{k} \cdot \mathbf{p}) (\mathbf{q} \cdot \mathbf{r})}{2 m^2 g_{G}(\mathbf{k} - \mathbf{p})} + \text{perm}^s.
\end{equation}
Similarly, one can derive the $3$-vertex collision integral, which in $2$-loop order reads:
\begin{equation}
\label{eq:3vertex1}
I_{3}[f](\mathbf{k},t) = 
I^{1 \leftrightarrow 2}[f](\mathbf{k},t) + 
I^{2 \leftrightarrow 1}[f](\mathbf{k},t) +  
I^{3 \leftrightarrow 0}[f](\mathbf{k},t),
\end{equation}
with 
\begin{subequations}
\label{eq:3vertex2}
\begin{align}
I^{1 \leftrightarrow 2}[f](\mathbf{k},t) 
&=
\int_{\mathbf{p},\mathbf{q}} 
|\gamma(\mathbf{k},-\mathbf{p},-\mathbf{q})|^2 \frac{g_{G}(\mathbf{k}) g_{G}(\mathbf{p}) g_{G}(\mathbf{q})}{2 \omega_{\mathbf{k}} 2\omega_{\mathbf{p}} 2 \omega_{\mathbf{q}}} (2 \pi)^{d+1}\,\delta (\omega_{\mathbf{k}} - \omega_{\mathbf{p}} - \omega_{\mathbf{q}}) \delta(\mathbf{k} - \mathbf{p} - \mathbf{q}) \nonumber\\
&\times 
\big[(f_{\mathbf{k}} + 1)f_{\mathbf{p}} f_{\mathbf{q}} - f_{\mathbf{k}} (f_{\mathbf{p}} + 1) (f_{\mathbf{q}} + 1)  \big],\\
I^{2 \leftrightarrow 1}[f](\mathbf{k},t) 
&=
\int_{\mathbf{p},\mathbf{q}} 
|\gamma(\mathbf{k},\mathbf{p},-\mathbf{q})|^2 \frac{g_{G}(\mathbf{k}) g_{G}(\mathbf{p}) g_{G}(\mathbf{q})}{2 \omega_{\mathbf{k}} 2\omega_{\mathbf{p}} 2 \omega_{\mathbf{q}}} (2 \pi)^{d+1}\,\delta (\omega_{\mathbf{k}} + \omega_{\mathbf{p}} - \omega_{\mathbf{q}}) \delta(\mathbf{k} + \mathbf{p} - \mathbf{q}) \nonumber\\
&\times 
\big[(f_{\mathbf{k}} + 1)(f_{\mathbf{p}} + 1) f_{\mathbf{q}} - f_{\mathbf{k}} f_{\mathbf{p}} (f_{\mathbf{q}} + 1)  \big] + (\mathbf{p} \to -\mathbf{p}, -\mathbf{q} \to \mathbf{q}),\\
I^{3 \leftrightarrow 0}[f](\mathbf{k},t)
&=
\int_{\mathbf{p},\mathbf{q}} 
|\gamma(\mathbf{k},\mathbf{p},\mathbf{q})|^2 \frac{g_{G}(\mathbf{k}) g_{G}(\mathbf{p}) g_{G}(\mathbf{q})}{2 \omega_{\mathbf{k}} 2\omega_{\mathbf{p}} 2 \omega_{\mathbf{q}}} (2 \pi)^{d+1}\,\delta (\omega_{\mathbf{k}} + \omega_{\mathbf{p}} + \omega_{\mathbf{q}}) \delta(\mathbf{k} + \mathbf{p} + \mathbf{q})  \nonumber\\
&\times 
\big[(f_{\mathbf{k}} + 1)(f_{\mathbf{p}} + 1) (f_{\mathbf{q}} + 1) - f_{\mathbf{k}} f_{\mathbf{p}} f_{\mathbf{q}}  \big].
\end{align}
\end{subequations}
Here, $|\gamma(\mathbf{k},\mathbf{p},\mathbf{q})|^2$ should be understood on-shell, i.e.,
\begin{equation}
\gamma(\mathbf{k},\mathbf{p},\mathbf{q}) = \frac{(\mathbf{k} \cdot \mathbf{p}) \omega_{\mathbf{q}}}{m g_G(\mathbf{q})} + \text{perm}^s.
\end{equation}

\section{Canonical scaling of the action}
\label{app:ActionScaling}
Let us determine the scaling dimension of each term of the effective action \eqref{eq:Seff4}. For instance, consider the quadratic part:
\begin{align}
S^{(2)}\left(s^{-1/\beta} t \right) 
&=
\int_{t_{\mathrm{ref}},\mathcal{C}}^{s^{-1/\beta}t} d t' \int \frac{d^d k}{(2\pi)^d} \, \theta_a(\mathbf{k},t') \i D_{a}^{-1}(\mathbf{k},t') \theta_a(-\mathbf{k},t')  \nonumber\\
&=
s^{-1/\beta + d} \int_{s^{1/\beta} t_{\mathrm{ref}},\mathcal{C}}^{t} d \tilde{t} \int \frac{d^d k'}{(2\pi)^d} \, \theta_a(s\mathbf{k}',s^{-1/\beta}\tilde{t}) i D_{a}^{-1}(s\mathbf{k}',s^{-1/\beta}\tilde{t}) \theta_a(-s\mathbf{k}',s^{-1/\beta}\tilde{t})\nonumber\\
&\approx
s^{-1/\beta + d - \alpha/\beta - \gamma + 2z}
\int_{t_{\mathrm{ref}},\mathcal{C}}^{t} d \tilde{t} \int \frac{d^d k}{(2\pi)^d} \, \theta_a(\mathbf{k},\tilde{t}) i D_{a}^{-1}(\mathbf{k},\tilde{t}) \theta_a(-\mathbf{k},\tilde{t}) = s^{d_{S,2}} S^{(2)}(t),  
\end{align}
\twocolumngrid
which yields exactly \eqref{eq:canScDimS2}. Here, in the last line we assumed that $t_{\mathrm{ref}}$ is sent to the remote past, such that the Schwinger-Keldysh integral is almost invariant under $t_{\mathrm{ref}} \to s^{1/\beta}t_{\mathrm{ref}}$, and used $\alpha = d \beta$.   

Repeating the same steps for interaction terms gives
\begin{equation}
S^{(3)}\left(s^{-1/\beta} t\right) \approx
s^{- 1/\beta + 2d - 3 \alpha/2\beta + z + 2 - \gamma} S^{(3)}(t) = s^{d_{S,3}} S^{(3)}(t)
\end{equation}
and
\begin{equation}
S^{(4)}\left(s^{-1/\beta} t\right) \approx
s^{-1/\beta + 3d - 2\alpha/\beta + 4 - \gamma} S^{(4)}(t) = s^{d_{S,4}} S^{(4)}(t),
\end{equation}
which results in \eqref{eq:canScDimS3} and \eqref{eq:canScDimS4}, respectively. It should be remarked that terms involving derivatives with respect to time lead to ambiguities. Indeed, on one hand, they should be rescaled with $t \to s^{-1/\beta} t$ following the definition \eqref{eq:dSdefinition}. On the other hand, the on-shell scaling of the frequency implies $t \to s^{-z} t$. Nevertheless, since in our case $z = 1/\beta$, this gives the same result. 

\section{Hydrodynamic decomposition}
\label{app:HydroDecomp}
In this Appendix, we provide the definition of the hydrodynamic decomposition used to obtain the data shown in \Fig{HydroDecompEvo}.

The three-component system is described, in a hydrodynamic formulation \cite{Yukawa2012a.PhysRevA.86.063614}, by the total density $\rho$, the spin vector $f_{\mu}$ and the nematic tensor $n_{\mu\nu}$:
\begin{align}
  \rho 
  &=  \sum_a \Phi_a^{\dagger}  \Phi_a\,
  \\
  f_{\mu} 
  &= \frac{1}{\rho} \sum_{a, a^\prime} \Phi_a^{\dagger} \left( \mathrm{f}_{\mu} \right)_{a a^\prime} \Phi_{a^\prime} ,    
  \label{eq:spinop}
  \\
  n_{\mu \nu} 
  &= \frac{1}{\rho} \sum_{a, a^\prime} \Phi_a^{\dagger} \left( \mathrm{n}_{\mu \nu} \right)_{a  a^\prime} \Phi_{a^\prime},
  \label{eq:nemop}
\end{align}
$\mu=x,y,z$, with $\mathrm{f}_{\mu}$ being the spin-1 matrices in the fundamental representation, and the nematic or quadrupole tensor representation $\mathrm{n}_{\mu \nu} =  ( \mathrm{f}_{\mu} \mathrm{f}_{\nu} + \mathrm{f}_{\nu} \mathrm{f}_{\mu} )/2$.
The above densities are related, via continuity equations \cite{Yukawa2012a.PhysRevA.86.063614}, to the mass current $\mathbf{j}$ defining the superfluid velocity field $\mathbf{v}$, as well as to the spin and nematic currents, $\mathbf{j}^{(s)}_{\mu}$ and $\mathbf{j}^{(n)}_{\mu\nu}$, respectively: 
\begin{align}
  \mathbf{j} 
  &=\rho\mathbf{v}
  = \frac{1}{2 m\,\i} \sum_a\left[ \Phi_a^{\dagger}  \left( \nabla \Phi_a \right) 
                                                        - \left(\nabla \Phi_a^{\dagger} \right) \Phi_a \right],
  \\
  \mathbf{j}^{(s)}_{\mu} 
  &= \frac{1}{2 m\,\i} \sum_{a,a'}\left( \mathrm{f}_{\mu} \right)_{a a^\prime}
                                      \left[ \Phi_a^{\dagger}  \left( \nabla \Phi_{a'} \right) 
                                            - \left(\nabla \Phi_a^{\dagger} \right) \Phi_{a'} \right],
  \\
  \mathbf{j}^{(n)}_{\mu\nu} 
  &= \frac{1}{2 m\,\i} \sum_{a,a'}\left( \mathrm{n}_{\mu\nu} \right)_{a a^\prime}
                                         \left[ \Phi_a^{\dagger}  \left( \nabla \Phi_{a'} \right) 
                                               - \left(\nabla \Phi_a^{\dagger} \right) \Phi_{a'} \right].
\end{align}
Note that the nematic tensor vanishes for a fully polarized (ferromagnetic) gas but becomes important in the situations considered in this work.
The spin and nematic currents can be expressed in terms of the hydrodynamic variables $\rho$, $f_{\mu}$, $n_{\mu\nu}$, and $\mathbf{v}$, as well as the gradients $\nabla f_{\mu}$ and $\nabla n_{\mu\nu}$. 
Furthermore, diagonalization of $f_{\mu}$ and $n_{\mu\nu}$ shows that the six independent variables are $\rho$, the total phase $\theta$ which defines the fluid momentum $m\mathbf{v}=\nabla\theta$, as well as three Euler angles and a polarization angle \cite{Yukawa2012a.PhysRevA.86.063614}.

For expressing the hydrodynamic energy it is useful to define the generalized velocities, corresponding to the quantum-pressure (q), the spin (s), the nematic (n), the incompressible (i), and compressible (c) parts:
\begin{align}
\vec{w}^{(q)} &= m^{-1}\nabla \sqrt{\rho},
&\vec{w}^{(i,c)} =&\ \sqrt{\rho}\, \vec{v}^{(i,c)},
\nonumber\\
\vec{w}_{\mu}^{(s)} &= (2m)^{-1}{\sqrt{\rho}}\,\nabla f_{\mu},
&\vec{w}_{\mu \nu}^{(n)} =&\ (2m)^{-1}{\sqrt{2\rho}}\,\nabla n_{\mu \nu}.
\end{align}
Here, $\vec{v}^{(i,c)}$ are obtained by a Helmholtz decomposition of $\vec{v} = \vec{v}^{(i)} + \vec{v}^{(c)}$, with the incompressible part having a vanishing divergence,  $\nabla \cdot\vec{v}^{(i)} = 0$, the compressible part a vanishing curl, $\nabla \times\vec{v}^{(c)} = 0$.

Using the hydrodynamic variables we can express the energy as 
\begin{equation}
E = E_\mathrm{kin}
+ \frac{g}{2} \int_{\vec{r}} \rho^2 ,
\end{equation}
where the kinetic part  reads as
\begin{equation}
E_\mathrm{kin} 
= \int_{\vec{r}} \left \{ \frac{1}{2 m} \left[ 
\left( \nabla \sqrt{\rho} \right)^2
+ \frac{\rho}{4} \left( \nabla f_{\mu} \right)^2 
+ \frac{\rho}{2} \left( \nabla n_{\mu \nu} \right)^2 
\right] 
+ \frac{m}{2}  \rho \vec{v}^2 \right\}.
\end{equation}
Hence, in Fourier space, the kinetic-energy spectrum is given by the correlation functions of the generalized velocities
\begin{equation}
\varepsilon_\mathrm{kin}(k) = \varepsilon^{(q)} (k) + \varepsilon^{(c)} (k) + \varepsilon^{(i)} (k)+ \varepsilon^{(s)} (k) + \varepsilon^{(n)} (k),
\end{equation}
averaged over the orientation of the momentum vector: 
\begin{align}
\varepsilon^{(\delta)} (k) & = \frac{m}{2} \int d \Omega_{\mathbf{k}} \langle \lvert  \vec{w}^{(\delta)} (\mathbf{k}) \rvert^2 \rangle,\quad (\delta=q,i,c)
\label{eq:epsq}
\\
\varepsilon^{(s)} (k)  &= \frac{m}{2} \int d \Omega_{\mathbf{k}} \langle   \vec{w}_{\mu} ^{(s)} (\mathbf{k}) \cdot \vec{w}_{\mu} ^{(s)} (\mathbf{k})  \rangle,
\label{eq:epss}
\\
\varepsilon^{(n)} (k)  &= \frac{m}{2} \int d \Omega_{\mathbf{k}} \langle   \vec{w}_{\mu \nu} ^{(n)} (\mathbf{k}) \cdot \vec{w}_{\mu \nu} ^{(n)} (\mathbf{k})  \rangle,
\label{eq:epsn}
\end{align}
where Einstein's sum convention is implied.
The respective total energies are obtained as $E^{(\delta)}=\int d^{d}x\int dk\, k^{d-1}\varepsilon^{(\delta)}(k)$.
The spectrum of the kinetic energy can then be used to calculate corresponding occupation numbers using the relation
\begin{equation}
n^{(\delta)} (k)  = 2m\,k^{-2} \varepsilon^ {(\delta)} (k),
\end{equation}
where $\delta = q, i, c, s, n$.
The total occupation number is then approximately given by 
\begin{equation}
n_\mathrm{tot} (k)  \approx 2m\,k^{-2} \varepsilon_\mathrm{kin}  (k)
\end{equation}
(see, e.g., \cite{Nowak:2011sk} for a discussion of deviations).

We emphasize that the spin $f_{\mu}$ and quadrupole operators $n_{\mu\nu}$ form $\mathrm{SU}(2)$ subgroups of the special unitary group $\mathrm{SU}(3)$, i.e., not only $\{f_{x},f_{y},f_{z}\}$ form an $\mathrm{su}(2)$ algebra of generators of $\mathrm{SU}(2)$ but also, e.g., $\{f_{x},n_{yz},n_{zz}-n_{yy}\}$ and $\{n_{xz},n_{yz},f_{z}\}$ \cite{Di2010a,Hamley2012a.NatPh.8.305H}.
On the basis of the field representations \eq{spinop} and \eq{nemop}, one can furthermore show that for the case of identical near-constant bosonic densities $\rho_{a}^{(0)}$ and comparatively large phase fluctuations, the fluctuations \eq{epss} and \eq{epsn} of the spin and nematic operators are determined by combinations of the relative phases between the Bose fields.


\end{appendix}
\twocolumngrid

\bibliographystyle{apsrev4-1}

%

\end{document}